\documentclass[pra,notitlepage,superscriptaddress,longbibliography,reprint]{revtex4-1}
\usepackage[english]{babel}
\usepackage{amsmath,amssymb,bbm,mathrsfs,bm,braket,graphicx,comment}
\usepackage[usenames]{xcolor}
\usepackage[colorlinks,linkcolor=blue,citecolor=blue,urlcolor=blue]{hyperref}

\allowdisplaybreaks

\newcommand{\abs}[1]{\left\lvert #1 \right\rvert}

\newcommand{\unitvec}[1]{\hat{\mathbf{#1}}}

\begin{document}

\title{Topological Defects in Anisotropic Driven Open Systems}






\author{L. M. Sieberer}
\email{lukas.sieberer@gmail.com}

\affiliation{Department of Physics, University of California, Berkeley,
  California 94720, USA}

\affiliation{Institute for Theoretical Physics, University of Innsbruck, A-6020
  Innsbruck, Austria}

\affiliation{Institute for Quantum Optics and Quantum Information of the
  Austrian Academy of Sciences, A-6020 Innsbruck, Austria}

\author{E. Altman}

\affiliation{Department of Physics, University of California, Berkeley,
  California 94720, USA}

\begin{abstract}
  We study the dynamics and unbinding transition of vortices in the compact
  anisotropic Kardar-Parisi-Zhang (KPZ) equation. The combination of
  non-equilibrium conditions and strong spatial anisotropy drastically affects
  the structure of vortices and amplifies their mutual binding forces, thus
  stabilizing the ordered phase. We find novel universal critical behavior in
  the vortex-unbinding crossover in finite-size systems. These results are
  relevant for a wide variety of physical systems, ranging from strongly coupled
  light-matter quantum systems to dissipative time crystals.
\end{abstract}

\maketitle

\textit{Introduction.---}The celebrated theory of Kosterlitz and Thouless (KT)
highlights the crucial role that is played by topological defects in the phase
transition of U(1)-symmetric and short-range interacting two-dimensional (2D)
systems in thermal equilibrium. At low temperatures, topological defects
(vortices) of opposite charge form tightly bound pairs, while they are free to
roam and destroy order at high temperatures. For the stability of the ordered
phase, it is crucial that vortices interact like charged particles, i.e., with a
Coulomb force that decays as $\sim 1/r$. In particular, any faster decay at
large distances would destabilize the ordered phase.

Interestingly, such a qualitative change of the vortex interaction can be induced
by taking the system out of thermal equilibrium. This has been studied
extensively in the context of the complex Ginzburg-Landau equation
(CGLE)~\cite{Aranson2002} and the compact KPZ (cKPZ) equation~\cite{Aranson1998,
  Wachtel2016} (the former can be reduced to the latter in the long-wavelength
limit~\cite{Grinstein1993,Grinstein1996}). The extent to which these equations
violate equilibrium conditions can be quantified in terms of a single parameter
that determines the strength of the characteristic non-linearity in the cKPZ
equation~\cite{Sieberer2013,Sieberer2014,Altman2015,Keeling2016,Sieberer2016a}.
Due to this non-linear term, the vortex interaction is exponentially screened at
large distances --- thus, the ordered phase ceases to exist. This finding is
particularly relevant, since the cKPZ equation is the long-wavelength
description of a vast variety of systems, ranging from ``polar active smectics''
or ``moving stripes''~\cite{Chen2013}, to driven-dissipative condensates such as
exciton-polaritons~\cite{Altman2015, Gladilin2014, Ji2015, He2015, Keeling2016,
  Sieberer2016a, Wachtel2016, Sieberer2016b, He2017, He2017b, Zamora2017,
  Squizzato2017}, synchronization in arrays of limit-cycle
oscillators~\cite{Lauter2016}, and limit-cycle phases that emerge from a Hopf
bifurcation~\cite{Lee2011,Ludwig2013a,Jin2013,Chan2015,Schiro2016} --- such
phases have attracted a lot of attention recently and could be coined
dissipative time crystals~\cite{Yao2018}.

\begin{figure}
  \centering
  \includegraphics[width=\linewidth]{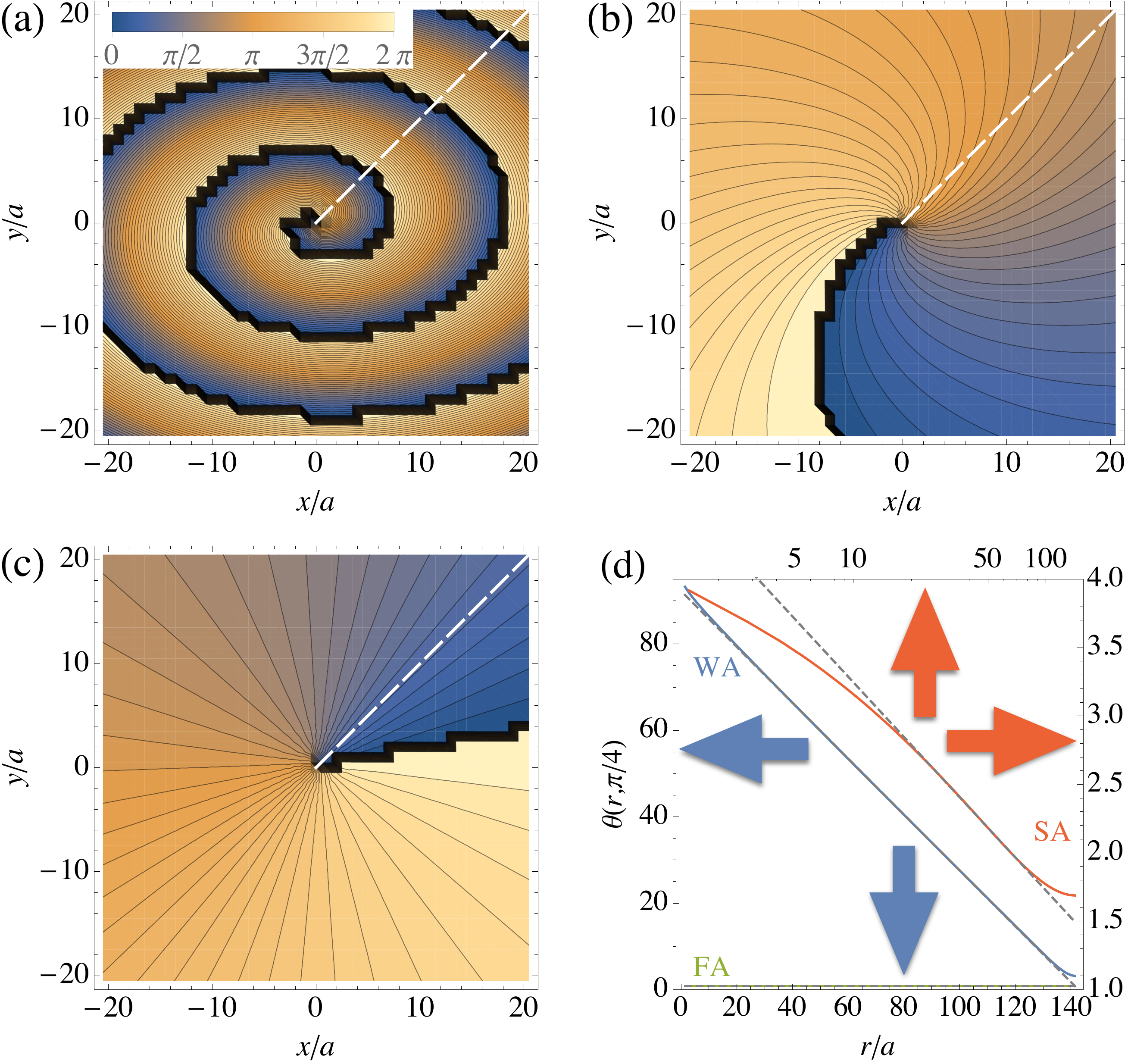}  
  \caption{(Color online) Single vortices in the caKPZ equation. (a) Weakly
    anisotropic (WA) regime with $\alpha_x = \lambda_x/(2D) \approx 0.9$ and
    $\alpha_y = \lambda_y/(2D) \approx 0.4$ in Eq.~\eqref{eq:vortex_caKPZ}. The
    vortex has a squeezed spiral structure with a clearly visible radially
    emitted wave. (b) Strongly anisotropic (SA), $\alpha_x \approx 0.9$,
    $\alpha_y \approx -0.4$.  The spiral structure is pronounced only at short
    distances from the vortex core. (c) In the fully anisotropic (FA) case, with
    $\alpha_x = -\alpha_y \approx 0.7$, there is no radial wave. We note, that
    (a-c) are equally non-linear in the sense that
    $\alpha_x^2 + \alpha_y^2 = 1$. (d) Radial dependence of the vortex field
    $\theta(r, \phi)$ along the dashed line in (a-c). $\theta(r,\pi/4)$ grows
    linearly in the WA regime (a), logarithmically in the the SA regime (b)
    (note the logarithmic $r$-axis), and is constant at the FA point. The dashed
    lines are linear fits which agree well with the data up to finite-size
    effects at large distances.}
  \label{fig:anisotropic_vortices}
\end{figure}

In this paper, we report that breaking \emph{rotational} symmetry has an equally
strong impact on the form of the vortex interaction, and acts to
\emph{stabilize} the ordered phase. This is highly significant for the systems
mentioned above, in which spatial anisotropy is either intrinsic or can be
imposed deliberately. The change in the vortex interaction can be understood
intuitively by considering the mere structure of a single topological defect
shown in Fig.~\ref{fig:anisotropic_vortices}. In the \emph{isotropic} cKPZ
equation, defects are ``radiative,'' i.e., they emit waves radially away from
the core, giving them a spiral structure
(Fig.~\ref{fig:anisotropic_vortices}(a)). Perturbations, e.g., due to the
presence of another defect, decay exponentially in the up-stream direction of
traveling waves --- heuristically, this explains why the interaction between
vortices is exponentially screened~\cite{Aranson1998}. As we show below, the
radially emitted wave decays away from the vortex core for sufficiently strong
anisotropy (Fig.~\ref{fig:anisotropic_vortices}(b)), and is completely absent in
a \emph{fully anisotropic} configuration
(Fig.~\ref{fig:anisotropic_vortices}(c)). Then, vortices in the
\emph{anisotropic} cKPZ (caKPZ) are similar to the ones in equilibrium systems
(up to an anisotropic deformation). We expect their interactions to be
long-range and, therefore, order to be stable, as also indicated by numerical
simulations~\cite{Grinstein1993, Grinstein1996}.

However, in a non-linear theory such as the caKPZ equation, single-vortex
solutions cannot simply be superposed to yield multi-vortex solutions, and in
the latter additional features may appear that cannot be anticipated from the
former. Below, we present an analytical calculation of the interaction between
defects, based on a recently developed mapping to a dual electrodynamics
problem~\cite{Sieberer2016b,Wachtel2016}. This perturbative calculation is valid
up to an exponentially large characteristic scale, and it shows that the
attraction between oppositely charged vortices is even enhanced as compared to
the linear (i.e., thermal equilibrium) isotropic case. For this enhancement to
occur, the combination of non-linearity and anisotropy is essential. In fact, as
explained above, the non-linearity alone would give a repulsive correction to
the interaction, while anisotropy in a linear theory can simply be absorbed in
an anisotropic rescaling of the units of length and does not have any relevant
qualitative effect on the form of the interaction.

Based on the modified vortex interaction, we derive renormalization group (RG)
equations that describe the vortex-unbinding crossover in systems that are fully
anisotropic and smaller than the characteristic scale. Since this scale is
parametrically large, we expect that the universal critical behavior we find
will be observed in experiments and numerical investigations of the caKPZ
equation. In particular, the divergence of the correlation length is in between
the essential singularity characteristic of the KT transition and true scaling
behavior as in usual continuous phase transitions.

Previous studies of the caKPZ equation~\cite{Chen2013, Altman2015, Keeling2016,
  Zamora2017} assumed that vortices do not proliferate on the characteristic
scale of the RG flow of the non-compact equation~\cite{Wolf1991, Chen2013}, and
can hence be included \textit{a posteriori} in an emergent equilibrium
description. This assumption, however, breaks down close to the unbinding
transition, where vortices are the dominant fluctuations. To access this region,
we focus on the combined impact of non-linearity and strong anisotropy on the
vortex dynamics.

\textit{Model.---}The caKPZ equation reads
\begin{equation}
  \label{eq:caKPZ}
  \partial_t \theta = \sum_{i = x, y} \left[ D_i \partial_i^2 \theta +
    \frac{\lambda_i}{2} \left( \partial_i \theta \right)^2 \right] + \eta,
\end{equation}
where $\eta$ is Gaussian noise with zero mean and correlations
$\left\langle \eta(\mathbf{r}, t) \eta(\mathbf{r}', t') \right\rangle = 2 \Delta
\delta(\mathbf{r} - \mathbf{r}') \delta(t - t')$.
$\theta$ is a compact variable, i.e., one that admits topological defects. In
physical realizations, $\theta$ may be the phase field in driven-open
condensates~\cite{Altman2015,Gladilin2014,Ji2015,He2015,Keeling2016,Sieberer2016a,Wachtel2016,Sieberer2016b,He2017,He2017b,Zamora2017},
in limit-cycle phases~\cite{Lee2011,Ludwig2013a,Jin2013,Chan2015,Schiro2016} or
in oscillator arrays. In the latter case, Eq.~\eqref{eq:caKPZ} emerges as the
continuum limit of the noisy Kuramoto-Sakaguchi
model~\cite{Lauter2016,Sakaguchi1986,Acebron2005} with anisotropic couplings
between the oscillators. $\theta$ may also represent the displacement field in
polar active smectics~\cite{Chen2013}.

For stability, we require $D_{x, y} > 0$, while
$\lambda_{x, y}$ are unrestricted; in the following, we set $D_x = D_y = D$,
which can always be achieved by an anisotropic rescaling of the units of
length. For $\lambda_x = \lambda_y = 0$, Eq.~\eqref{eq:caKPZ} reduces to the
(continuum limit of the) $XY$ model with dissipative dynamics~\footnote{Then,
  the stationary distribution of the field $\theta$ is given by the Gibbs weight
  $\mathcal{P} \sim e^{-\mathcal{H}_{XY}}$ with
  $\mathcal{H}_{XY} = \frac{D}{\Delta} \int d^2 \mathbf{r} \left( \nabla \theta
  \right)^2$,
  and the usual equilibrium KT theory applies.}. We denote $\lambda_{x, y}$
having the same and opposite signs as \textit{weakly anisotropic} (WA) and
\textit{strongly anisotropic} (SA) regimes, respectively. In particular, we
denote the configuration with $\lambda_x = - \lambda_y$ as \textit{fully
  anisotropic} (FA). We shall restrict ourselves to small values of
$\abs{\lambda_{x, y}}$ in order to avoid a dynamical instability of
Eq.~\eqref{eq:caKPZ}~\cite{Lauter2016,He2017}. In systems described by
Eq.~\eqref{eq:caKPZ} with $\lambda_x = \lambda_y$, the ordered phase is always
destroyed in the thermodynamic limit by the proliferation of
vortices~\cite{Aranson1998,Aranson1998b,Wachtel2016}. Here, we investigate
whether order can be stable if $\lambda_x \neq \lambda_y$.

\textit{Structure of a single vortex.---}A vortex is a solution of
Eq.~\eqref{eq:caKPZ} without noise that is stationary up to uniform oscillations
with a frequency $\omega_0$,
\begin{equation}
  \label{eq:vortex_caKPZ}
  \partial_t \theta = D \nabla^2 \theta + \frac{\lambda_x}{2} \left( \partial_x
    \theta \right)^2 + \frac{\lambda_y}{2} \left( \partial_y \theta \right)^2 -
  \omega_0 = 0,
\end{equation}
and obeys $\oint d \mathbf{l} \cdot \nabla \theta = 2 \pi$ for any integration
path that surrounds the vortex core. We solve Eq.~\eqref{eq:vortex_caKPZ}
numerically by discretizing it on a lattice (see~\cite{supplemental_material}
for details). In addition, we determine analytically the asymptotic behavior of
$\theta(r, \phi)$ for $r \to \infty$ at fixed polar angle $\phi$, which takes
the form
\begin{equation}
  \label{eq:vortex_asymptotic}
  \theta(r, \phi) = k_0(\phi) r + b(\phi) \ln(r/a) + \Phi(\phi) + O(1/r),
\end{equation}
where $k_0(\phi)$ is the (anisotropic) asymptotic wave number, $b(\phi)$ the
coefficient of the first sub-leading correction, and $\Phi(\phi)$ contains the
topological part of the vortex field. $a$ is a microscopic cutoff scale such as
the lattice spacing in oscillator arrays, or the healing length in driven-open
condensates. As illustrated in Fig.~\ref{fig:anisotropic_vortices}(a,d), in the
WA regime, we find a squeezed spiral structure with~\cite{supplemental_material}
\begin{equation}
  \label{eq:k_0}
  k_0(\phi) = - \sqrt{\frac{2 \omega_0}{\lambda_x \cos(\phi)^2 + \lambda_y \sin(\phi)^2}},
\end{equation}
where $\omega_0$ is determined by the regularization at short
distances~\cite{Aranson1998}. In the SA regime shown in
Fig.~\ref{fig:anisotropic_vortices}(b,d), when one of $\lambda_{x,y}$ is
negative, Eq.~\eqref{eq:k_0} implies $k_0(\phi) = \omega_0 = 0$. The leading
asymptotic behavior is then given by the logarithmic term in
Eq.~\eqref{eq:vortex_asymptotic} with $b(\phi) = b_0 = \mathrm{const.}$ Finally,
for FA parameters $\lambda_x = - \lambda_y$, also the coefficient $b_0$
vanishes. Indeed, then the exact solution takes the form
$\theta(r, \phi) = \Phi_0(\phi)$~\cite{supplemental_material}, and the absence
of any radial dependence is evident in
Fig.~\ref{fig:anisotropic_vortices}(c,d). For $\lambda_x = - \lambda_y \to 0$,
this solution smoothly deforms into an ``ordinary'' $XY$ vortex with
$\Phi_0(\phi) = \phi$, which is in stark contrast to the isotropic case, where
the transition from from the linear to the non-linear problem is highly
non-analytic. Since turning on the non-linearity in a fully anisotropic system
does not alter the \emph{radial dependence} of the far field of a single vortex,
we conclude that the interaction of vortices at large distances is not screened
as in the isotropic case, and thus the ordered phase is indeed stable in the
thermodynamic limit. It is an interesting open question how the logarithmic
dependence of the vortex field~\eqref{eq:vortex_asymptotic} in the SA regime
affects the interaction at asymptotic distances.

\textit{Electrodynamic duality and vortex interaction.---}The vortex interaction
can be calculated explicitly within a dual electrodynamic
formalism~\cite{Sieberer2016b,Wachtel2016}. This calculation treats the
non-linearity in Eq.~\eqref{eq:caKPZ} perturbatively and is valid up to a finite
but exponentially large scale we determine below. The duality defines the
electric field as $\mathbf{E} = - \unitvec{z} \times \nabla \theta$.  For the
overdamped dynamics described by Eq.~\eqref{eq:caKPZ}, fluctuations of the
magnetic field are gapped and can be integrated out. The basic equations are
then Gauss' law $\nabla \cdot \mathbf{E} = 2 \pi n/\varepsilon$, and
\begin{equation}
  \label{eq:dual_caKPZ}
  \varepsilon \partial_t \mathbf{E} = - D \nabla \times \left( \nabla \times
    \mathbf{E} \right) - 2 \pi \mathbf{j} - \unitvec{z} \times \nabla \Biggl(
  \sum_{i = x, y} \frac{\lambda_i}{2} E_i^2 + \eta \Biggr).
\end{equation}
The dielectric constant $\varepsilon$ accounts for screening of the electric
field due to fluctuations consisting of bound vortex pairs. It takes the
microscopic value $\varepsilon = 1$, and is renormalized upon coarse-graining as
described below. $n$ and $\mathbf{j}$ are the vortex density and current,
respectively, which obey the continuity equation
$\partial_t n = - \nabla \cdot \mathbf{j}$. For $n = \mathbf{j} = 0$,
Eq.~\eqref{eq:dual_caKPZ} reduces to the non-compact anisotropic KPZ
equation. The vortex density is controlled by the fugacity $y$. Close to the
putative unbinding transition $y \ll 1$, and we restrict ourselves to consider a
dipole,
$n(\mathbf{r}) = \sum_{\sigma = \pm} \sigma \delta(\mathbf{r} -
\mathbf{r}_{\sigma})$,
where $\sigma = \pm$ are the charges of the vortices. They are assumed to
undergo diffusive motion~\cite{Ambegaokar1978,Ambegaokar1980,Aranson1998b}
according to
\begin{equation}
  \label{eq:vortex_eom}
  \frac{d \mathbf{r}_{\sigma}}{d t} = \mu \sigma \mathbf{E}(\mathbf{r}_{\sigma}) +
  \boldsymbol{\xi}_{\sigma},
\end{equation}
and the correlations of the zero-mean Gaussian noise sources
$\boldsymbol{\xi}_{\sigma}$ are given by
$\langle \xi_{\sigma, i}(t) \xi_{\sigma', j}(t') \rangle = 2 \mu T
\delta_{\sigma \sigma'} \delta_{i j} \delta(t - t'),$
where the vortex ``temperature'' $T$ is related to the noise strength $\Delta$
in Eq.~\eqref{eq:caKPZ}~\cite{Wachtel2016}. The vortex mobility $\mu$ is
introduced phenomenologically, and we consider the limit of low mobility
$\mu \ll D$. Then, retardation effects due to the vortices' motion are
negligible, and $\mathbf{E}(\mathbf{r}_{\sigma})$ in Eq.~\eqref{eq:vortex_eom}
can be approximated by the instantaneous electrostatic field which is determined
by Eq.~\eqref{eq:dual_caKPZ} with
$\partial_t \mathbf{E} = \mathbf{j} = \eta = 0$~\cite{Wachtel2016}. Details of
this calculation are given in the Supplement~\cite{supplemental_material}, and
here we only point out key features of the solution. To address the possibility
of a bound state, we focus on the dynamics of the dipole moment
$\mathbf{r} = \sum_{\sigma = \pm} \sigma \mathbf{r}_{\sigma} = \mathbf{r}_+ -
\mathbf{r}_-$.
We parametrize the non-linearity as
$\alpha_{\pm} = (\lambda_x \pm \lambda_y)/(2 D)$, with $\alpha_- = 0$ and
$\alpha_+ = 0$ corresponding to isotropic and fully anisotropic systems,
respectively. The ``isotropic'' second-order correction $\propto \alpha_+^2$ was
obtained previously~\cite{Wachtel2016}. It is a central, conservative, and,
crucially, \emph{repulsive} force. In the ``anisotropic'' second-order
correction $\propto \alpha_-^2$, the leading contribution is also central and
conservative, but \emph{attractive}. Additionally, it features sub-leading terms
that cannot be derived from a potential. The ``mixed'' correction
$\propto \alpha_+ \alpha_-$ includes terms $\propto \left( x, - y \right)$ that
favor alignment of the dipole along one of the principal axes, in line with
numerical simulations of the anisotropic
CGLE~\cite{Faller1998,ROLANDFALLERandLORENZKRAMER1999}.

The isotropic, anisotropic, and mixed corrections are power series in logarithms
$\ln(r/a)$. Thus, perturbation theory breaks down at a scale
$L_v \sim a e^{1/\alpha_{\mathrm{max}}}$ where
$\alpha_{\mathrm{max}} = \max \{ \abs{\alpha_{\pm}} \}$. In weakly
out-of-equilibrium (and thus weakly non-linear) systems, the scale $L_v$ can
easily be much larger than any experimentally relevant system size. Then, we
expect the dynamics of vortices to be described by the perturbatively obtained
interaction. In the following, we discuss how the usual KT theory is modified
due to non-equilibrium conditions and anisotropy. We focus on the FA
configuration, where anisotropy has the most profound impact. Moreover, and as
discussed in detail in~\cite{supplemental_material}, a vast simplification
occurs in the FA case on scales
$r \ll L_T = a e^{\left( T/ \alpha_{\mathrm{max}} \right)^{1/3}}$: Then,
fluctuations of the orientation of the dipole lead to an angular averaging,
rendering the problem effectively isotropic. Strong anisotropy is nevertheless
manifest in the result of the angular average.

\textit{Vortex unbinding crossover.---}The noise in Eqs.~\eqref{eq:dual_caKPZ}
and~\eqref{eq:vortex_eom} creates pairs of vortices and antivortices which then
diffuse under the influence of their interaction and eventually recombine. Such
fluctuations on short scales between the microscopic cutoff $a$ and a running
cutoff scale $a e^{\ell}$ renormalize the parameters that enter an effective
description on larger scales. This is described by the following RG flow
equations~\cite{supplemental_material}:
\begin{equation}
  \label{eq:RG}
  \begin{aligned}
    \frac{d \varepsilon}{d \ell} & = \frac{2 \pi^2 y^2}{T}, & \frac{d y}{d \ell}
    & = \frac{1}{2} \left( 4 - \frac{1}{\varepsilon T} +
      \frac{c \alpha_-^2}{3 \varepsilon^2} \right) y, \\
    \frac{d T}{d \ell} & = \frac{c \alpha_-^2 T}{3 \varepsilon^2}, & \frac{d
      c}{d \ell} & = -3,
  \end{aligned}
\end{equation}
where $c$ is the running coefficient of the term $\propto \ln(r/a)^2$ in the
effective dipole distribution with microscopic value $c = 3/2$
(see~\cite{supplemental_material}; recall that also the microscopic value
$\varepsilon = 1$ is fixed). Integrating the flow equation for $c$ yields
$c = 3 (1 - 2 \ell)/2$, i.e., the logarithmic scale $\ell$ appears explicitly in
the flow equations for the remaining couplings. This again necessarily
invalidates the perturbative flow equations at large scales --- however, the
condition $r \ll L_v$, which we assumed in the derivation of the flow equations,
is always more stringent. Note also that the characteristic KPZ scale on which
the renormalization of the (suitably rescaled) non-linearity in
Eq.~\eqref{eq:caKPZ} due to non-topological fluctuations becomes substantial, is
generically much larger than $L_v, L_T$~\cite{Wachtel2016}. Hence, analyzing
Eqs.~\eqref{eq:RG} we can consider $\alpha_-$ as a fixed parameter.

\begin{figure}
  \centering
  \includegraphics[width=\linewidth]{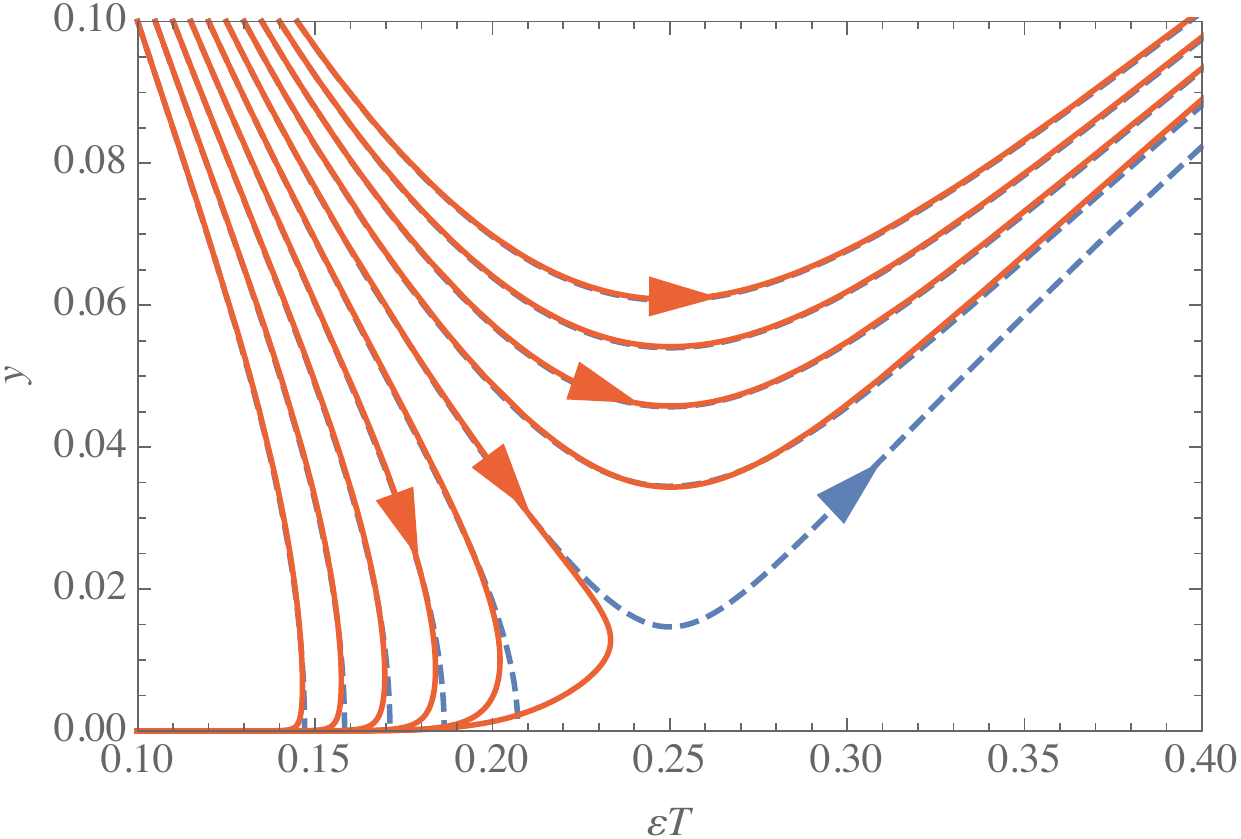}
  \caption{(Color online) RG flow~\eqref{eq:RG}. Dashed, blue: KT flow for
    $\alpha_-^2 = 0$; Solid, red: $\alpha_-^2 = 0.01$. There are two phases with
    $y, \varepsilon T \to 0$ ($\varepsilon T \to \mathrm{const.}$ for
    $\alpha_-^2 = 0$) and $y, \varepsilon T \to \infty$, respectively. For
    $\alpha_-^2 = 0.01$, the critical temperature is $T_c \approx 0.13$, which
    is slightly larger than the KT critical temperature. The microscopic value
    of the fugacity is chosen as $y = 0.1$, and the temperature is varied in the
    range $T = 0.1, \dotsc, 0.145$.}
  \label{fig:RG_flow}
\end{figure}

The RG flow is shown in Fig.~\ref{fig:RG_flow}. Remarkably, it is qualitatively
different from both the equilibrium KT flow and the RG flow in an isotropic
non-equilibrium system.
The most striking feature is the existence of a low-temperature phase in which
vortices remain bound and fluctuations are anomalously suppressed since both
$y, T \to 0$. In contrast, in isotropic systems vortices unbind at any finite
temperature~\cite{Wachtel2016}; the low-temperature ordered phase in thermal
equilibrium, on the other hand, is different in that $T$ is conserved by the RG
flow. The strong suppression of vortex fluctuations can be traced back to the
dominant correction to the vortex interaction being \emph{attractive} in the FA
case. Consequently, the fundamental difference to the flow equations for
isotropic systems in Ref.~\cite{Wachtel2016} is that here $c$ flows to
\emph{negative} values, and therefore the terms $\propto c$ in the equations for
$T$ and $y$ renormalize these quantities to \emph{lower} values, thus
antagonizing the unbinding of vortices. This leads to \emph{increased} stability
of the ordered phase as compared to the equilibrium case: The critical
temperature $T_c$ is higher for the same value of $y$. Heuristically, the lower
the probability for vortex pairs to be created at a microscopic scale, the
stronger noise-induced fluctuations the system can afford and still remain
ordered. This is true also in equilibrium, but here we found that for $y \to 0$
the critical temperature \emph{diverges} whereas it remains finite in KT theory.

While at low temperatures the flow $T \to 0$ is presumably cut at large scales
when the flow equations~\eqref{eq:RG} become invalid, at high temperatures the
rapid growth of $\varepsilon$, indicating the screening of vortex interactions,
stops the flow of $T$. Then, at larger scales, the flow in the disordered phase
is the same as in equilibrium~\cite{supplemental_material} (in particular,
$y, \varepsilon \to \infty$).

The existence of two distinct phases points to the existence of a fixed point
that controls critical behavior at the transition. Even if the ``true'' critical
behavior at the largest scales is not captured by the flow
equations~\eqref{eq:RG}, they still entail the finite-size criticality that is
observable up to parametrically large scales. However, in contrast to usual
continuous phase transitions, the flow equations~\eqref{eq:RG} cannot have
a true fixed point since $c$ grows steadily. A ``flowing fixed point'' can be
found by the change of variables $\tilde{\varepsilon} = \varepsilon/x$,
$\tilde{y} = \sqrt{x} y$, and $\tilde{T} = x T$, where $x = -c$ (note that $c <
0$ in the regime of interest at large $\ell$), which recasts the flow equations
as
\begin{equation}
  \label{eq:RG_x}
  \begin{split}
    \frac{d \tilde{\varepsilon}}{d x} & = \frac{1}{x} \left( \frac{2 \pi^2
        \tilde{y}^2}{3 \tilde{T}} - \tilde{\varepsilon} \right), \quad \frac{d
      \tilde{T}}{d x} = \frac{1}{3 x} \left( 3 - \frac{\alpha_-^2}{3
        \tilde{\varepsilon}^2} \right) \tilde{T}, \\ \frac{d \tilde{y}}{d x} & =
    \frac{1}{6} \left[ 4 - \frac{1}{\tilde{\varepsilon} \tilde{T}} + \frac{1}{x}
      \left( 3 - \frac{\alpha_-^2}{3 \tilde{\varepsilon}^2} \right) \right]
    \tilde{y}.
  \end{split}
\end{equation}
These equations have a fixed point at
$\tilde{\varepsilon}_{*} = \lvert \alpha_- \rvert/3,$
$\tilde{y}_{*} = \sqrt{3/8} (1/\pi),$
$\tilde{T}_{*} = 3/(4 \lvert \alpha_- \rvert)$. The existence of this fixed
point implies that the correlation length diverges at the transition: As $T$ is
tuned closer to its critical value, the RG flow stays close to the fixed point
up to larger scales before it eventually runs off to the ordered or disordered
phase. In the disordered high-temperature phase, the correlation length is given
by the scale at which the flow trajectory departing from the vicinity of the
fixed point reaches $y = 1$. It can be calculated from an asymptotic analysis of
the linearized flow equations, which yields~\cite{supplemental_material}
\begin{equation}
  \label{eq:xi}
  4 \sqrt{\ln(\xi/a)} + \ln(\ln(\xi/a))/4 \sim - \ln(t),
\end{equation}
where $t = (T - T_c)/T_c$ is the reduced temperature. This peculiar universal
divergence of $\xi$ is stronger than conventional scaling $\xi/a \sim t^{-\nu}$
but weaker than the essential singularity $\xi/a \sim e^{C/\sqrt{t}}$ at the
equilibrium KT transition. Experimentally or numerically, it will be challenging
to confirm the precise type of singularity~\eqref{eq:xi} --- especially, since
the asymptotic seems to be approached only for $T$ very close to $T_c$ when
$\xi$ becomes extremely large~\cite{supplemental_material}. A more easily
accessible feature that distinguishes the transition is the absence of the
characteristic jump of the superfluid stiffness $\sim 1/\varepsilon$ with the
universal value $1/(\varepsilon T) \to 4$ for $T$ approaching $T_c$ from below
in the KT transition. Here, in contrast, the renormalized value of $\varepsilon$
diverges at $T_c$, leading to a smoothly vanishing superfluid stiffness at the
transition.

\textit{Conclusions.---}We studied the effect of strong spatial anisotropy on
the structure and dynamics of vortices in 2D out-of-equilibrium systems with
U(1) symmetry. These are described by the caKPZ equation~\eqref{eq:caKPZ}, and
physical realizations include active systems~\cite{Chen2013}, driven-dissipative
condensates~\cite{Altman2015,Gladilin2014,Ji2015,He2015,Keeling2016,Sieberer2016a,Wachtel2016,Sieberer2016b,He2017,He2017b,Zamora2017},
oscillator arrays~\cite{Lauter2016}, and limit-cycle
phases~\cite{Lee2011,Ludwig2013a,Jin2013,Chan2015,Schiro2016}. To address the
thermodynamic stability of the ordered phase in which vortices exist only as
tightly bound pairs, we considered the structure of single vortices. The absence
of a radially emitted wave in the FA configuration indicates that the
interaction is long-range and thus the ordered phase could be stable. Our
perturbative calculation of the vortex interaction shows that in FA and up to an
exponentially large scale, the vortex dynamics is dominated by Coulomb
interactions with \emph{attractive} corrections. Consequently, the
characteristic KT behavior such as an essential singularity of the correlation
length and an universal jump (rounded by finite size) of the superfluid density
gives way to novel universal behavior. This prediction could be checked directly
in experiments (e.g., with exciton-polaritons, see~\cite{Zamora2017} for
relevant parameter regimes) or numerics~\cite{Gladilin2017}. The modification of
the vortex interaction should also have directly observable effects on
phase-ordering kinetics~\cite{Ryskin1991,Yurke1993,Bray2000,Bray2000a}, which is
an interesting problem for further studies. Another interesting question is
whether the dynamical instability reported in Refs.~\cite{Lauter2016,He2017} and
that would lead to explosive desynchronization in arrays of limit-cycle
oscillators could be mitigated by going to the SA regime. Moreover, here we
focused on the FA configuration. The extension of our analysis to arbitrary
anisotropy is an open problem.

We thank I. Carusotto, S. Diehl, S. Gazit, L. He, A. Kamenev, M. Szyma\'nska,
G. Wachtel, and A. Zamora for helpful discussions, and we acknowledge funding
through the ERC synergy grant UQUAM.

\bibliography{anisotropic_KPZ}

\end{document}


\title{Supplemental Material: Topological Defects in Anisotropic Driven Open
  Systems}

\begin{abstract}
  In the Supplemental Material, we present details of our analysis of
  single-vortex solutions in the compact anisotropic KPZ equation. We study
  these solutions using asymptotic analysis and numerics. Moreover, we calculate
  the interaction between a vortex and an antivortex in the limit of low
  mobility of topological defects. In this limit, the interaction can be
  obtained from a dual electrostatic description. We solve the electrostatic
  problem perturbatively in the non-linearity of the KPZ equation. Based on this
  calculation, we derive RG equations that describe the vortex-unbinding
  crossover in parametrically large systems. Our analysis of the critical RG
  flow trajectories reveals a peculiar universal divergence of the correlation
  length as the transition is approached from the disordered side.
\end{abstract}

\maketitle
\tableofcontents{}


\section{A single vortex in the compact anisotropic KPZ equation}
\label{sec:single-vort-comp}

Here we present some details of our analysis of the field generated by a single
topological defect in the compact anisotropic KPZ (caKPZ) equation. As explained
in the main text, to judge whether the ordered phase is stable in the
thermodynamic limit, the crucial question is how the interaction of vortices
behaves at asymptotically large distances. This question cannot be addressed
within the perturbative treatment of the non-linearity we formulate below in the
framework of the electrodynamic duality, since the perturbative expansion breaks
down at large distances. We can nevertheless gain some insight by considering
the simpler problem of a single vortex: In the isotropic KPZ equation, these
vortices emit waves in the radial direction, and the exponential screening of
the vortex interaction can be traced back to the shocks which are created when
the emitted waves collide. Thus --- at least heuristically --- we conclude that
the interaction is not screened if there are no waves emitted from the vortex
cores in the (strongly) anisotropic KPZ equation. (Recall that due to the
non-linearity a multi-vortex solution cannot simply be constructed by linear
superposition of single vortices.) Below, using a combination of analytical
asymptotic analysis and numerics we show that this is indeed the case in the
fully anisotropic KPZ equation. We find that in the full weakly anisotropic (WA)
regime, topological defects emit (deformed) radial waves and we would expect
their interactions to be exponentially screened. Radial waves correspond to the
asymptotic behavior $\theta(r, \phi) \sim k_0(\phi) r$ for $r \to \infty$ of the
field generated by a topological defect at the origin, where $k_0(\phi)$ is the
asymptotic wave number that depends on the polar angle $\phi$. By contrast, in
the strongly anisotropic (SA) regime, the leading asymptotic behavior of the
far-field of a topological defect is
$\theta(r, \phi) \sim b_0 \ln(r/a) + \Phi(\phi)$, and the coefficient $b_0$
vanishes at the fully anisotropic (FA) point ($\lambda_x = - \lambda_y$). Thus,
vortices in the fully anisotropic KPZ equation are qualitatively very similar to
ordinary vortices in the $XY$ model. Away from the fully anisotropic point, the
asymptotics $\theta(r, \phi) \sim b_0 \ln(r/a)$ can be interpreted as a radial
wave with a wave number that vanishes as $1/r$. Further studies are required to
test whether this behavior leads to sufficient screening to destabilize the
ordered phase.

We thus want to find solutions to the caKPZ equation without noise,
\begin{equation}
  \label{eq:vortex_caKPZ}
  \partial_t \theta = D \nabla^2 \theta + \frac{\lambda_x}{2} \left( \partial_x
    \theta \right)^2 + \frac{\lambda_y}{2} \left( \partial_y \theta \right)^2,
\end{equation}
subject to the topological constraint
$\oint d \mathbf{l} \cdot \nabla \theta = 2 \pi$, where the line integral
encircles the vortex core. As we show below, in the WA regime, such vortex
solutions oscillate uniformly, i.e., they take the form
$\theta(\mathbf{r}, t) = \theta_0(\mathbf{r}) + \omega_0 t$, where
$\omega_0 > 0$ for $\lambda_{x, y} > 0$. We thus find it convenient to rewrite
Eq.~\eqref{eq:vortex_caKPZ} in a rotating frame by the transformation $\theta
\to \theta - \omega_0 t$, such that
\begin{equation}
  \label{eq:vortex_caKPZ_rf}
  \partial_t \theta = D \nabla^2 \theta + \frac{\lambda_x}{2} \left( \partial_x
    \theta \right)^2 + \frac{\lambda_y}{2} \left( \partial_y \theta \right)^2 -
  \omega_0 = 0,
\end{equation}
which is the equation stated in the main text [Eq.~(2)]. This is a non-linear
partial differential equation, and in the absence of rotational symmetry, the
solution cannot be separated into parts that depend only on the radial
coordinate or polar angle, respectively. As a further complication, the
continuum compact KPZ (cKPZ) equation has to be regularized at short distances
(e.g., by considering the cKPZ equation as the far-field phase equation derived
from the complex Ginzburg-Landau equation (CGLE), or by discretizing the cKPZ
equation on a lattice). In particular, the value of the oscillation frequency is
determined by the regularization. Nevertheless, some progress can be made if we
are modest and consider the asymptotic far-field behavior only. Below, we check
our analytical results for the far field with numerics for the full solution of
Eq.~\eqref{eq:vortex_caKPZ}.

We begin the discussion of the analytical approach by reviewing vortices in the
isotropic KPZ equation~\cite{Aranson1998} (see~\cite{Aranson2002} and references
therein for vortices in the CGLE). Hence, we set
$\lambda_x = \lambda_y = \lambda_+$ in Eq.~\eqref{eq:vortex_caKPZ_rf},
\begin{equation}
  \partial_t \theta = D \nabla^2 \theta + \frac{\lambda_+}{2} \left( \nabla \theta
  \right) - \omega_0 = 0.
\end{equation}
A vortex sitting at the origin is described by a solution of the form
$\theta(r, \phi) = \phi + R(r)$, where due to the rotational symmetry the
function $R(r)$ depends only on the radius. For the radial dependence we find
the equation ($\alpha_+ = \lambda_+/(2 D)$)
\begin{equation}
  \label{eq:R}
  \left( R'' + \frac{R'}{r} \right) + \alpha_+ \left[ \frac{1}{r^2} + R^{\prime 2}
  \right] - \frac{\omega_0}{D} = 0,
\end{equation}
which can be linearized by means of a Cole-Hopf transformation,
$w = e^{\alpha_+ R}$. We note that this implies that $w$ takes values in
$\R_{>0}$ since $R \in \R$. The Cole-Hopf transformation brings Eq.~\eqref{eq:R}
to the form of a modified Bessel equation,
\begin{equation}
  \label{eq:w}
  w'' + \frac{w'}{r} + \frac{\alpha_+^2 w}{r^2} = \kappa_0^2 w,
\end{equation}
where $\kappa_0^2 = \alpha_+ \omega_0/D$. For $\omega_0 = 0$, two linearly
independent solutions to this equation are given by $w = r^{\pm i \alpha_+}$,
and accordingly real-valued solutions take the form
$w = w_0 \cos(\alpha_+ \ln(r/a) + b)$ with $w_0, b \in \R$. However, this oscillating
function does not have an inverse Cole-Hopf transformation
$\forall r \in \R_{>0}$ and thus it does not yield a valid solution for a
vortex. For finite $\omega_0$, the solution to Eq.~\eqref{eq:w} which is bounded
at large $r$ is the modified Bessel function $w(r) = K_{i \alpha_+}(\kappa_0 r)$
At large scales it behaves as $w(r) \sim e^{-\kappa_0 r}/\sqrt{r}$ and it
assumes a maximum at $r_0 = e^{-\pi/(2 \alpha_+)}/\kappa_0$, while at shorter
scales it starts to oscillate. Hence, the KPZ equation can describe vortices
only for $r > r_0$, and some regularization is required at shorter scales. The
precise value of $\omega_0$ is determined by the
regularization~\cite{Aranson1998}. In the CGLE, one finds
$\omega_0 \sim \lambda_+ e^{-\pi D/\lambda_+}/(2 a)$, where $a$ is the vortex
core radius~\cite{Aranson1998}. Before we move on to discuss vortices in the
anisotropic case, we note that the asymptotic behavior of the Bessel function
implies the following asymptotic behavior of $\theta$:
\begin{equation}
  \label{eq:vortex_asymptotic_isotropic}
  \begin{split}
    \theta(r, \phi) & = \phi + \frac{1}{\alpha_+} \ln(K_{i \alpha_+}(\kappa_0
    r)) \\ & = -k_0 r - \frac{1}{2 \alpha_+} \ln(2 \alpha_+ k_0 r/\pi) + \phi +
    O(1/r),
  \end{split}
\end{equation}
where $k_0 = \kappa_0/\alpha_+ = \sqrt{\omega_0/(\alpha_+ D)}$.

\subsection{Far field of a single anisotropic vortex}

Next, we consider vortices in the anisotropic KPZ equation, i.e., we seek
solutions to
\begin{equation}
  \label{eq:vortex_caKPZ_rf_rescaled}
  \nabla^2 \theta + \alpha_x \left( \partial_x \theta \right)^2 + \alpha_y
  \left( \partial_y \theta \right)^2 - \varpi_0 = 0,
\end{equation}
where $\varpi_0 = \omega_0/D$, $\alpha_{x,y} = \lambda_{x,y}/(2 D)$, and with
$\alpha_x \neq \alpha_y$.  The asymptotic behavior in the isotropic
case~\eqref{eq:vortex_asymptotic_isotropic} motivates the following ansatz for $r \to \infty$:
\begin{equation}
  \label{eq:vortex_asymptotic}
  \theta(r, \phi) = k_0(\phi) r + b(\phi) \ln(r/a) + \Phi(\phi) + O(1/r).
\end{equation}
Here, $\Phi(\phi)$ contains the topological part, i.e.,
$\int_0^{2 \pi} d \phi \, \Phi'(\phi) = 2 \pi$, where $\Phi' = d \Phi/d\phi$. We
note, that a constant contribution to the vortex field can be added
arbitrarily. It is not determined by the caKPZ equation, since the latter
contains only derivatives of $\theta$. With the above ansatz, the gradient of
the phase behaves at large $r$ as
\begin{multline}  
  \nabla \theta(r, \phi) = \unitvec{e}_r \left( k_0(\phi) + \frac{b(\phi)}{r}
  \right) \\ + \frac{\unitvec{e}_{\phi}}{r} \left( k_0'(\phi) r + b'(\phi)
    \ln(r/a) + \Phi'(\phi) \right) + O(1/r^2),
\end{multline}
where $\unitvec{e}_r = \left( \cos(\phi), \sin(\phi) \right)$, and
$\unitvec{e}_{\phi} = \left( -\sin(\phi), \cos(\phi) \right)$. Inserting this
expression in Eq.~\eqref{eq:vortex_caKPZ_rf_rescaled}, we find by matching the
leading terms for $r \to \infty$ (for $\alpha_{x,y} > 0$, we take the negative
square root to match our numerical findings, see Sec.~\ref{sec:numerics};
$k(\phi)$ is positive for $\alpha_{x, y} < 0$):
\begin{equation}
  \label{eq:k_0}
  k_0(\phi) = - \sqrt{\frac{\varpi_0}{\alpha_x \cos(\phi)^2 + \alpha_y \sin(\phi)^2}},
\end{equation}
which describes a vortex emitting a slightly deformed radial wave. Clearly, this
solution is well-behaved as a function of $\phi$ as long as both $\alpha_x$ and
$\alpha_y$ are positive. As in the isotropic case, we expect that $\varpi_0$ is
determined by matching the asymptotic solution to the (regularized) solution in
the core region. However, we cannot perform a Cole-Hopf transformation as
before, and therefore it is not easily possible to find a solution that is valid
at short distances. Crucially, within the WA regime, the structure of a single
vortex is qualitatively unchanged, which implies that the vortex interaction is
screened and the ordered phase is unstable whenever $\alpha_{x, y}$ have the
same sign.

\begin{figure}
  \centering
  \includegraphics[width=0.48\linewidth]{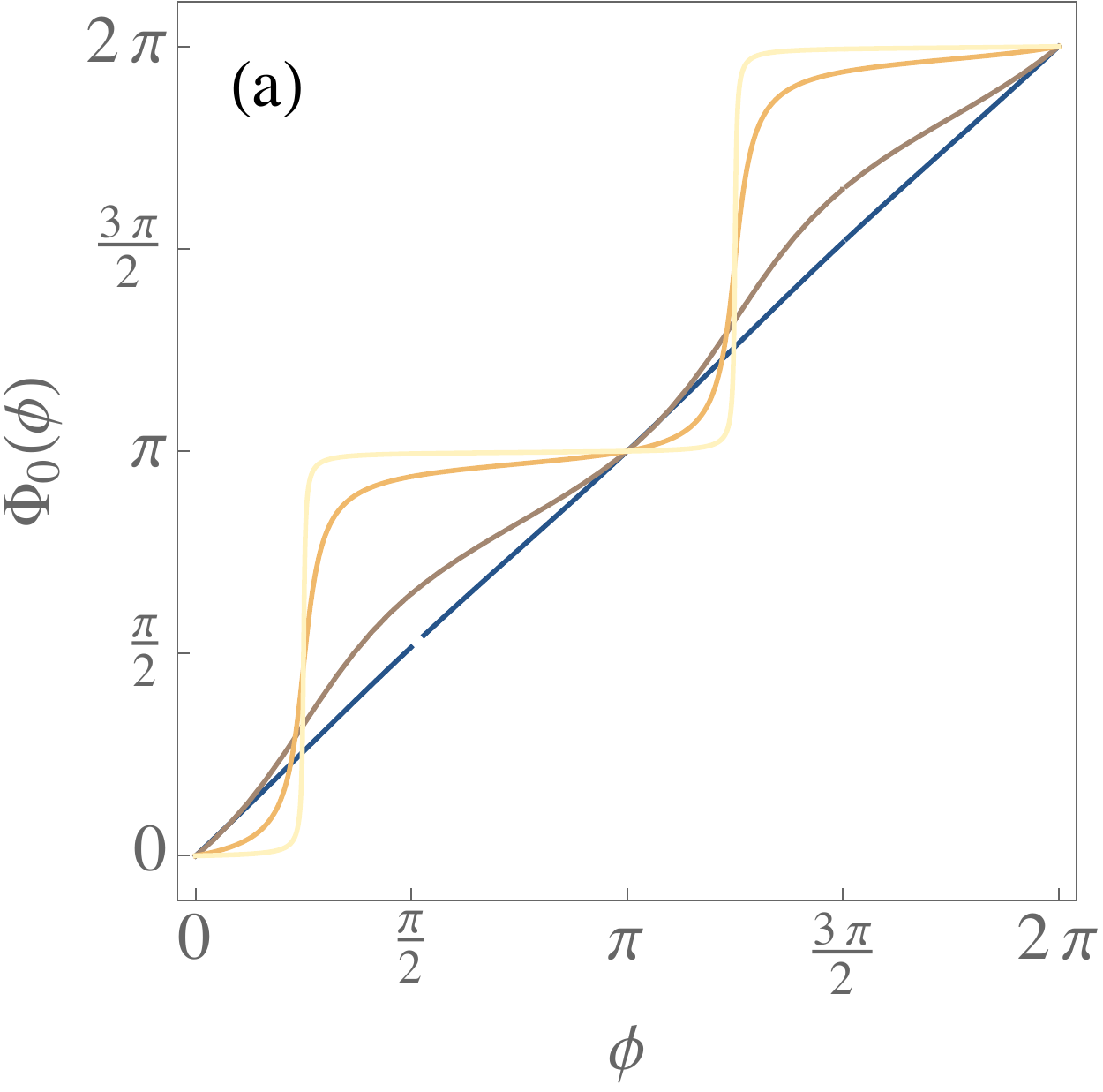}
  \hfill
  \includegraphics[width=0.48\linewidth]{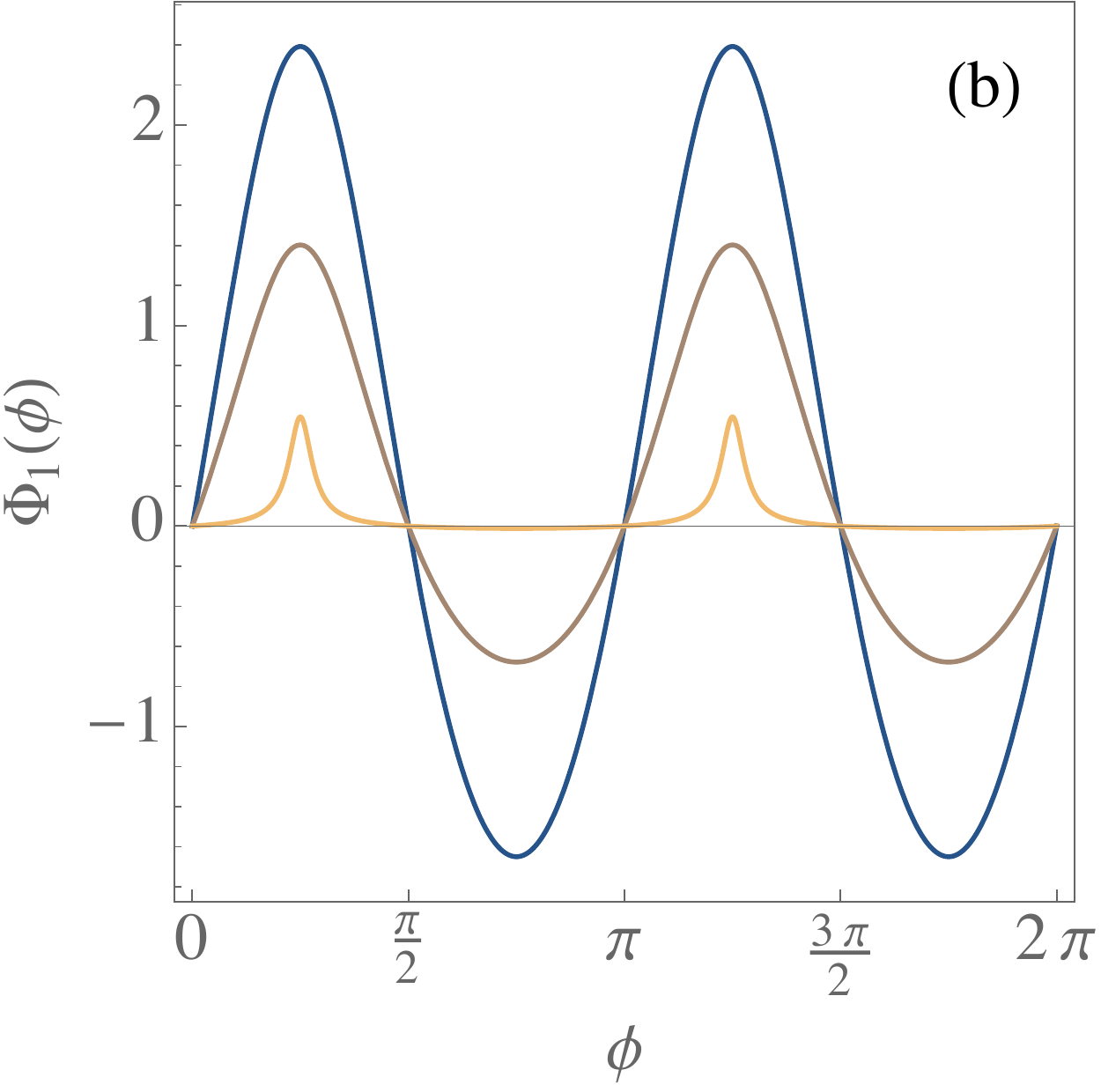}
  \caption{(a) Angular dependence of the vortex field
    in the fully anisotropic KPZ equation for $\alpha_- = 0.1, 1, 10, 100$ (blue
    to light orange). (b) First-order correction to the fully anisotropic vortex for
    $\alpha_- = 0.5, 1, 10$ (blue to orange).}
  \label{fig:Phi_01}
\end{figure}

In the SA regime, when $\alpha_x$ and $\alpha_y$ have opposite signs,
Eq.~\eqref{eq:k_0} indicates that $\varpi_0$ and $k_0(\phi)$ vanish --- any
non-zero $\varpi_0$ would lead to imaginary values of $k_0(\phi)$ and can be
discarded for this reason. This matches our numerical findings, see
Sec.~\ref{sec:numerics}. Dropping the leading term from
Eq.~\eqref{eq:vortex_asymptotic}, we find by matching terms $O((\ln(r/a)/r)^2)$ in
Eq.~\eqref{eq:vortex_caKPZ_rf_rescaled}:
\begin{equation}  
  \left( \alpha_+ - \alpha_- \cos(2 \phi) \right) b'(\phi)^2 = 0,
\end{equation}
where $\alpha_{\pm} = (\alpha_x \pm \alpha_-)/2$. It follows that $b'(\phi) = 0$
and hence $b(\phi) = b_0$. All terms at $O(\ln(r/a)/r^2)$ vanish for
$b'(\phi) = 0$, and the next non-trivial contribution comes at $O(1/r^2)$:
\begin{multline}
  \label{eq:Phi}
  \Phi''(\phi) + \alpha_+ \left( b_0^2 + \Phi'(\phi)^2 \right) + \alpha_- \left(
    b_0^2 \cos(2 \phi) \right. \\ \left. - 2 b_0 \Phi'(\phi) \sin(2 \phi) -
    \Phi'(\phi)^2 \cos(2 \phi) \right) = 0.
\end{multline}
This is an ordinary yet non-linear differential equation, and there is no
constructive way to find a solution. However, a vast simplification occurs in
the FA limit $\alpha_x = - \alpha_y$. There, the numerical solution shown in
Fig.~1 of the main text indicates that $\theta(r, \phi)$ does not depend on $r$
at all, i.e., the ansatz $\theta(r, \phi) = \Phi_0(\phi)$ yields an exact
solution. Going back to Eq.~\eqref{eq:vortex_caKPZ_rf}, with this ansatz we find
\begin{multline}  
  \nabla^2 \theta + \alpha_- \left[ \left( \partial_x \theta \right)^2 -
    \left( \partial_x \theta \right)^2 \right] \\ =
  \frac{1}{r^2} \left( \Phi_0'' - \alpha_- \cos(2 \phi) \Phi_0^{\prime 2} \right) = 0,
\end{multline}
This equation can be integrated trivially once,
\begin{equation}  
  \begin{split}
    \int_0^{\phi} d \phi' \, \frac{\Phi_0''(\phi')}{\Phi_0'(\phi')^2} & = -
    \int_0^{\phi} d \phi' \, \frac{d}{d \phi'} \frac{1}{\Phi_0'(\phi')} \\ & = -
    \left( \frac{1}{\Phi_0'(\phi)} - \frac{1}{\Phi_0'(0)} \right) \\ & =
    \alpha_- \int_0^{\phi} d \phi' \cos(2 \phi') \\ & = \frac{\alpha_-}{2} \sin(2
    \phi).
  \end{split}
\end{equation}
We thus find
\begin{equation}  
  \Phi_0'(\phi) = \frac{\Phi_0'(0)}{1 - \frac{\alpha_-}{2} \Phi_0'(0) \sin(2
    \phi)},
\end{equation}
and another integration yields the result
\begin{multline}
  \label{eq:Phi_0}
  \Phi_0(\phi) = 2 \nu \Phi_0'(0) \left( \arctan(\nu \left( 2 \tan(\phi) -
      \alpha_- \Phi_0'(0) \right)) \right. \\ \left. + \arctan(\alpha_- \nu
    \Phi_0'(0)) \right),
\end{multline}
where a constant of integration is added such that $\Phi_0(0) = 0$; in order to
obtain a smooth solution we have to choose different branches of the $\arctan$
in the intervals $0 < \phi < \pi/2, \pi/2 < \phi < 3 \pi/2$, and
$3 \pi/2 < \phi < 2 \pi$. The constant $\nu$ in Eq.~\eqref{eq:Phi_0} is given by
\begin{equation}  
  \nu = \frac{1}{\sqrt{4 - \left( \alpha_- \Phi_0'(0) \right)^2}}.
\end{equation}
Finally, $\Phi_0'(0)$ is determined by the condition that for a singly-charged
vortex the function $\Phi_0(\phi)$ should wind once around the unit circle, which
yields
\begin{equation}  
  \Phi_0'(0) = \frac{2}{\sqrt{4 + \alpha_-^2}}.
\end{equation}
Vortex solutions for different values of $\alpha_-$ are shown in
Fig.~\ref{fig:Phi_01}(a).
For $\alpha_- \to 0$ the solution smoothly deforms
into an ``ordinary'' $XY$-type vortex with $\Phi_0(\phi) = \phi$, while in the
opposite limit it approaches a step-like form. In fact, Eq.~\eqref{eq:Phi_0} is
analytic in $\alpha_-$ at $\alpha_- = 0$.  This is in stark contrast to the
isotropic case, where the transition from from the linear to the non-linear
problem is highly non-analytic (see the above expression for $\omega_0$). Since
turning on the non-linearity in a fully anisotropic system does not alter the
\emph{radial dependence} of the far field of a single vortex, we conclude that
the interaction of vortices at large distances is not screened as in the
isotropic case, and thus the ordered phase can be stable.

Now let's reinstate $\alpha_+$. First, note that for $b_0 = 0$,
Eq.~\eqref{eq:Phi} becomes
\begin{equation}  
  \Phi''(\phi) + \left( \alpha_+ - \alpha_- \cos(2 \phi) \right) \Phi'(\phi)^2 = 0.
\end{equation}
As before, this equation can be integrated and we find
\begin{equation}  
  - \left( \frac{1}{\Phi'(\phi)} - \frac{1}{\Phi'(0)} \right) = - \alpha_+ \phi
  + \frac{\alpha_-}{2} \sin(2 \phi).
\end{equation}
Since the resulting expression for $\Phi'(\phi)$ is not periodic in $\phi$,
evidently this cannot be a valid solution, and we have to allow for a finite
value of $b_0$, i.e., away from the FA configuration vortices do have a
non-trivial radial dependence. We restrict ourselves to small values of
$\alpha_+$ for which we set
\begin{equation}  
  \begin{split}
    \Phi(\phi) & = \Phi_0(\phi) + \alpha_+ \Phi_1(\phi) + O(\alpha_+^2), \\
    b_0 & = \alpha_+ b_{01} + O(\alpha_+^2),
  \end{split}
\end{equation}
where the zeroth-order solution $\Phi_0(\phi)$ is given by
Eq.~\eqref{eq:Phi_0}. Inserting this ansatz in Eq.~\eqref{eq:Phi} leads to a
linear second-order differential equation for $\Phi_1(\phi)$; $b_{01}$ is
determined by the condition that $\Phi_0'(\phi)$ has to be periodic, which
yields $b_{01} = -2/\alpha_-^2$. The first constant of integration $\Phi_1'(0)$
has to be chosen such that
$\lim_{\phi \searrow 0} \Phi_1(\phi) = \lim_{\phi \nearrow 2 \pi} \Phi_1(\phi)$,
while w.l.o.g.\ we choose the second constant of integration $\Phi_1(0)$ such
that $\Phi(0) = 0$. The resulting solution $\Phi_1(\phi)$ is shown in
Fig.~\ref{fig:Phi_01}(b). It is an interesting question for future research how the logarithmic dependence
of the vortex field~\eqref{eq:vortex_asymptotic} on the distance from the core
--- corresponding to an emitted wave with wave number $\sim 1/r$ --- affects the
interaction at asymptotic distances.

\subsection{Numerics}
\label{sec:numerics}

To confirm the results of the previous section numerically, we discretize
Eq.~\eqref{eq:vortex_caKPZ} on a lattice with sites
$\mathbf{r} = \left( x, y \right)$, i.e., we replace spatial derivatives with
finite differences according to (cf.\ Ref.~\cite{Sieberer2016b})
\begin{equation}
  \label{eq:finite_differences}
  \begin{split}
    \partial_x^2 \theta & \to - \sum_{\sigma = \pm} \sin(\theta_{\mathbf{r}} -
    \theta_{\mathbf{r} + \sigma \unitvec{x}}), \\ \left( \partial_x \theta
    \right)^2 & \to - \sum_{\sigma = \pm} \left( \cos(\theta_{\mathbf{r}} -
      \theta_{\mathbf{r} + \sigma \unitvec{x}}) - 1 \right).
  \end{split}
\end{equation}
$\unitvec{x}$ and $\unitvec{y}$ are unit vectors, and for convenience we choose
the lattice spacing as $a = 1$. With the above prescription,
Eq.~\eqref{eq:vortex_caKPZ} becomes
\begin{multline}
  \label{eq:discrete_vortex_caKPZ}  
  \partial_t \theta_{\mathbf{r}} = - \sum_{\sigma = \pm} \left[ D \left(
      \sin(\theta_{\mathbf{r}} - \theta_{\mathbf{r} + \sigma \unitvec{x}}) +
      \sin(\theta_{\mathbf{r}} - \theta_{\mathbf{r} + \sigma \unitvec{y}})
    \right) \vphantom{\frac{\lambda_x}{2}} \right. \\ + \frac{\lambda_x}{2}
  \left( \cos(\theta_{\mathbf{r}} - \theta_{\mathbf{r} + \sigma \unitvec{x}}) -
    1 \right) \\ \left. + \frac{\lambda_y}{2} \left( \cos(\theta_{\mathbf{r}} -
      \theta_{\mathbf{r} + \sigma \unitvec{y}}) - 1 \right) \right].
\end{multline}
We choose initial conditions corresponding to an ordinary $XY$ vortex that is
displaced by half a lattice spacing from the origin, i.e.,
$\theta_{\mathbf{r}}(0) = \tan((y - 1/2)/(x - 1/2))$, and evolve this
configuration in time. For open boundary conditions, we found that the core of
the topological defect remains stationary. (For large values of the
non-linearities $\lambda_{x, y}$, the vortex starts to move, and new vortices
are generated dynamically. In the simulations presented here, we always stay
below this instability.) Since the evolution
equation~\eqref{eq:discrete_vortex_caKPZ} is dissipative and the topological
charge is conserved, the field configuration converges to a vortex solution of
the non-linear problem. For the plots in Fig.~1 of the main text, we evolved
Eq.~\eqref{eq:discrete_vortex_caKPZ} on a lattice of $200 \times 200$ sites.

\begin{figure*}
  \centering
  \includegraphics[width=0.4\linewidth]{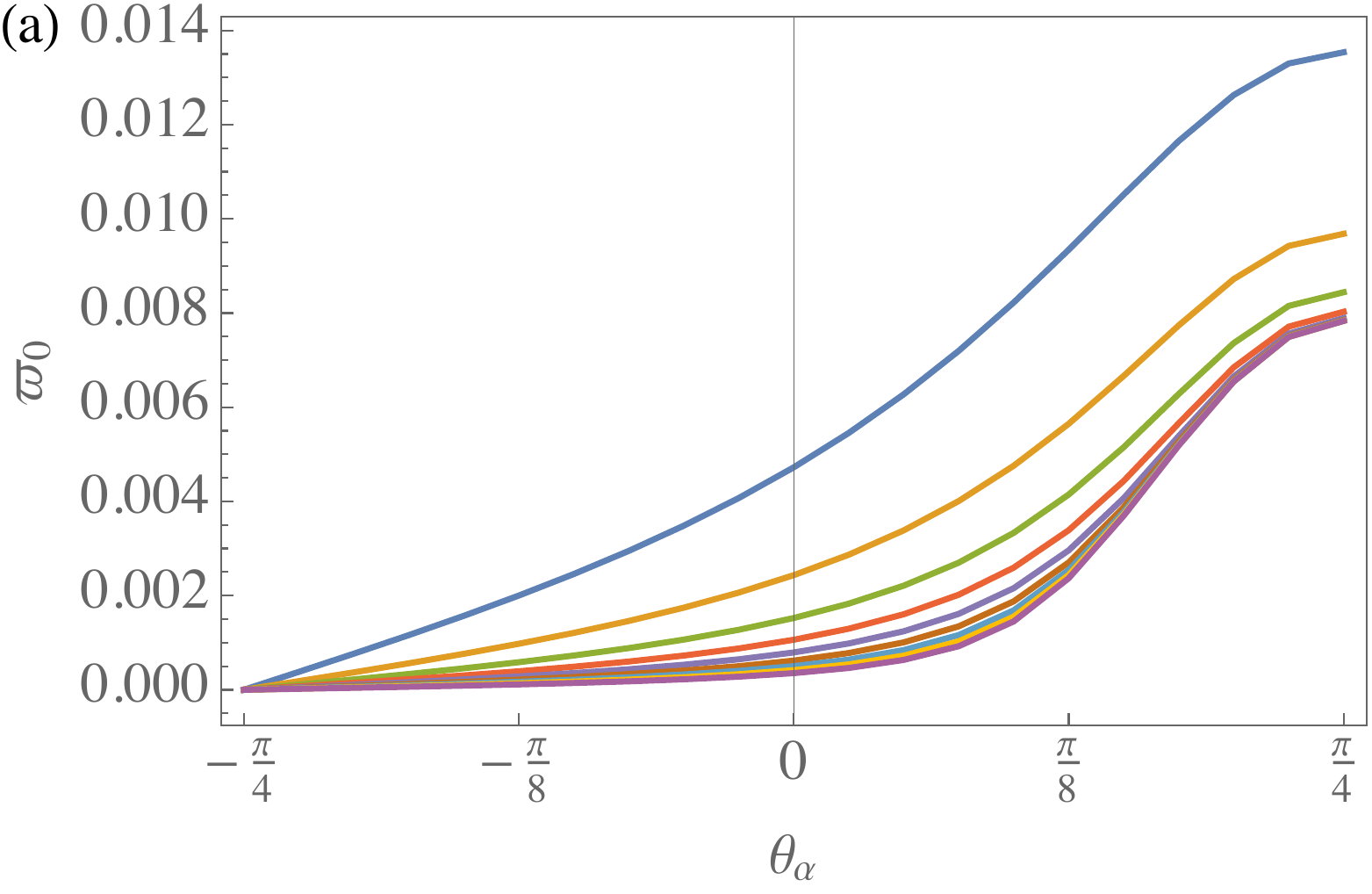}
  \hspace{1cm}
  \includegraphics[width=0.4\linewidth]{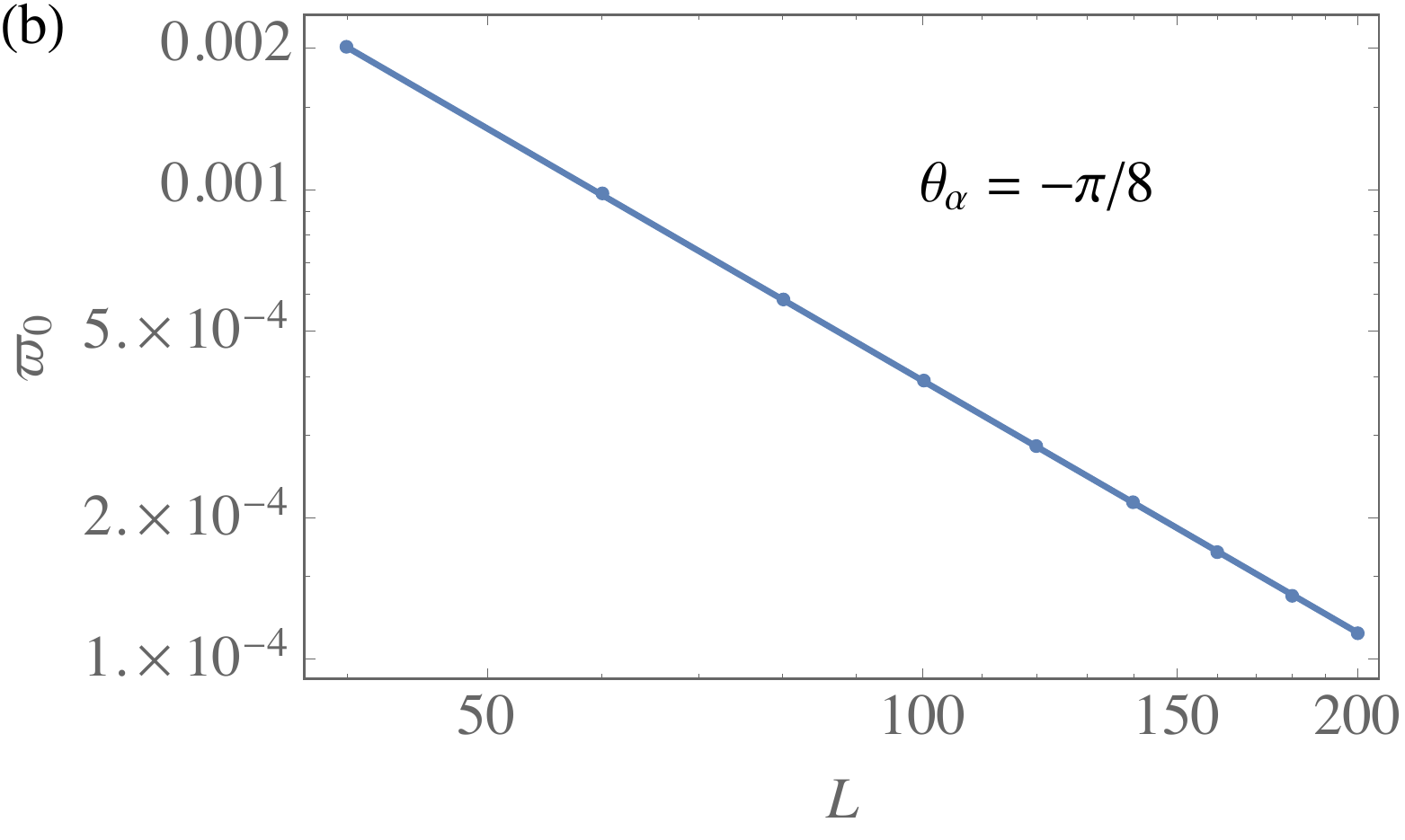} \\
  \includegraphics[width=0.4\linewidth]{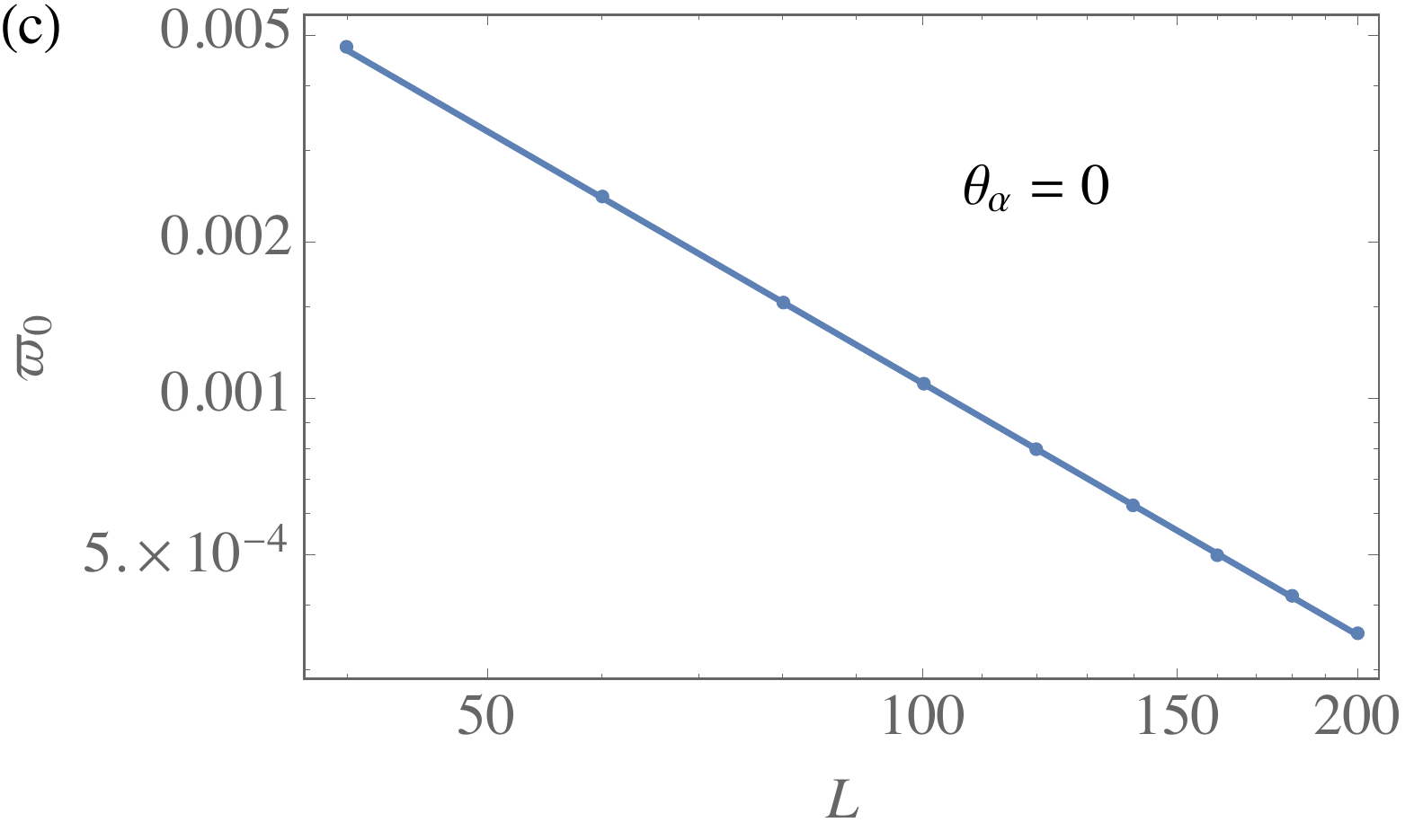}
  \hspace{1cm}
  \includegraphics[width=0.4\linewidth]{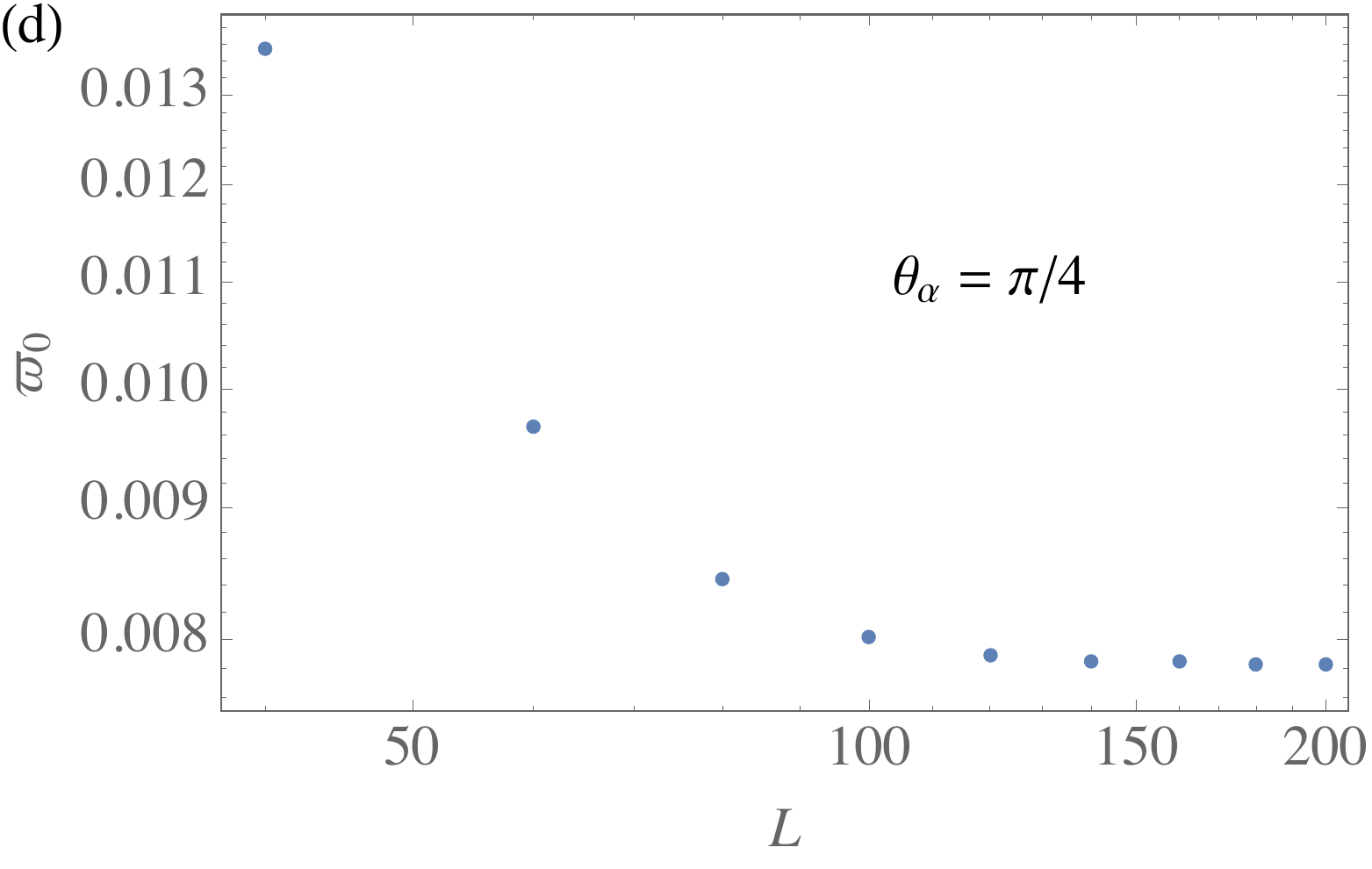}
  \caption{(a) Oscillation frequency $\varpi_0 = \omega_0/D$ as a function of
    $\theta_{\alpha}$. The latter is defined by
    $\left( \alpha_x, \alpha_y \right) = 1/2 \left( \cos(\theta_{\alpha}),
      \sin(\theta_{\alpha}) \right)$,
    i.e., for all values of $\theta_{\alpha}$ the system is equally non-linear
    in the sense that $\sqrt{\alpha_x^2 + \alpha_y^2} = 1/2$, and only the
    degree of anisotropy is varied. Specifically, for $\theta_{\alpha} = -\pi/4$
    the system is fully anisotropic; the separation between the SA and WA
    regimes is at $\theta_{\alpha} = 0$, and at $\theta_{\alpha} = \pi/4$ the
    system is isotropic. The different curves correspond to system sizes
    $L = 40, 60, 80, \dotsc, 200$ (top to bottom). Numerically, $\varpi_0$
    vanishes only at the FA point, while it is finite due to finite-size effects
    everywhere else. (b,c) In the SA regime we found that $\varpi_0$ vanishes
    algebraically with increasing system size, while (d) it converges to a
    finite value in the isotropic case with $\theta_{\alpha} = \pi/4$ (and also
    for other values of $\theta_{\alpha}$ in the WA regime).}
  \label{fig:omega0}
\end{figure*}

As discussed in the previous section, vortices in the caKPZ equation oscillate
uniformly in the WA regime. From the numerical solution of
Eq.~\eqref{eq:discrete_vortex_caKPZ}, the oscillation frequency can be obtained
by fitting the steady linear growth of $\theta_{\mathbf{r}}(t)$ at late times
(i.e., when convergence is reached). We find a vanishing oscillation frequency
$\varpi_0$ only exactly at the FA point, and small but finite oscillation
frequencies throughout the SA region. However, as illustrated in
Fig.~\ref{fig:omega0}, this is just a finite-size effect.

\section{Interaction of vortices in the compact anisotropic KPZ equation}

In linear theories, the superposition of two solutions gives another valid
solution. This is no longer true in the presence of a non-linearity as in the
case of the caKPZ equation. In particular, the superposition of two
single-vortex solutions does not yield a two-vortex solution. This is the main
difficulty in trying to find the vortex interaction. Only for very large
separations of topological defects, we could gain some insight in the asymptotic
behavior of the vortex interaction as discussed in the previous section. Here,
we present an alternative approach that works up to parametrically large
distances, and is based on a recently developed formulation of the caKPZ
equation as non-linear electrodynamics~\cite{Sieberer2016b,Wachtel2016}.

In Sec.~\ref{sec:electr-dual}, we briefly review the electrodynamic duality for
the compact KPZ equation~\cite{Sieberer2016b,Wachtel2016} and its extension to
the anisotropic case. Within this framework, we calculate the interaction
between a vortex and an antivortex perturbatively. This rather tedious
calculation, and some numerical checks of the result, are presented in
Sec.~\ref{sec:pert-calc-vort}.

\subsection{Electrodynamic duality}
\label{sec:electr-dual}

In the following, we derive a dual description of the caKPZ equation [Eq.~(1) of
the main text]~\cite{Sieberer2016b, Wachtel2016},
\begin{equation}  
  \partial_t \theta = \sum_{i = x, y} \left[ D_i \partial_i^2 \theta +
    \frac{\lambda_i}{2} \left( \partial_i \theta \right)^2 \right] + \eta.
\end{equation}
As in the main text, we set $D_x = D_y = D$, which corresponds simply to an
anisotropic rescaling of the units of length. We find it convenient to rewrite
the non-linear terms in the following way:
\begin{equation}  
  \sum_{i = x, y} \lambda_i \left( \partial_i \theta \right)^2 = \lambda_+
  \left( \nabla \theta \right)^2 + \lambda_- \left[ \left( \partial_x \theta
    \right)^2 - \left( \partial_y \theta \right)^2 \right],
\end{equation}
which separates an isotropic contribution that is proportional to
$\lambda_+ = \frac{\lambda_x + \lambda_y}{2}$ from a purely anisotropic
contribution with coefficient $\lambda_- = \frac{\lambda_x - \lambda_y}{2}$. To
keep the notation compact, we define a $\odot$ product of vectors
$\mathbf{a} = \left( a_x, a_y \right)$ and
$\mathbf{b} = \left( b_x, b_y \right)$ through the relation.
$\mathbf{a} \odot \mathbf{b} = a_x b_x - a_y b_y$. We also use the abbreviation
$\mathbf{a}^{\odot 2} = \mathbf{a} \odot \mathbf{a} = a_x^2 - a_y^2$. It is
straightforward to check that this product is commutative and distributive,
i.e., in calculations it can be handled like the usual scalar product. With this
notation, the caKPZ equation can be written as
\begin{equation}
  \label{eq:caKPZ_odot}
  \partial_t \theta = D \nabla^2 \theta + \frac{\lambda_+}{2} \left( \nabla \theta
  \right)^2 + \frac{\lambda_-}{2} \left( \nabla \theta \right)^{\odot 2} + \eta.
\end{equation}

To explicitly incorporate vortices in the caKPZ equation, we reformulate it in
terms of the electric field, which is defined as
\begin{equation}
  \label{eq:electric_field}
  \mathbf{E} = - \unitvec{z} \times \nabla \theta.
\end{equation}
$\unitvec{z} = \left( 0, 0, 1 \right)$ is a unit vector pointing in the
direction perpendicular to the $xy$-plane on which $\theta$ and $\mathbf{E}$ are
defined. We note that in the presence of topological defects in the KPZ equation
without noise, $\theta$ has a contribution that depends linearly on time and is
uniform in space, see Sec.~\ref{sec:single-vort-comp}. This contribution
corresponds to oscillations of vortices in the complex Ginzburg-Landau equation,
and drops out if we consider $\mathbf{E}$ instead of $\theta$.

By cyclic permutation of the vectors in the defining
relation~\eqref{eq:electric_field} for $\mathbf{E}$ we obtain
$\nabla \theta = \unitvec{z} \times \mathbf{E}$, which leads to the following
expressions for the non-linear terms in the caKPZ equation:
\begin{equation}  
  \left( \nabla \theta \right)^2 = E^2, \qquad \left( \nabla \theta
  \right)^{\odot 2} = - \mathbf{E}^{\odot 2}.
\end{equation}
Thus, Eq.~\eqref{eq:caKPZ_odot} can be written as
\begin{equation}
  \label{eq:KPZ_E}
  \partial_t \mathbf{E} = D \nabla^2 \mathbf{E} - \unitvec{z} \times \left(
    \frac{\lambda_+}{2} E^2 - \frac{\lambda_-}{2} \mathbf{E}^{\odot 2} + \eta
  \right). 
\end{equation}
This equation has to be extended to explicitly account for vortices. To this
end, we first note that the circulation of the gradient of $\theta$ around a
closed loop is determined by the number of enclosed vortices. This statement,
recast in differential form, can be written as
\begin{equation}
  \label{eq:Gauss}
  \nabla \cdot \mathbf{E} = 2 \pi n,
\end{equation}
where $n$ is the vortex density. Second, since vortices are created only in
pairs (or at the boundary of the sample), the vortex density $n$ and current
$\mathbf{j}$ obey on equation of continuity:
\begin{equation}  
  \label{eq:continuity}
  \partial_t n + \nabla \cdot \mathbf{j} = 0.
\end{equation}
Since Eqs.~\eqref{eq:Gauss} and~\eqref{eq:continuity} can be combined to read
\begin{equation}  
  \nabla \cdot \left( \partial_t \mathbf{E} + 2 \pi \mathbf{j} \right) = 0,
\end{equation}
we see that
\begin{equation}
  \label{eq:u}
  \partial_t \mathbf{E} = - 2 \pi \mathbf{j} + \nabla \times \mathbf{u},
\end{equation}
where $\mathbf{u}$ is a vector field that is determined by the condition that
Eq.~\eqref{eq:u} should reproduce the KPZ equation~\eqref{eq:KPZ_E} in the
absence of vortices, i.e., for $n = \mathbf{j} = 0$. This condition is fulfilled
by setting
\begin{equation}
  \label{eq:156}
  \mathbf{u} = - D \nabla \times \mathbf{E} + \unitvec{z} \left(
    \frac{\lambda_+}{2} E^2 - \frac{\lambda_-}{2}
    \mathbf{E}^{\odot 2} + \eta \right).
\end{equation}
The required extension of the KPZ equation~\eqref{eq:KPZ_E} to include vortices
is thus given by
\begin{multline}
  \label{eq:dual_KPZ}
  \partial_t \mathbf{E} = - D \nabla \times \left( \nabla \times \mathbf{E}
  \right) - 2 \pi \mathbf{j} \\ - \unitvec{z} \times \nabla \left(
    \frac{\lambda_+}{2} E^2 - \frac{\lambda_-}{2} \mathbf{E}^{\odot 2} + \eta
  \right).
\end{multline}
As a final step, as customary in ``macroscopic'' electrodynamics, we separate
the contributions due to free and bound vortices by decomposing the vortex
density and current as
\begin{equation}  
  \begin{split}
    n & = n_f + n_b, \\
    \mathbf{j} & = \mathbf{j}_f + \mathbf{j}_b.
  \end{split}
\end{equation}
Bound vortices lead to polarization of the medium, which can be described by a
polarization density $\mathbf{P}$ that satisfies
\begin{equation}  
  \begin{split}
    \nabla \cdot \mathbf{P} & = - 2 \pi n_b, \\
    \partial_t \mathbf{P} & = 2 \pi \mathbf{j}_b.
  \end{split}
\end{equation}
Adding the polarization to the electric field we obtain the displacement field
\begin{equation}
  \label{eq:displacement_field}
  \mathbf{D} = \mathbf{E} + 2 \pi \mathbf{P} = \left( 1 + 2 \pi \chi \right) \mathbf{E} = \varepsilon \mathbf{E},
\end{equation}
where we set $\mathbf{P} = \chi \mathbf{E}$ with the susceptibility $\chi$, and
the last relation defines the dielectric constant
$\varepsilon = 1 + 2 \pi \chi$. We show below the even though we are considering
an anisotropic system, to the lowest perturbative order in the KPZ non-linearity
it is sufficient to consider a single isotropic dielectric constant instead of a
dielectric tensor. With these definitions, Eqs.~\eqref{eq:Gauss}
and~\eqref{eq:dual_KPZ} can be written as
\begin{equation}
  \label{eq:macroscopic_Gauss}
  \nabla \cdot \mathbf{E} = \frac{2 \pi}{\varepsilon} n_f,
\end{equation}
and
\begin{multline}
  \label{eq:macroscopic_dual_KPZ}
  \varepsilon \partial_t \mathbf{E} = - D \nabla \times \left( \nabla \times
    \mathbf{E} \right) - 2 \pi \mathbf{j}_f \\ - \unitvec{z} \times \nabla
  \left( \frac{\lambda_+}{2} E^2 - \frac{\lambda_-}{2} \mathbf{E}^{\odot 2} +
    \eta \right).
\end{multline}
In the following, we drop the subscript $f$.

\subsection{Perturbative calculation of the vortex interaction}
\label{sec:pert-calc-vort}

As we are interested in the interaction of vortices, which is encoded in the
deterministic dynamics, we set the noise to zero, $\eta = 0$. Moreover, we
assume that the mobility of vortices is small. This allows us to consider the
static limit~\cite{Wachtel2016} in which
$\partial_t \mathbf{E} = \mathbf{j} = 0$. Then,
Eq.~\eqref{eq:macroscopic_dual_KPZ} becomes
\begin{equation}
  \label{eq:16}
  0 = D \left( \nabla^2 \mathbf{E} - 2 \pi \nabla n
  \right) - \unitvec{z} \times \nabla \left( \frac{\lambda_+}{2} E^2 - \frac{\lambda_-}{2}
    \mathbf{E}^{\odot 2} \right).
\end{equation}
This equation and Eq.~\eqref{eq:macroscopic_Gauss} determine the electrostatic
field generated by a collection of free vortices with local density $n$. It is
convenient to represent the electric field in terms of scalar and vector
potentials,
\begin{equation}
  \label{eq:phi_A}
  \mathbf{E} = - \nabla \phi - \mathbf{A}.
\end{equation}
As usual, this decomposition does not uniquely define the potentials. It is
invariant under gauge transformations of the form $\phi' = \phi - \chi$ and
$\mathbf{A}' = \mathbf{A} + \nabla \chi$. This redundancy can be eliminated by
working in a particular gauge. Here, we impose the Lorenz gauge condition
$\partial_t \phi + D \nabla \cdot \mathbf{A} = 0$, which in the static limit
reduces to $\nabla \cdot \mathbf{A} = 0$. Then, the scalar potential encodes the
longitudinal part of the electric field, and the vector potential the transverse
part. It can be written in terms of another potential $\psi$ as
\begin{equation}
  \label{eq:A_psi}
  \mathbf{A} = - \unitvec{z} \times \nabla \psi.
\end{equation}
Inserting Eqs.~\eqref{eq:phi_A} and~\eqref{eq:A_psi} in
Eqs.~\eqref{eq:macroscopic_Gauss} and~\eqref{eq:16}, we obtain
\begin{align}
  \label{eq:phi_psi_diff_equ}
  - \nabla^2 \phi & = \frac{2 \pi}{\varepsilon} n, \\
  D \nabla^2 \psi & = \frac{\lambda_+}{2} \left( \nabla \phi - \unitvec{z} \times
                    \nabla \psi \right)^2 - \frac{\lambda_-}{2} \left( \nabla \phi - \unitvec{z} \times
                    \nabla \psi \right)^{\odot 2}.
\end{align}
These equations can be integrated with the aid of the fundamental solution of
the Laplacian (i.e., the electrostatic potential generated by a point charge),
\begin{equation}
  \label{eq:G}
  G(\mathbf{r}) = - \nabla^{-2} \delta(\mathbf{r}) = - \frac{1}{2 \pi}
  \ln(r/a).
\end{equation}
As usual, $a$ is to be understood as a microscopic cutoff. Here and in the
following, the potential $G(\mathbf{r})$ should be set to zero for $r < a$. With
the aid of the fundamental solution~\eqref{eq:G},
Eqs.~\eqref{eq:phi_psi_diff_equ} can be rewritten as convolution integrals,
\begin{equation}
  \label{eq:phi}
  \phi(\mathbf{r}) = \frac{2 \pi}{\varepsilon} \int_{\mathbf{r}'} G(\mathbf{r} -
  \mathbf{r}') n(\mathbf{r}'),  
\end{equation}
and
\begin{multline}  
  \label{eq:psi}
  \psi(\mathbf{r}) = - \frac{1}{2 D} \int_{\mathbf{r}'} G(\mathbf{r} -
  \mathbf{r}') \left[ \lambda_+ \left( \nabla \phi(\mathbf{r}') - \unitvec{z}
      \times \nabla \psi(\mathbf{r}') \right)^2 \right. \\ \left. - \lambda_-
    \left( \nabla \phi(\mathbf{r}') - \unitvec{z} \times \nabla
      \psi(\mathbf{r}') \right)^{\odot 2} \right],
\end{multline}
where we write $\int_{\mathbf{r}} = \int d^2 \mathbf{r}$, and the integration
extends over the area occupied by the system. While Eq.~\eqref{eq:phi} fully
determines the potential $\phi$ for a given charge distribution $n$,
Eq.~\eqref{eq:psi} is an integro-differential equation for $\psi$. By iterating
this equation, we obtain a solution in the form of a perturbative expansion in
$\lambda_+$ and $\lambda_-$,
\begin{equation}
  \label{eq:psi_expansion}
  \psi = \frac{1}{\varepsilon} \psi^{(0)} + \frac{1}{\varepsilon^2} \psi^{(1)} +
  \frac{1}{\varepsilon^3} \psi^{(2)} + \dotsb
\end{equation}
(we explicitly specify factors of $1/\varepsilon$ to get simpler expressions
below). More concretely, setting $\lambda_+ = \lambda_- = 0$ in
Eq.~\eqref{eq:psi}, we find that the zeroth-order contribution vanishes,
$\psi^{(0)} = 0$. The first-order contribution $\psi^{(1)}$ can be obtained by
inserting $\psi = \psi^{(0)} = 0$ on the RHS of Eq.~\eqref{eq:psi}. Another
iteration yields the lowest-order terms,
\begin{equation}
  \label{eq:psi_12}
  \begin{split}
    \psi^{(1)} & = - \sum_{\sigma = \pm} \sigma
    \alpha_{\sigma} \psi_{\sigma}^{(1)}, \\
    \psi^{(2)} & = - \sum_{\sigma = \pm} \alpha_{\sigma}^2 \psi_{\sigma}^{(2)} +
    \alpha_+ \alpha_- \psi_{+-}^{(2)}.
  \end{split}  
\end{equation}
As in the previous section, we denote $\alpha_{\pm} = \lambda_{\pm}/(2 D)$. The
explicit expressions for the lowest-order terms read
\begin{equation}  
  \label{eq:psi^1}
  \psi_{\sigma}^{(1)}(\mathbf{r}) = \varepsilon^2 \int_{\mathbf{r}'} G(\mathbf{r} - \mathbf{r}')
  \left( \nabla \phi(\mathbf{r}') \right)^{\odot 2},
\end{equation}
(here and in the following, for $\sigma = +$ the $\odot$ product should be
replaced by the usual scalar product)
\begin{equation}
  \label{eq:psi^2_sigma}
  \psi_{\sigma}^{(2)}(\mathbf{r}) = 2 \varepsilon \int_{\mathbf{r}'}
  G(\mathbf{r} - \mathbf{r}') \left[ \nabla \phi(\mathbf{r}') \odot \left(
      \unitvec{z} \times \nabla \psi_{\sigma}^{(1)}(\mathbf{r}') \right)
  \right],
\end{equation}
and the mixed term is given by
\begin{multline}
  \label{eq:psi^2_+-_foo}
  \psi_{+-}^{(2)}(\mathbf{r}) = 2 \varepsilon \int_{\mathbf{r}'} G(\mathbf{r}
  - \mathbf{r}') \left[ \nabla \phi(\mathbf{r}') \cdot \left( \unitvec{z} \times
      \nabla \psi_-^{(1)}(\mathbf{r}') \right) \right. \\ \left. + \nabla
    \phi(\mathbf{r}') \odot \left( \unitvec{z} \times \nabla
      \psi_+^{(1)}(\mathbf{r}') \right) \right].
\end{multline}
The above expressions are valid for any charge distribution $n$. In the
following, we consider a dipole which is described by
\begin{equation}  
  n(\mathbf{r}) = \delta(\mathbf{r} - \mathbf{r}_+) - \delta(\mathbf{r} - \mathbf{r}_-).
\end{equation}
Inserting this in Eq.~\eqref{eq:phi} yields the scalar electrostatic potential
generated by the dipole:
\begin{equation}    
  \phi(\mathbf{r}) = \frac{2 \pi}{\varepsilon} \left( G(\mathbf{r} -
    \mathbf{r}_+) - G(\mathbf{r} - \mathbf{r}_-) \right)
  = \frac{1}{\varepsilon} \ln \! \abs{\frac{\mathbf{r} -
      \mathbf{r}_+}{\mathbf{r} - \mathbf{r}_-}}.
\end{equation}
Equations~\eqref{eq:psi^1}, \eqref{eq:psi^2_sigma}, and~\eqref{eq:psi^2_+-_foo}
require the gradient of the scalar potential. In terms of
\begin{equation}
  \label{eq:f}
  \mathbf{f}(\mathbf{r}) = 2 \pi \nabla G(\mathbf{r}) = -
  \frac{\mathbf{r}}{r^2},
\end{equation}
we find
\begin{equation}
  \label{eq:grad_phi}
  \begin{split}
    \nabla \phi(\mathbf{r}) & = \frac{1}{\varepsilon} \left(
      \mathbf{f}(\mathbf{r} - \mathbf{r}_+) - \mathbf{f}(\mathbf{r} -
      \mathbf{r}_-) \right) \\ & = - \frac{1}{\varepsilon} \left(
      \frac{\mathbf{r} - \mathbf{r}_+}{\abs{\mathbf{r} - \mathbf{r}_+}^2} -
      \frac{\mathbf{r} - \mathbf{r}_-}{\abs{\mathbf{r} - \mathbf{r}_-}^2}
    \right).
  \end{split}
\end{equation}
Inserting this in Eqs.~\eqref{eq:psi^1} and~\eqref{eq:psi^2_sigma} yields
\begin{equation}
  \label{eq:psi_f}
  \psi_{\sigma}^{(1)}(\mathbf{r}) = \int_{\mathbf{r}'} G(\mathbf{r} - \mathbf{r}')
  \left( \mathbf{f}(\mathbf{r}' - \mathbf{r}_+) -
  \mathbf{f}(\mathbf{r}' - \mathbf{r}_-) \right)^{\odot 2},  
\end{equation}
and
\begin{multline}
  \label{eq:psi^2_sigma_f}
  \psi_{\sigma}^{(2)}(\mathbf{r}) = 2 \int_{\mathbf{r}'} G(\mathbf{r} -
  \mathbf{r}') \left[ \left( \mathbf{f}(\mathbf{r}' - \mathbf{r}_+) -
      \mathbf{f}(\mathbf{r}' - \mathbf{r}_-) \right) \vphantom{\left(
        \unitvec{z} \times \nabla \psi_{\sigma}^{(1)}(\mathbf{r}') \right)}
  \right. \\ \left. \odot \left( \unitvec{z} \times \nabla
    \psi_{\sigma}^{(1)}(\mathbf{r}') \right) \right],
\end{multline}
and from Eq.~\eqref{eq:psi^2_+-_foo} we obtain
\begin{multline}
  \label{eq:psi^2_+-}
  \psi_{+-}^{(2)}(\mathbf{r}) = 2 \int_{\mathbf{r}'} G(\mathbf{r} - \mathbf{r}')
  \left( \mathbf{f}(\mathbf{r}' - \mathbf{r}_+) - \mathbf{f}(\mathbf{r}' -
    \mathbf{r}_-) \right)^T \\ \left[ \left( \unitvec{z} \times \nabla
      \psi_-^{(1)}(\mathbf{r}') \right) + \sigma_z \left( \unitvec{z} \times
      \nabla \psi_+^{(1)}(\mathbf{r}') \right) \right],
\end{multline}
where we used that with the Pauli matrix $\sigma_z$ the $\odot$ product can be
written as
$\mathbf{a} \odot \mathbf{b} = a_x b_x - a_y b_y = \mathbf{a}^T \sigma_z
\mathbf{b}$.

The integrals in the above expressions for the lowest-order contributions to
$\psi$ are divergent in the thermodynamic limit. This complication is resolved
upon taking the gradient as required by Eq.~\eqref{eq:A_psi}. Therefore, in the
following we find it convenient to consider $\mathbf{a} = \nabla \psi$. In
analogy to Eqs.~\eqref{eq:psi_expansion} and~\eqref{eq:psi_12} we write
\begin{equation}  
  \mathbf{a} = \frac{1}{\varepsilon^2} \mathbf{a}^{(1)} +
  \frac{1}{\varepsilon^3} \mathbf{a}^{(2)} + \dotso,
\end{equation}
(recall that $\psi^{(0)} = 0$) and
\begin{equation}
  \begin{split}
    \mathbf{a}^{(1)} & = - \sum_{\sigma = \pm} \sigma
    \alpha_{\sigma} \mathbf{a}_{\sigma}^{(1)}, \\
    \mathbf{a}^{(2)} & = - \sum_{\sigma = \pm} \alpha_{\sigma}^2
    \mathbf{a}_{\sigma}^{(2)} + \alpha_+ \alpha_- \mathbf{a}_{+-}^{(2)},
  \end{split}
\end{equation}
where
\begin{equation}
  \label{eq:a^1_sigma}
  \mathbf{a}_{\sigma}^{(1)}(\mathbf{r}) = \frac{1}{2 \pi} \int_{\mathbf{r}'} \mathbf{f}(\mathbf{r} - \mathbf{r}')
  \left( \mathbf{f}(\mathbf{r}' - \mathbf{r}_+) -
    \mathbf{f}(\mathbf{r}' - \mathbf{r}_-) \right)^{\odot 2},  
\end{equation}
and
\begin{multline}
  \label{eq:a^2_sigma}
  \mathbf{a}_{\sigma}^{(2)}(\mathbf{r}) = \frac{1}{\pi} \int_{\mathbf{r}'}
  \mathbf{f}(\mathbf{r} - \mathbf{r}') \left[ \left( \mathbf{f}(\mathbf{r}' -
      \mathbf{r}_+) - \mathbf{f}(\mathbf{r}' - \mathbf{r}_-) \right)
    \vphantom{\left( \unitvec{z} \times \mathbf{a}_{\sigma}^{(1)}(\mathbf{r}')
      \right)} \right. \\ \left. \odot \left( \unitvec{z} \times
      \mathbf{a}_{\sigma}^{(1)}(\mathbf{r}') \right) \right],
\end{multline}
and finally
\begin{multline}
  \label{eq:a^2_+-}
  \mathbf{a}_{+-}^{(2)}(\mathbf{r}) = \frac{1}{\pi} \int_{\mathbf{r}'}
  \mathbf{f}(\mathbf{r} - \mathbf{r}') \left( \mathbf{f}(\mathbf{r}' -
    \mathbf{r}_+) - \mathbf{f}(\mathbf{r}' - \mathbf{r}_-) \right)^T \\ \left[
    \left( \unitvec{z} \times \mathbf{a}_-^{(1)}(\mathbf{r}') \right) + \sigma_z
    \left( \unitvec{z} \times \mathbf{a}_+^{(1)}(\mathbf{r}') \right) \right].
\end{multline}

In terms of the quantities defined above, the vector potential is given by
$\mathbf{A} = - \unitvec{z} \times \mathbf{a}$, and up to second order in the
KPZ non-linearity the static electric field can be written as
\begin{equation}
  \label{eq:E_expansion}
  \mathbf{E} = \frac{1}{\varepsilon} \mathbf{E}^{(0)} - \sum_{\sigma = \pm}
  \left[ \frac{\sigma \alpha_{\sigma}}{\varepsilon^2} \mathbf{E}_{\sigma}^{(1)}
    + \frac{\alpha_{\sigma}^2}{\varepsilon^3} \mathbf{E}_{\sigma}^{(2)} \right]
   + \frac{\alpha_+ \alpha_-}{\varepsilon^3} \mathbf{E}_{+-}^{(2)},
\end{equation}
where
\begin{equation}
  \label{eq:E_012}
  \begin{split}
    \mathbf{E}^{(0)}(\mathbf{r}) & = - \varepsilon \nabla \phi(\mathbf{r}), \\
    \mathbf{E}^{(1,2)}_{\sigma}(\mathbf{r}) & = \unitvec{z} \times
    \mathbf{a}^{(1,2)}_{\sigma}(\mathbf{r}), \\
    \mathbf{E}^{(2)}_{+-}(\mathbf{r}) & = \unitvec{z} \times
    \mathbf{a}^{(2)}_{+-}(\mathbf{r}).
  \end{split}
\end{equation}

In the following sections, we evaluate the integrals in
Eqs.~\eqref{eq:a^1_sigma}, \eqref{eq:a^2_sigma}, and~\eqref{eq:a^2_+-} both
analytically and --- to check the rather lengthy analytical calculations ---
numerically. For completeness, we also repeat the calculation of the purely
isotropic corrections, which can also be found in
Ref.~\cite{Wachtel2016}. Actually, we do not need to find the electric field at
arbitrary points in space, since the force acting on the charges is determined
by the electric field at the position of one of the charges. Therefore, we
evaluate the second-order corrections~\eqref{eq:a^2_sigma} and~\eqref{eq:a^2_+-}
only at $\mathbf{r} = \mathbf{r}_+$. The first-order
correction~\eqref{eq:a^1_sigma}, however, has to be calculated for any
$\mathbf{r}$, since it is required in~\eqref{eq:a^2_sigma}
and~\eqref{eq:a^2_+-}.

Equations~\eqref{eq:a^1_sigma}, \eqref{eq:a^2_sigma}, and~\eqref{eq:a^2_+-} are
integrals over products of the function $\mathbf{f}(\mathbf{r})$ defined in
Eq.~\eqref{eq:f} with different arguments. The pole of
$\mathbf{f}(\mathbf{r}) = -\mathbf{r}/r^2$ at $\mathbf{r} = 0$, which is cut off
at the scale $a$, can lead to logarithmic contributions to the integrals in the
limit $a \to 0$. In some cases, these singular contributions are lifted by the
angular integration. The main theme of the calculation we present in the
following is therefore to identify the poles that do give singular
contributions. Once these poles have been identified, the integrals can be
evaluated by shifting the integration variable such that the ``dangerous'' poles
are at the origin $\mathbf{r} = 0$, and the corresponding integration has to be
cut at $r = a$, while the remaining integrals can be extended over the entire
plane.

\subsubsection{First order correction}
\label{sec:first-order-corr}

We split the first order correction in Eq.~\eqref{eq:a^1_sigma} into three contributions
\begin{equation}
  \label{eq:a_1}
  \mathbf{a}_{\sigma}^{(1)}(\mathbf{r}) = \mathbf{a}^{(1)}_{\sigma, +}(\mathbf{r}) - 2
  \mathbf{a}^{(1)}_{\sigma, +-}(\mathbf{r}) +
  \mathbf{a}^{(1)}_{\sigma, -}(\mathbf{r}),
\end{equation}
where (recall that for $\sigma = +$ the $\odot$ product should be replaced by
the usual scalar product)
\begin{align}
  \mathbf{a}^{(1)}_{\sigma, \pm}(\mathbf{r})
  & = \frac{1}{2 \pi}
    \int_{\mathbf{r}'} \mathbf{f}(\mathbf{r} - \mathbf{r}')
    \mathbf{f}(\mathbf{r}' - \mathbf{r}_{\pm})^{\odot 2}, \\
  \label{eq:a^1_pm}
  \mathbf{a}^{(1)}_{\sigma, +-}(\mathbf{r})
  & = \frac{1}{2 \pi}
    \int_{\mathbf{r}'} \mathbf{f}(\mathbf{r} - \mathbf{r}') \left(
    \mathbf{f}(\mathbf{r}' - \mathbf{r}_+) \odot \mathbf{f}(\mathbf{r}' -
    \mathbf{r}_-) \right).
\end{align}
Let's consider $\mathbf{a}^{(1)}_{\sigma, \pm}(\mathbf{r})$ first. Shifting the
integration variable as $\mathbf{r}' \to \mathbf{r}' + \mathbf{r}_{\pm}$ and
denoting $\mathbf{R}_{\pm} = \mathbf{r} - \mathbf{r}_{\pm}$, we obtain
\begin{equation}  
  \mathbf{a}^{(1)}_{\sigma, \pm}(\mathbf{r}) = \frac{1}{2 \pi} \int_{\mathbf{r}'}
  \mathbf{f}(\mathbf{R}_{\pm} - \mathbf{r}') \mathbf{f}(\mathbf{r}')^{\odot 2}.
\end{equation}
This and many of the following integrals are conveniently performed using {\sc
  Mathematica}, resulting in
\begin{align}
  \label{eq:a^1_+_result}
  \mathbf{a}^{(1)}_{+, \pm}(\mathbf{r}) & = \mathbf{f}(\mathbf{R}_{\pm})
                                          \ln(R_{\pm}/a), \\
  \label{eq:a^1_-_result}
  \mathbf{a}^{(1)}_{-, \pm}(\mathbf{r}) & = \mathbf{e}(\mathbf{R}_{\pm}),
\end{align}
where
\begin{multline}
  \mathbf{e}(\mathbf{r}) = -\frac{\cos(\theta)
    \sin(\theta)}{r}
  \begin{pmatrix}
    - \sin(\theta) \\ \cos(\theta)
  \end{pmatrix}
  \\ = - \cos(\theta) \sin(\theta) \frac{\unitvec{z} \times \mathbf{r}}{r^2} = -
  x y \frac{\unitvec{z} \times \mathbf{r}}{r^4}.
\end{multline}
Note that these expressions are valid for $R_{\pm} > a$ and should be set to
zero below the cutoff $a$.

The calculation of $\mathbf{a}^{(1)}_{\sigma, +-}(\mathbf{r})$ is more
involved. In Eq.~\eqref{eq:a^1_pm}, we replace
$\mathbf{r}' \to \mathbf{r}' + \mathbf{r}$ and as above we write
$\mathbf{R}_{\pm} = \mathbf{r} - \mathbf{r}_{\pm}$, which yields
\begin{equation}  
  \begin{split}
    \mathbf{a}^{(1)}_{\sigma, +-}(\mathbf{r}) & = - \frac{1}{2 \pi}
    \int_{\mathbf{r}'} \mathbf{f}(\mathbf{r}') \left(
      \mathbf{f}(\mathbf{r}' + \mathbf{R}_+) \odot \mathbf{f}(\mathbf{r}' + \mathbf{R}_-) \right)
    \\ & = \frac{1}{2 \pi} \int_{\mathbf{r}'} \frac{\mathbf{r}'}{r^{\prime 2}}
    \frac{\left( \mathbf{r}' + \mathbf{R}_+ \right) \odot
      \left( \mathbf{r}' + \mathbf{R}_- \right)}{\abs{\mathbf{r}' + \mathbf{R}_+}^2
      \abs{\mathbf{r}' + \mathbf{R}_-}^2}.
  \end{split}
\end{equation}
To evaluate these integrals, we switch to polar coordinates for
$\mathbf{r}'$ and $\mathbf{R}_{\pm}$:
\begin{equation}  
  \mathbf{r}' = r'
  \begin{pmatrix}
    \cos(\theta' + \theta_+) \\ \sin(\theta' + \theta_+)
  \end{pmatrix}
  , \quad
  \mathbf{R}_{\pm} = R_{\pm}
  \begin{pmatrix}
    \cos(\theta_{\pm}) \\ \sin(\theta_{\pm})
  \end{pmatrix}
  .
\end{equation}
Moreover, we use the following Fourier-cosine series:
\begin{multline}  
  \frac{1}{\abs{\mathbf{r}' + \mathbf{R}_-}^2} = \frac{1}{\abs{r^{\prime
        2} - R_-^2}} \sum_{n = 0}^{\infty} \left( 2 - \delta_{n,0} \right) \\
  \times \left( - \frac{r_{<}}{r_{>}} \right)^n \cos(n (\theta' + \theta_+ -
  \theta_-)),
\end{multline}
where $r_{<}$ and $r_{>}$ are the lesser and greater, respectively, of $r'$ and
$R_-$. Finally, we set
\begin{equation}  
  \frac{1}{\abs{\mathbf{r}' + \mathbf{R}_+}^2} = \frac{1}{r^{\prime 2} + R_+^2} \frac{1}{1 + s'
  \cos(\theta')}, \quad s' = \frac{2 r' R_+}{r^{\prime 2} + R_+^2}.
\end{equation}
Then, Eq.~\eqref{eq:a^1_pm} becomes
\begin{multline}  
  \mathbf{a}^{(1)}_{\sigma, +-}(\mathbf{r}) = \frac{1}{2 \pi} \sum_{n =
    0}^{\infty} \left( 2 - \delta_{n, 0} \right) \int_{\mathbf{r}'}
  \frac{1}{r^{\prime 2}} \frac{1}{r^{\prime 2} + R_+^2} \\ \times
  \frac{1}{\abs{r^{\prime 2} - R_-^2}} \left( - \frac{r_{<}}{r_{>}} \right)^n
  \cos(n (\theta' + \theta_+ - \theta_-)) \\ \times \mathbf{r}' \frac{\left(
      \mathbf{r}' + \mathbf{R}_+ \right) \odot \left( \mathbf{r}' + \mathbf{R}_-
    \right)}{1 + s' \cos(\theta')}.
\end{multline}
After some lengthy but straightforward algebra, the angular integrals can be
preformed using the relation~\cite{Gradshteyn2007}
\begin{equation}
  \label{eq:angular_int}
  \int_0^{2 \pi} d \theta \frac{\cos(n \theta)}{1 + s \cos(\theta)} = \frac{2
    \pi}{\sqrt{1 - s^2}} \left( \frac{\sqrt{1 - s^2} - 1}{s} \right)^n,  
\end{equation}
which holds for $s^2 < 1$ and $n \geq 0$. In the resulting expression, the
summation over $n$ can be carried out, and finally performing the integral over
$r'$ yields the result:
\begin{equation}  
  \mathbf{a}^{(1)}_{+, +-}(\mathbf{r}) =  - \frac{1}{2} \left(
    \mathbf{f}(\mathbf{R}_+) \ln(R/R_-) +
    \mathbf{f}(\mathbf{R}_-) \ln(R/R_+)
  \right),
\end{equation}
where $R$ is the magnitude of $\mathbf{R} = \mathbf{r}_+ - \mathbf{r}_-$. We
omit the cumbersome expression for $\mathbf{a}^{(1)}_{-, +-}(\mathbf{r})$.
Combining these results with Eqs.~\eqref{eq:a^1_+_result} and
\eqref{eq:a^1_-_result} gives the first order correction
$\mathbf{E}^{(1)}(\mathbf{r})$. Again, we omit the rather lengthy expression.

The calculation of $\mathbf{a}^{(1)}_{\sigma, +-}(\mathbf{r})$ is actually much
simpler for the special case $\mathbf{r} = \mathbf{r}_+$ that gives the electric
field acting on the charge at $\mathbf{r}_+$. Then, shifting
$\mathbf{r}' \to \mathbf{r} + \mathbf{r}_+$, Eq.~\eqref{eq:a^1_pm} becomes
\begin{equation}  
  \mathbf{a}^{(1)}_{\sigma, +-}(\mathbf{r}_+)
  = - \frac{1}{2 \pi}
  \int_{\mathbf{r}} \mathbf{f}(\mathbf{r}) \left(
    \mathbf{f}(\mathbf{r}) \odot \mathbf{f}(\mathbf{r} +
    \mathbf{R}) \right),
\end{equation}
where we used $\mathbf{f}(-\mathbf{r}) = - \mathbf{f}(\mathbf{r})$. Again,
\textsc{Mathematica} does the job, and combining the result with
Eqs.~\eqref{eq:a^1_+_result} and~\eqref{eq:a^1_-_result} we obtain
\begin{align}
  \label{eq:a^1_+_r_+}
  \mathbf{a}_+^{(1)}(\mathbf{r}_+) & = \frac{1}{2} \mathbf{f}(\mathbf{R}) \left( 4 \ln(R/a) - 1
                                     \right), \\
  \label{eq:a^1_-_r_+}
  \mathbf{a}_-^{(1)}(\mathbf{r}_+) & = \frac{3}{2} \mathbf{f}(\mathbf{R}) \cos(2
                                     \theta_{\mathbf{R}}) - \frac{1}{R^2}
                             \begin{pmatrix}
                               R_x \\ - R_y
                             \end{pmatrix}
  \left( \ln(R/a) - 1 \right),
\end{align}
where we used the polar representation of $\mathbf{R}$,
\begin{equation}  
  \mathbf{R} =
  \begin{pmatrix}
    R_x \\ R_y
  \end{pmatrix}
  = R
  \begin{pmatrix}
    \cos(\theta_{\mathbf{R}}) \\ \sin(\theta_{\mathbf{R}})
  \end{pmatrix}.
\end{equation}

Equations~\eqref{eq:a^1_+_r_+} and~\eqref{eq:a^1_-_r_+} give the first order
corrections to the electric field at the position of the positive charge,
\begin{equation}
  \mathbf{E}_+^{(1)}(\mathbf{r}_+) = \frac{1}{2} \unitvec{z} \times \mathbf{f}(\mathbf{R}) \left( 4 \ln(R/a) - 1
  \right),
\end{equation}
and
\begin{multline}  
  \mathbf{E}_-^{(1)}(\mathbf{r}_+) = \frac{3}{2} \unitvec{z} \times
  \mathbf{f}(\mathbf{R}) \cos(2 \theta_{\mathbf{R}}) \\ - \frac{1}{R^2}
                             \begin{pmatrix}
                               R_y \\ R_x
                             \end{pmatrix}
  \left( \ln(R/a) - 1 \right).
\end{multline}

\begin{widetext}
\subsubsection{Second order correction: diagonal terms}
\label{sec:second-order-corr-diag}

For $\mathbf{r} = \mathbf{r}_+$, the second-order correction
Eq.~\eqref{eq:a^2_sigma} becomes
\begin{equation}  
  \mathbf{a}_{\sigma}^{(2)}(\mathbf{r}_+) = - \frac{1}{\pi} \int_{\mathbf{r}} \mathbf{f}(\mathbf{r})
  \left[ \left( \mathbf{f}(\mathbf{r}) - \mathbf{f}(\mathbf{r} + \mathbf{R})
    \right) \odot \left( \unitvec{z}
      \times \mathbf{a}_{\sigma}^{(1)}(\mathbf{r} + \mathbf{r}_+) \right) \right].
\end{equation}
We decompose the second order correction in two contributions,
\begin{equation}  
  \mathbf{a}_{\sigma}^{(2)}(\mathbf{r}_+) = \mathbf{a}^{(2)}_{\sigma, 1}(\mathbf{r}_+) +
  \mathbf{a}^{(2)}_{\sigma, 2}(\mathbf{r}_+), 
\end{equation}
where (cf.\ Eq.~\eqref{eq:a_1})
\begin{equation}
  \label{eq:a^2_sigma_12}
  \begin{split}
    \mathbf{a}^{(2)}_{\sigma, 1}(\mathbf{r}_+) & = - \frac{1}{\pi} \int_{\mathbf{r}} \mathbf{f}(\mathbf{r})
                                     \left\{ \left( \mathbf{f}(\mathbf{r}) - \mathbf{f}(\mathbf{r} + \mathbf{R})
                                     \right) \odot \left[ \unitvec{z}
                                     \times \left( \mathbf{a}_{\sigma, +}^{(1)}(\mathbf{r} +
                                     \mathbf{r}_+) +  \mathbf{a}_{\sigma, -}^{(1)}(\mathbf{r} +
                                     \mathbf{r}_+) \right) \right] \right\}, \\
  \mathbf{a}^{(2)}_{\sigma, 2}(\mathbf{r}_+) & = \frac{2}{\pi} \int_{\mathbf{r}} \mathbf{f}(\mathbf{r})
                                     \left[ \left( \mathbf{f}(\mathbf{r}) - \mathbf{f}(\mathbf{r} + \mathbf{R})
                                     \right) \odot \left( \unitvec{z}
                                     \times \mathbf{a}^{(1)}_{\sigma, +-}(\mathbf{r} +
                                     \mathbf{r}_+) \right) \right].
  \end{split}
\end{equation}
To proceed with the calculation of $\mathbf{a}_{\sigma, 1}^{(2)}(\mathbf{r}_+)$,
we have to specify whether we are dealing with the isotropic or fully
anisotropic case. Before going into that, let us simplify the expression for
$\mathbf{a}^{(2)}_{\sigma, 2}(\mathbf{r}_+)$. Here, ``simplifying'' refers to
splitting into two parts,
\begin{equation}  
  \mathbf{a}^{(2)}_{\sigma, 2}(\mathbf{r}_+) = \mathbf{a}^{(2)}_{\sigma, 2, 1}(\mathbf{r}_+) +
  \mathbf{a}^{(2)}_{\sigma, 2 ,2}(\mathbf{r}_+),
\end{equation}
where
\begin{align}
  \label{eq:a^2_sigma_21_foo}
  \mathbf{a}^{(2)}_{\sigma, 2, 1}(\mathbf{r}_+) & = \frac{2}{\pi} \int_{\mathbf{r}} \mathbf{f}(\mathbf{r})
                                          \left[ \mathbf{f}(\mathbf{r}) \odot \left( \unitvec{z}
                                          \times \mathbf{a}^{(1)}_{\sigma, +-}(\mathbf{r} +
                                          \mathbf{r}_+) \right) \right], \\
  \label{eq:a^2_sigma_22_foo}
  \mathbf{a}^{(2)}_{\sigma, 2, 2}(\mathbf{r}_+) & = - \frac{2}{\pi} \int_{\mathbf{r}} \mathbf{f}(\mathbf{r})
                                          \left[ \mathbf{f}(\mathbf{r} + \mathbf{R})
                                          \odot \left( \unitvec{z}
                                          \times \mathbf{a}^{(1)}_{\sigma, +-}(\mathbf{r} +
                                          \mathbf{r}_+) \right) \right].
\end{align}
Copying from Eq.~\eqref{eq:a^1_pm}, we find
\begin{equation}
  \label{eq:a^1_sigma_+-_rr_+}  
    \mathbf{a}^{(1)}_{\sigma, +-}(\mathbf{r} + \mathbf{r}_+) = \frac{1}{2 \pi}
    \int_{\mathbf{r}'} \mathbf{f}(\mathbf{r} + \mathbf{r}_+ - \mathbf{r}')
    \left( \mathbf{f}(\mathbf{r}' - \mathbf{r}_+) \odot \mathbf{f}(\mathbf{r}' -
      \mathbf{r}_-) \right) = \frac{1}{2 \pi} \int_{\mathbf{r}'}
    \mathbf{f}(\mathbf{r} - \mathbf{r}') \left(
      \mathbf{f}(\mathbf{r}') \odot \mathbf{f}(\mathbf{r}' + 
      \mathbf{R}) \right).
\end{equation}
Then, we can write Eq.~\eqref{eq:a^2_sigma_21_foo} as
\begin{equation}
  \label{eq:a^2_sigma_21}  
  \mathbf{a}^{(2)}_{\sigma, 2, 1}(\mathbf{r}_+) = \frac{1}{\pi^2}
  \int_{\mathbf{r}, \mathbf{r}'} \mathbf{f}(\mathbf{r}) \left[
    \mathbf{f}(\mathbf{r}) \odot \left( \unitvec{z} \times
      \mathbf{f}(\mathbf{r} - \mathbf{r}') \right) \right] \left(
    \mathbf{f}(\mathbf{r}') \odot \mathbf{f}(\mathbf{r}' + \mathbf{R}) \right)
  = \frac{2}{\pi} \int_{\mathbf{r}'} \mathbf{c}_{\sigma}(\mathbf{r}') \left(
    \mathbf{f}(\mathbf{r}') \odot \mathbf{f}(\mathbf{r}' + \mathbf{R})
  \right),
\end{equation}
where
\begin{equation}
  \label{eq:c_sigma}
  \mathbf{c}_{\sigma}(\mathbf{r}') = \frac{1}{2 \pi} \int_{\mathbf{r}}
  \mathbf{f}(\mathbf{r}) \left[ \mathbf{f}(\mathbf{r}) \odot
    \left( \unitvec{z} \times \mathbf{f}(\mathbf{r} - \mathbf{r}') \right)
  \right].
\end{equation}
Finally, plugging Eq.~\eqref{eq:a^1_sigma_+-_rr_+} into Eq.~\eqref{eq:a^2_sigma_22_foo}, the latter becomes
\begin{equation}
  \label{eq:a^2_sigma_22}
  \mathbf{a}^{(2)}_{\sigma, 2, 2}(\mathbf{r}_+) = - \frac{1}{\pi^2} \int_{\mathbf{r},
    \mathbf{r}'} \mathbf{f}(\mathbf{r}) \left[ \mathbf{f}(\mathbf{r} +
    \mathbf{R}) \odot \left( \unitvec{z} \times \mathbf{f}(\mathbf{r} -
        \mathbf{r}') \right) \right] \left( \mathbf{f}(\mathbf{r}')
    \odot \mathbf{f}(\mathbf{r}' + \mathbf{R}) \right).
\end{equation}
So far, we have split the second order correction into three contributions,
given by Eqs.~\eqref{eq:a^2_sigma_12},~\eqref{eq:a^2_sigma_21}, and~\eqref{eq:a^2_sigma_22}. In the
following, we calculate those, first for $\sigma = +$ and then for $\sigma = -$.

\paragraph{$\sigma = +$}
\label{sec:isotropic-case}

We start with $\mathbf{a}^{(2)}_{+, 1}(\mathbf{r}_+)$ defined in
Eq.~\eqref{eq:a^2_sigma_12}. Using Eq.~\eqref{eq:a^1_+_result} we obtain
\begin{equation}
  \label{eq:a^1_++}
  \mathbf{a}_{+,+}^{(1)}(\mathbf{r} + \mathbf{r}_+) +  \mathbf{a}_{+,-}^{(1)}(\mathbf{r}
  + \mathbf{r}_+) = \mathbf{f}(\mathbf{r}) \ln(r/a) + \mathbf{f}(\mathbf{r} +
  \mathbf{R}) \ln(\abs{\mathbf{r} + \mathbf{R}}/a),
\end{equation}
and inserting this relation in Eq.~\eqref{eq:a^2_sigma_12} leaves us with
\begin{equation}
  \label{eq:a^2_+1}
  \begin{split}
    \mathbf{a}^{(2)}_{+, 1}(\mathbf{r}_+) & = \frac{1}{\pi} \int_{\mathbf{r}}
    \mathbf{f}(\mathbf{r}) \left\{ \unitvec{z} \cdot \left[ \left(
          \mathbf{f}(\mathbf{r}) - \mathbf{f}(\mathbf{r} + \mathbf{R}) \right)
        \times \left( \mathbf{f}(\mathbf{r}) \ln(r/a) + \mathbf{f}(\mathbf{r} +
          \mathbf{R}) \ln(\abs{\mathbf{r} + \mathbf{R}}/a) \right) \right]
    \right\} \\ & = \frac{1}{\pi} \int_{\mathbf{r}} \mathbf{f}(\mathbf{r})
    \left[ \unitvec{z} \cdot \left( \mathbf{f}(\mathbf{r}) \times
        \mathbf{f}(\mathbf{r} + \mathbf{R}) \right) \right] \ln(r
    \abs{\mathbf{r} + \mathbf{R}}/a^2) \\ & = \mathbf{b}_1(\mathbf{R}) +
    \mathbf{b}_2(\mathbf{R}),
  \end{split}
\end{equation}
where
\begin{align}  
  \mathbf{b}_1(\mathbf{R}) & = \frac{1}{\pi} \int_{\mathbf{r}} \mathbf{f}(\mathbf{r})
                             \left[ \unitvec{z} \cdot \left( \mathbf{f}(\mathbf{r}) \times
                             \mathbf{f}(\mathbf{r} + \mathbf{R}) \right) \right]
                             \ln(r R /a^2), \\
  \mathbf{b}_2(\mathbf{R}) & = \frac{1}{\pi} \int_{\mathbf{r}}
                             \mathbf{f}(\mathbf{r})
                             \left[ \unitvec{z} \cdot \left( \mathbf{f}(\mathbf{r}) \times
                             \mathbf{f}(\mathbf{r} + \mathbf{R}) \right) \right]
                             \ln(\abs{\mathbf{r} + \mathbf{R}}/R) =
                             \frac{1}{\pi} \int_{\mathbf{r}}
                             \mathbf{f}(\mathbf{r} - \mathbf{R})
                             \left[ \unitvec{z} \cdot \left(
                             \mathbf{f}(\mathbf{r} - \mathbf{R}) \times
                             \mathbf{f}(\mathbf{r}) \right) \right]
                             \ln(r/R).
\end{align}
The results read as follows:
\begin{align}  
  \mathbf{b}_1(\mathbf{R}) & = - \frac{1}{4} \unitvec{z} \times
                             \mathbf{f}(\mathbf{R}) \left( 6 \ln(R/a)^2 + 4
                             \ln(R/a) + 1 \right), \\
  \mathbf{b}_2(\mathbf{R}) & = 0,
\end{align}
and hence we obtain for the first part of the second order correction:
\begin{equation}  
  \mathbf{a}^{(2)}_{+, 1}(\mathbf{r}_+) = \mathbf{b}_1(\mathbf{R}).
\end{equation}

We move on to calculate $\mathbf{a}^{(2)}_{+, 2, 1}(\mathbf{r}_+)$, given by
Eq.~\eqref{eq:a^2_sigma_21}, and find
\begin{equation}  
  \mathbf{a}^{(2)}_{+, 2, 1}(\mathbf{r}_+) = - \frac{1}{2} \unitvec{z} \times
  \mathbf{f}(\mathbf{R}) \left( \ln(R/a)^2 - 1 \right).
\end{equation}

The nastiest part by far is $\mathbf{a}^{(2)}_{+, 2, 2}(\mathbf{r}_+)$, which can be
written as
\begin{equation}  
  \begin{split}
    \mathbf{a}^{(2)}_{+, 2, 2}(\mathbf{r}_+) & = - \frac{1}{\pi^2}
    \int_{\mathbf{r}, \mathbf{r}'} \mathbf{f}(\mathbf{r}) \left[
      \mathbf{f}(\mathbf{r} + \mathbf{R}) \cdot \left( \unitvec{z} \times
        \mathbf{f}(\mathbf{r} - \mathbf{r}') \right) \right] \left(
      \mathbf{f}(\mathbf{r}') \cdot \mathbf{f}(\mathbf{r}' + \mathbf{R}) \right)
    \\ & = - \frac{1}{\pi^2} \int_{\mathbf{r}, \mathbf{r}'}
    \mathbf{f}(\mathbf{r}) \left[ \unitvec{z} \cdot \left( \mathbf{f}(\mathbf{r}
        - \mathbf{r}') \times \mathbf{f}(\mathbf{r} + \mathbf{R}) \right)
    \right] \left( \mathbf{f}(\mathbf{r}') \cdot \mathbf{f}(\mathbf{r}' +
      \mathbf{R}) \right) \\ & = \frac{1}{\pi^2} \int_{\mathbf{r},
      \mathbf{r}'} \frac{\mathbf{r}}{r^2} \frac{\unitvec{z} \cdot \left[
        \mathbf{r} \times \mathbf{R} - \mathbf{r}' \times \left( \mathbf{r} +
          \mathbf{R} \right) \right]}{\abs{\mathbf{r} - \mathbf{r}'}^2
      \abs{\mathbf{r} + \mathbf{R}}^2} \frac{\mathbf{r}' \cdot \left(
        \mathbf{r}' + \mathbf{R} \right)}{r^{\prime 2} \abs{\mathbf{r}' +
        \mathbf{R}}^2}.
  \end{split}
\end{equation}
To evaluate these integrals, we switch to polar coordinates for
$\mathbf{r}, \mathbf{r}',$ and $\mathbf{R}$:
\begin{equation}  
  \mathbf{r} = r
  \begin{pmatrix}
    \cos(\theta + \theta_{\mathbf{R}}) \\ \sin(\theta + \theta_{\mathbf{R}})
  \end{pmatrix}
  , \quad \mathbf{r}' = r'
  \begin{pmatrix}
    \cos(\theta' + \theta_{\mathbf{R}}) \\ \sin(\theta' + \theta_{\mathbf{R}})
  \end{pmatrix}
  , \quad \mathbf{R} = R
  \begin{pmatrix}
    \cos(\theta_{\mathbf{R}}) \\ \sin(\theta_{\mathbf{R}})
  \end{pmatrix}.
\end{equation}
Moreover, we use the following Fourier-cosine series:
\begin{equation}
  \label{eq:cos_r_rp}
  \frac{1}{\abs{\mathbf{r} - \mathbf{r}'}^2} = \frac{1}{\abs{r^2 - r^{\prime
        2}}} \sum_{n = 0}^{\infty} \left( 2 - \delta_{n,0} \right) \left(
    \frac{r_{<}}{r_{>}} \right)^n \cos(n (\theta - \theta')),
\end{equation}
where $r_{<}$ and $r_{>}$ are the lesser and greater, respectively, of $r$ and
$r'$. Finally, we write
\begin{equation}
  \label{eq:s}
  \frac{1}{\abs{\mathbf{r} + \mathbf{R}}^2} = \frac{1}{r^2 + R^2} \frac{1}{1 + s
  \cos(\theta)}, \qquad s = \frac{2 r R}{r^2 + R^2},
\end{equation}
and we use an analogous representation with $\mathbf{r}$ replaced by
$\mathbf{r}'$. This leads us to
\begin{multline}  
  \mathbf{a}^{(2)}_{+, 2, 2}(\mathbf{r}_+) = \frac{1}{\pi^2} \sum_{n = 0}^{\infty}
  \left( 2 - \delta_{n, 0} \right) \int_{\mathbf{r}, \mathbf{r}'} \frac{1}{r^2 +
    R^2} \frac{1}{r^{\prime 2} + R^2} \frac{1}{\abs{r^2 - r^{\prime 2}}} \left(
    \frac{r_{<}}{r_{>}} \right)^n \cos(n (\theta - \theta')) \\ \times
  \frac{\mathbf{r}}{r^2 r^{\prime 2}} \frac{\unitvec{z} \cdot \left[ \mathbf{r}
      \times \mathbf{R} - \mathbf{r}' \times \left( \mathbf{r} + \mathbf{R}
      \right) \right]}{1 + s \cos(\theta)} \frac{\mathbf{r}' \cdot \left(
      \mathbf{r}' + \mathbf{R} \right)}{1 + s' \cos(\theta')}.
\end{multline}
We then proceed to symmetrize the integrand with respect to $\theta \to -\theta$
and $\theta' \to -\theta'$, and to rearrange the trigonometric functions in the
numerator such that the angular integrals can be preformed using
Eq.~\eqref{eq:angular_int}. In the result, the summation over $n$ can be carried
out straightforwardly. Performing the integrals over $r$ and $r'$ leads us to
\begin{equation}  
  \mathbf{a}^{(2)}_{+, 2, 2}(\mathbf{r}_+) = 0.
\end{equation}

Hence, the second order correction to the electric field at the position of the
positive charge is given by
\begin{equation}
  \label{eq:E^2_+_r_+}
  \mathbf{E}_+^{(2)}(\mathbf{r}_+) = \unitvec{z} \times \mathbf{a}^{(2)}_{+, 1}(\mathbf{r}_+) +
  \unitvec{z} \times \mathbf{a}^{(2)}_{+, 2, 1}(\mathbf{r}_+) = \frac{1}{4} \mathbf{f}(\mathbf{R}) \left( 8
    \ln(R/a)^2 + 4 \ln(R/a) - 1 \right).
\end{equation}

\paragraph{$\sigma = -$}
\label{sec:fully-anis-case}

According to Eq.~\eqref{eq:a^1_-_result},
\begin{equation}
  \label{eq:z_cross_a^1}
  \begin{split}
    \unitvec{z} \times \left( \mathbf{a}_{-,+}^{(1)}(\mathbf{r} + \mathbf{r}_+)
      + \mathbf{a}_{-, -}^{(1)}(\mathbf{r} + \mathbf{r}_+) \right) & =
    \unitvec{z} \times \left( \mathbf{e}(\mathbf{r}) + \mathbf{e}(\mathbf{r} +
      \mathbf{R}) \right) \\ & = - \cos(\theta) \sin(\theta)
    \mathbf{f}(\mathbf{r}) - \cos(\theta_{\mathbf{r} + \mathbf{R}})
    \sin(\theta_{\mathbf{r} + \mathbf{R}}) \mathbf{f}(\mathbf{r} + \mathbf{R}),
  \end{split}
\end{equation}
where we used
\begin{equation}  
  \unitvec{z} \times \mathbf{e}(\mathbf{r}) = \cos(\theta) \sin(\theta)
  \frac{\mathbf{r}}{r^2} = - \cos(\theta) \sin(\theta)
  \mathbf{f}(\mathbf{r}).
\end{equation}
Inserting Eq.~\eqref{eq:z_cross_a^1} in Eq.~\eqref{eq:a^2_sigma_12}, we obtain
\begin{equation}  
  \begin{split}
    \mathbf{a}^{(2)}_{-, 1}(\mathbf{r}_+) & = \frac{1}{\pi} \int_{\mathbf{r}}
    \mathbf{f}(\mathbf{r}) \left[ \left( \mathbf{f}(\mathbf{r}) -
        \mathbf{f}(\mathbf{r} + \mathbf{R}) \right) \odot \left( \cos(\theta)
        \sin(\theta) \mathbf{f}(\mathbf{r}) + \cos(\theta_{\mathbf{r} +
          \mathbf{R}}) \sin(\theta_{\mathbf{r} +
          \mathbf{R}}) \mathbf{f}(\mathbf{r} + \mathbf{R}) \right) \right] \\ &
    = \mathbf{k}_1(\mathbf{R}) + \mathbf{k}_2(\mathbf{R}) +
    \mathbf{k}_3(\mathbf{R}),
  \end{split}
\end{equation}
where
\begin{align}  
  \mathbf{k}_1(\mathbf{R})
  & = \frac{1}{\pi} \int_{\mathbf{r}} \cos(\theta) \sin(\theta)
    \mathbf{f}(\mathbf{r}) \left[ \left(
    \mathbf{f}(\mathbf{r}) - \mathbf{f}(\mathbf{r} + \mathbf{R}) \right)
    \odot \mathbf{f}(\mathbf{r}) \right], \\ \mathbf{k}_2(\mathbf{R})
  & = - \frac{1}{\pi} \int_{\mathbf{r}} \cos(\theta_{\mathbf{r} + \mathbf{R}})
    \sin(\theta_{\mathbf{r} + \mathbf{R}}) \mathbf{f}(\mathbf{r}) \left(
    \mathbf{f}(\mathbf{r} + \mathbf{R}) \right)^{\odot 2}
  \\
  & = - \frac{1}{\pi} \int_{\mathbf{r}} \cos(\theta) \sin(\theta)
    \mathbf{f}(\mathbf{r} - \mathbf{R}) \left( \mathbf{f}(\mathbf{r})
    \right)^{\odot 2}, \\ \mathbf{k}_3(\mathbf{R})
  & = \frac{1}{\pi} \int_{\mathbf{r}} \cos(\theta_{\mathbf{r} +
    \mathbf{R}}) \sin(\theta_{\mathbf{r} + \mathbf{R}}) \mathbf{f}(\mathbf{r})
    \left( \mathbf{f}(\mathbf{r}) \odot \mathbf{f}(\mathbf{r} + \mathbf{R})
    \right).
\end{align}
Both $\mathbf{k}_{1,2}(\mathbf{R})$ can be calculated directly as before; to
simplify $\mathbf{k}_3(\mathbf{R})$, we parameterize $\mathbf{r}$ as
\begin{equation}  
  \mathbf{r} = r
  \begin{pmatrix}
    \cos(\theta + \theta_{\mathbf{R}}) \\ \sin(\theta + \theta_{\mathbf{R}})
  \end{pmatrix},
\end{equation}
and use
\begin{equation}  
  \mathbf{r} + \mathbf{R} = \abs{\mathbf{r} + \mathbf{R}}
  \begin{pmatrix}
    \cos(\theta_{\mathbf{r} + \mathbf{R}}) \\ \sin(\theta_{\mathbf{r} + \mathbf{R}})
  \end{pmatrix}
  = r
  \begin{pmatrix}
    \cos(\theta + \theta_{\mathbf{R}}) \\ \sin(\theta + \theta_{\mathbf{R}})
  \end{pmatrix}
  + R
  \begin{pmatrix}
    \cos(\theta_{\mathbf{R}}) \\ \sin(\theta_{\mathbf{R}})
  \end{pmatrix}.
\end{equation}
Taking the product of the components of the last equation, we find
\begin{equation}  
  \cos(\theta_{\mathbf{r} + \mathbf{R}}) \sin(\theta_{\mathbf{r} + \mathbf{R}})
  = \frac{\left( r \cos(\theta + \theta_{\mathbf{R}}) + R
      \cos(\theta_{\mathbf{R}}) \right) \left( r \sin(\theta +
      \theta_{\mathbf{R}}) + R \sin(\theta_{\mathbf{R}})
    \right)}{r^2 + R^2 + 2 r R \cos(\theta)}.
\end{equation}
Using this relation, also $\mathbf{k}_3(\mathbf{R})$ can be calculated by
\textsc{Mathematica}. We omit the cumbersome results for $\mathbf{k}_{1,2,3}$,
and also the result for $\mathbf{a}^{(2)}_{+, 2, 1}(\mathbf{r}_+)$, given by
Eq.~\eqref{eq:a^2_sigma_21}.

It remains to calculate $\mathbf{a}^{(2)}_{-,2,2}(\mathbf{r}_+)$,
\begin{equation}  
  \mathbf{a}^{(2)}_{-, 2, 2}(\mathbf{r}_+) = \frac{1}{\pi^2} \int_{\mathbf{r},
    \mathbf{r}'} \frac{\mathbf{r}}{r^2} \frac{\left( \mathbf{r} + \mathbf{R}
    \right) \odot \left[ \unitvec{z} \times \left( \mathbf{r} - \mathbf{r}'
      \right) \right]}{\abs{\mathbf{r} + \mathbf{R}}^2 \abs{\mathbf{r} -
      \mathbf{r}'}^2} \frac{\mathbf{r}' \odot \left( \mathbf{r}' + \mathbf{R}
    \right)}{r^{\prime 2} \abs{\mathbf{r}' + \mathbf{R}}^2}.  
\end{equation}
Using Eqs.~\eqref{eq:cos_r_rp} and~\eqref{eq:s}, this can be written as
\begin{multline}
  \label{eq:a^2_-_22}
  \mathbf{a}^{(2)}_{-, 2, 2}(\mathbf{r}_+) = \frac{1}{\pi^2} \sum_{n = 0}^{\infty}
  \left( 2 - \delta_{n, 0} \right) \int_{\mathbf{r}, \mathbf{r}'} \frac{1}{r^2 +
    R^2} \frac{1}{r^{\prime 2} + R^2} \frac{1}{\abs{r^2 - r^{\prime 2}}} \left(
    \frac{r_{<}}{r_{>}} \right)^n \cos(n (\theta - \theta')) \\ \times
  \frac{\mathbf{r}}{r^2 r^{\prime 2}} \frac{\left( \mathbf{r} + \mathbf{R}
    \right) \odot \left[ \unitvec{z} \times \left( \mathbf{r} - \mathbf{r}'
      \right) \right]}{1 + s \cos(\theta)} \frac{\mathbf{r}' \odot \left(
      \mathbf{r}' + \mathbf{R} \right)}{1 + s' \cos(\theta')}.
\end{multline}
Repeating similar steps as above, we can get \textsc{Mathematica} to calculate
the final result, which is surprisingly simple:
\begin{equation}  
  \mathbf{a}^{(2)}_{-, 2, 2}(\mathbf{r}_+) = \frac{1}{2 R}
  \begin{pmatrix}
    \sin(3 \theta_{\mathbf{R}}) \\ \cos(3 \theta_{\mathbf{R}})
  \end{pmatrix}.
\end{equation}

For the anisotropic second order correction to the electric field we thus obtain
\begin{equation}  
  \begin{split}
    \mathbf{E}_-^{(2)}(\mathbf{r}_+) & = - \frac{1}{16} \left[
      \mathbf{f}(\mathbf{R}) \left( 8 \ln(R/a)^2 - 20 \ln(R/a) + 15 - 8 \cos(4
        \theta_{\mathbf{R}}) \right) - \frac{6}{R^2}
    \begin{pmatrix}
      R_x \\ - R_y
    \end{pmatrix}
    \cos(2 \theta_{\mathbf{R}}) \left( 4 \ln(R/a) - 5 \right) \right] \\ & \sim -
  \frac{1}{2} \mathbf{f}(\mathbf{R}) \ln(R/a)^2,
  \end{split}
\end{equation}
where the asymptotic expansion corresponds to $R \to \infty$. Remarkably, the
dominant contribution at large distances has the same form as in the isotropic
case~\eqref{eq:E^2_+_r_+}. In particular, it is central and potential.

\subsubsection{Second order correction: mixed terms}
\label{sec:second-order-corr-mixed}

Setting $\mathbf{r} = \mathbf{r}_+$ in Eq.~\eqref{eq:a^2_+-} we obtain
\begin{equation}
  \label{eq:a^2_+-_foo}
  \mathbf{a}_{+-}^{(2)}(\mathbf{r}_+) = - \frac{1}{\pi} \int_{\mathbf{r}}
  \mathbf{f}(\mathbf{r}) \left( \mathbf{f}(\mathbf{r}) - \mathbf{f}(\mathbf{r}
    + \mathbf{R}) \right)^T \left[ \left( \unitvec{z} \times
      \mathbf{a}_-^{(1)}(\mathbf{r} + \mathbf{r}_+) \right) + \sigma_z \left(
      \unitvec{z} \times \mathbf{a}_+^{(1)}(\mathbf{r} + \mathbf{r}_+) \right)
  \right] = \mathbf{h}_1(\mathbf{R}) + \mathbf{h}_2(\mathbf{R}).
\end{equation}
Here, we defined
\begin{align}  
  \mathbf{h}_1(\mathbf{R}) & = - \frac{1}{\pi} \int_{\mathbf{r}}
                             \mathbf{f}(\mathbf{r}) \left[ \left( \mathbf{f}(\mathbf{r}) - \mathbf{f}(\mathbf{r}
                             + \mathbf{R}) \right) \cdot \left( \unitvec{z} \times
                             \mathbf{a}_-^{(1)}(\mathbf{r} + \mathbf{r}_+)
                             \right) \right], \\
  \mathbf{h}_2(\mathbf{R}) & = - \frac{1}{\pi} \int_{\mathbf{r}}
                             \mathbf{f}(\mathbf{r}) \left[ \left( \mathbf{f}(\mathbf{r}) - \mathbf{f}(\mathbf{r}
                             + \mathbf{R}) \right) \odot \left(
                             \unitvec{z} \times \mathbf{a}_+^{(1)}(\mathbf{r} +
                             \mathbf{r}_+) \right) \right].
\end{align}
Some further definitions: For $i = 1,2$ we set
\begin{equation}  
  \mathbf{h}_i(\mathbf{R}) = \mathbf{h}_{i, 1} + \mathbf{h}_{i, 2},
\end{equation}
where
\begin{align}  
  \mathbf{h}_{1,1}(\mathbf{R}) & = - \frac{1}{\pi} \int_{\mathbf{r}}
                                 \mathbf{f}(\mathbf{r}) \left\{ \left( \mathbf{f}(\mathbf{r}) - \mathbf{f}(\mathbf{r}
                                 + \mathbf{R}) \right) \cdot \left[ \unitvec{z} \times
                                 \left( \mathbf{a}_{-, +}^{(1)}(\mathbf{r} +
                                 \mathbf{r}_+) + \mathbf{a}_{-, -}^{(1)}(\mathbf{r} + \mathbf{r}_+)
                                 \right) \right] \right\}, \\
  \mathbf{h}_{1,2}(\mathbf{R}) & = \frac{2}{\pi} \int_{\mathbf{r}}
                                 \mathbf{f}(\mathbf{r}) \left[ \left( \mathbf{f}(\mathbf{r}) - \mathbf{f}(\mathbf{r}
                                 + \mathbf{R}) \right) \cdot \left( \unitvec{z} \times
                                 \mathbf{a}_{-, +-}^{(1)}(\mathbf{r} +
                                 \mathbf{r}_+) \right) \right], \\
  \mathbf{h}_{2,1}(\mathbf{R}) &  = - \frac{1}{\pi} \int_{\mathbf{r}}
                                 \mathbf{f}(\mathbf{r}) \left\{ \left( \mathbf{f}(\mathbf{r}) - \mathbf{f}(\mathbf{r}
                                 + \mathbf{R}) \right) \odot \left[ \unitvec{z} \times
                                 \left( \mathbf{a}_{+, +}^{(1)}(\mathbf{r} +
                                 \mathbf{r}_+) + \mathbf{a}_{+, -}^{(1)}(\mathbf{r} + \mathbf{r}_+)
                                 \right) \right] \right\}, \\
  \mathbf{h}_{2,2}(\mathbf{R}) & = \frac{2}{\pi} \int_{\mathbf{r}}
                                 \mathbf{f}(\mathbf{r}) \left[ \left( \mathbf{f}(\mathbf{r}) - \mathbf{f}(\mathbf{r}
                                 + \mathbf{R}) \right) \odot \left(
                                 \unitvec{z} \times \mathbf{a}_{+,
                                 +-}^{(1)}(\mathbf{r} + \mathbf{r}_+) \right) \right].
\end{align}

In $\mathbf{h}_{1,1}(\mathbf{R})$, we can use Eq.~\eqref{eq:z_cross_a^1}, which yields
\begin{equation}  
  \begin{split}
    \mathbf{h}_{1,1}(\mathbf{R}) & = \frac{1}{\pi} \int_{\mathbf{r}}
    \mathbf{f}(\mathbf{r}) \left[ \left( \mathbf{f}(\mathbf{r}) -
        \mathbf{f}(\mathbf{r} + \mathbf{R}) \right) \cdot \left( \cos(\theta) \sin(\theta) \mathbf{f}(\mathbf{r}) +
          \cos(\theta_{\mathbf{r} + \mathbf{R}}) \sin(\theta_{\mathbf{r} +
            \mathbf{R}}) \mathbf{f}(\mathbf{r} + \mathbf{R}) \right) \right] \\
      & = \mathbf{l}_1(\mathbf{R}) + \mathbf{l}_2(\mathbf{R}) + \mathbf{l}_3(\mathbf{R}),
  \end{split}
\end{equation}
where
\begin{align}  
  \mathbf{l}_1(\mathbf{R}) & = \frac{1}{\pi} \int_{\mathbf{r}} \cos(\theta) \sin(\theta)
                             \mathbf{f}(\mathbf{r}) \left[ \left( \mathbf{f}(\mathbf{r}) -
                             \mathbf{f}(\mathbf{r} + \mathbf{R}) \right) \cdot
                             \mathbf{f}(\mathbf{r}) \right], \\
  \mathbf{l}_2(\mathbf{R}) & = - \frac{1}{\pi} \int_{\mathbf{r}} 
                             \cos(\theta_{\mathbf{r} + \mathbf{R}}) \sin(\theta_{\mathbf{r} +
                             \mathbf{R}})
                             \mathbf{f}(\mathbf{r}) \left( \mathbf{f}(\mathbf{r}
                             + \mathbf{R}) \right)^2 \\
                           & = - \frac{1}{\pi} \int_{\mathbf{r}} 
                             \cos(\theta) \sin(\theta)
                             \mathbf{f}(\mathbf{r} - \mathbf{R}) \left( \mathbf{f}(\mathbf{r}) \right)^2, \\
  \mathbf{l}_3(\mathbf{R}) & = \frac{1}{\pi} \int_{\mathbf{r}} \cos(\theta_{\mathbf{r} + \mathbf{R}}) \sin(\theta_{\mathbf{r} +
                             \mathbf{R}})
                             \mathbf{f}(\mathbf{r}) \left(
                             \mathbf{f}(\mathbf{r}) \cdot 
                             \mathbf{f}(\mathbf{r} + \mathbf{R}) \right). \\
\end{align}
These quantities are the same as $\mathbf{k}_{1,2,3}$ defined in the previous
section up to the replacement of the $\odot$ product with the usual scalar
product. The computation of the integrals goes along the same lines as above.

Next, we consider $\mathbf{h}_{1, 2}(\mathbf{R})$, which we split into two
components,
\begin{equation}  
  \mathbf{h}_{1,2}(\mathbf{R}) = \mathbf{h}_{1,2,1}(\mathbf{R}) + \mathbf{h}_{1,2,2}(\mathbf{R}),
\end{equation}
where
\begin{align}  
  \mathbf{h}_{1,2,1}(\mathbf{R}) & = \frac{2}{\pi} \int_{\mathbf{r}}
                                   \mathbf{f}(\mathbf{r}) \left[
                                   \mathbf{f}(\mathbf{r}) \cdot \left( \unitvec{z} \times
                                   \mathbf{a}_{-, +-}^{(1)}(\mathbf{r} +
                                   \mathbf{r}_+) \right) \right], \\
  \mathbf{h}_{1,2,2}(\mathbf{R}) & = - \frac{2}{\pi} \int_{\mathbf{r}}
                                   \mathbf{f}(\mathbf{r}) \left[ \mathbf{f}(\mathbf{r}
                                   + \mathbf{R}) \cdot \left( \unitvec{z} \times
                                   \mathbf{a}_{-, +-}^{(1)}(\mathbf{r} +
                                   \mathbf{r}_+) \right) \right].
\end{align}
In the first contribution, we use Eq.~\eqref{eq:a^1_sigma_+-_rr_+} and find
\begin{equation}
  \mathbf{h}_{1,2,1}(\mathbf{R}) = \frac{1}{\pi^2} \int_{\mathbf{r},
    \mathbf{r}'} \mathbf{f}(\mathbf{r}) \left[ \mathbf{f}(\mathbf{r}) \cdot
    \left( \unitvec{z} \times \mathbf{f}(\mathbf{r} - \mathbf{r}') \right)
  \right] \left( \mathbf{f}(\mathbf{r}') \odot \mathbf{f}(\mathbf{r}' +
    \mathbf{R}) \right) = \frac{2}{\pi} \int_{\mathbf{r}'}
  \mathbf{c}_+(\mathbf{r}') \left( \mathbf{f}(\mathbf{r}') \odot
    \mathbf{f}(\mathbf{r}' + \mathbf{R}) \right),
\end{equation}
where $\mathbf{c}_+(\mathbf{r}')$ is defined in Eq.~\eqref{eq:c_sigma}.

We move on to $\mathbf{h}_{1,2,2}(\mathbf{R})$, given by
\begin{equation}  
  \mathbf{h}_{1,2,2}(\mathbf{R}) = - \frac{1}{\pi^2} \int_{\mathbf{r},
    \mathbf{r}'} \mathbf{f}(\mathbf{r}) \left[ \mathbf{f}(\mathbf{r} +
    \mathbf{R}) \cdot \left( \unitvec{z} \times \mathbf{f}(\mathbf{r} -
      \mathbf{r}') \right) \right] \left( \mathbf{f}(\mathbf{r}') \odot
    \mathbf{f}(\mathbf{r}' + \mathbf{R}) \right).
\end{equation}
Comparison with Eq.~\eqref{eq:a^2_sigma_22} shows that $\mathbf{h}_{1,2,2}(\mathbf{R})$ is
given by Eq.~\eqref{eq:a^2_-_22} with the first of the $\odot$ products replaced by
the usual scalar product, which yields
\begin{equation}  
  \mathbf{h}_{1,2,2}(\mathbf{R}) = \frac{1}{2 R}
  \begin{pmatrix}
    -\sin(3 \theta_{\mathbf{R}}) \\ \cos(3 \theta_{\mathbf{R}})
  \end{pmatrix}.
\end{equation}

The next on the list is $\mathbf{h}_{2,1}(\mathbf{R})$, which we write --- using
Eq.~\eqref{eq:a^1_++} --- as
\begin{equation}
  \mathbf{h}_{2,1}(\mathbf{R}) = - \frac{1}{\pi} \int_{\mathbf{r}}
  \mathbf{f}(\mathbf{r}) \left\{ \left( \mathbf{f}(\mathbf{r}) - \mathbf{f}(\mathbf{r}
      + \mathbf{R}) \right) \odot \left[ \unitvec{z} \times
      \left( \mathbf{f}(\mathbf{r}) \ln(r/a) + \mathbf{f}(\mathbf{r} +
        \mathbf{R}) \ln(\abs{\mathbf{r} + \mathbf{R}}/a)
      \right) \right] \right\}.
\end{equation}
Unfortunately, this can not be simplified as we did above in Eq.~\eqref{eq:a^2_+1}
because in general
$\mathbf{a} \odot \left( \unitvec{z} \times \mathbf{a} \right) \neq 0$. Hence,
we have to invent something new:
\begin{equation}  
  \mathbf{h}_{2,1}(\mathbf{R}) = \mathbf{m}_1(\mathbf{R}) +
  \mathbf{m}_2(\mathbf{R}) + \mathbf{m}_3(\mathbf{R}) + \mathbf{m}_4(\mathbf{R}),
\end{equation}
where
\begin{align}
  \mathbf{m}_1(\mathbf{R}) & = - \frac{1}{\pi} \int_{\mathbf{r}}
                             \mathbf{f}(\mathbf{r}) \left[
                             \mathbf{f}(\mathbf{r}) \odot \left( \unitvec{z} \times
                             \mathbf{f}(\mathbf{r}) \right) \right] \ln(r/a), \\
  \mathbf{m}_2(\mathbf{R}) & = - \frac{1}{\pi} \int_{\mathbf{r}}
                             \mathbf{f}(\mathbf{r}) \left[ \left( \mathbf{f}(\mathbf{r}) - \mathbf{f}(\mathbf{r}
                             + \mathbf{R}) \right) \odot \left( \unitvec{z} \times
                             \mathbf{f}(\mathbf{r} +
                             \mathbf{R}) 
                             \right) \right] \ln(\abs{\mathbf{r} +
                             \mathbf{R}}/R) \\
                           & = - \frac{1}{\pi} \int_{\mathbf{r}}
                             \mathbf{f}(\mathbf{r} - \mathbf{R}) \left[ \left(
                             \mathbf{f}(\mathbf{r} - \mathbf{R}) - \mathbf{f}(\mathbf{r}) \right) \odot \left( \unitvec{z} \times
                             \mathbf{f}(\mathbf{r}) 
                             \right) \right] \ln(r/R), \\
  \label{eq:m3}
  \mathbf{m}_3(\mathbf{R}) & = - \frac{1}{\pi} \ln(R/a) \int_{\mathbf{r}}
                             \mathbf{f}(\mathbf{r}) \left[ \left( \mathbf{f}(\mathbf{r}) - \mathbf{f}(\mathbf{r}
                             + \mathbf{R}) \right) \odot \left( \unitvec{z} \times
                             \mathbf{f}(\mathbf{r} +
                             \mathbf{R}) 
                             \right) \right], \\
  \mathbf{m}_4(\mathbf{R}) & = \frac{1}{\pi} \int_{\mathbf{r}}
                             \mathbf{f}(\mathbf{r}) \left[ \mathbf{f}(\mathbf{r}
                             + \mathbf{R}) \odot \left( \unitvec{z} \times
                             \mathbf{f}(\mathbf{r}) \right) \right] \ln(r/a).
\end{align}
$\mathbf{m}_1(\mathbf{R})$ vanishes because the integrand is antisymmetric under
reflections $\mathbf{r} \to -\mathbf{r}$.

Now comes $\mathbf{h}_{2,2}(\mathbf{R})$, which again consists of two
contributions,
\begin{equation}  
  \mathbf{h}_{2,2}(\mathbf{R}) = \mathbf{h}_{2,2,1}(\mathbf{R}) + \mathbf{h}_{2,2,2}(\mathbf{R}),
\end{equation}
where
\begin{align}  
  \mathbf{h}_{2,2,1}(\mathbf{R}) & = \frac{2}{\pi} \int_{\mathbf{r}}
                                   \mathbf{f}(\mathbf{r}) \left[
                                   \mathbf{f}(\mathbf{r}) \odot \left(
                                   \unitvec{z} \times \mathbf{a}_{+,
                                   +-}^{(1)}(\mathbf{r} + \mathbf{r}_+) \right)
                                   \right], \\
  \mathbf{h}_{2,2,2}(\mathbf{R}) & = - \frac{2}{\pi} \int_{\mathbf{r}}
                                   \mathbf{f}(\mathbf{r}) \left[ \mathbf{f}(\mathbf{r}
                                   + \mathbf{R}) \odot \left(
                                   \unitvec{z} \times \mathbf{a}_{+,
                                   +-}^{(1)}(\mathbf{r} + \mathbf{r}_+) \right) \right].
\end{align}
In the first contribution, we use Eq.~\eqref{eq:a^1_sigma_+-_rr_+} and find
\begin{equation}  
  \begin{split}
    \mathbf{h}_{2,2,1}(\mathbf{R}) & = \frac{1}{\pi^2} \int_{\mathbf{r},
      \mathbf{r}'} \mathbf{f}(\mathbf{r}) \left[ \mathbf{f}(\mathbf{r}) \odot
      \left( \unitvec{z} \times \mathbf{f}(\mathbf{r} - \mathbf{r}') \right)
    \right] \left( \mathbf{f}(\mathbf{r}') \cdot \mathbf{f}(\mathbf{r}' +
      \mathbf{R}) \right) \\ & = \frac{2}{\pi} \int_{\mathbf{r}'}
    \mathbf{c}_-(\mathbf{r}') \left( \mathbf{f}(\mathbf{r}') \cdot
      \mathbf{f}(\mathbf{r}' + \mathbf{R}) \right),
  \end{split}
\end{equation}
where $\mathbf{c}_-(\mathbf{r}')$ is defined in Eq.~\eqref{eq:c_sigma}.

Finally, we consider
\begin{equation}  
  \mathbf{h}_{2,2,2}(\mathbf{R}) = - \frac{1}{\pi^2} \int_{\mathbf{r}, \mathbf{r}'}
  \mathbf{f}(\mathbf{r}) \left[ \mathbf{f}(\mathbf{r}
    + \mathbf{R}) \odot \left(
      \unitvec{z} \times \mathbf{f}(\mathbf{r} - \mathbf{r}') \right) \right]
  \left( \mathbf{f}(\mathbf{r}' \cdot \mathbf{f}(\mathbf{r}' + \mathbf{R})
  \right).
\end{equation}
Comparing this with Eq.~\eqref{eq:a^2_sigma_22} shows that
$\mathbf{h}_{2,2,2}(\mathbf{R})$ is given by Eq.~\eqref{eq:a^2_-_22} with the
second of the $\odot$ products replaced by the usual scalar product. We find
\begin{equation}  
  \mathbf{h}_{2,2,2}(\mathbf{R}) = 0.
\end{equation}

Combining all of the above results (many of which we haven't stated for
brevity), the mixed second order correction to the electric field is thus given
by
\begin{equation}
  \mathbf{E}^{(2)}_{+-}(\mathbf{r}_+) = \frac{1}{16} \left[
    2 \mathbf{f}(\mathbf{R}) \cos(2 \theta_{\mathbf{R}}) \left( 28 \ln(R/a) + 3 \right) + \frac{1}{R^2}
    \begin{pmatrix}
      R_x \\ - R_y
    \end{pmatrix}
    \left( 8 \ln(R/a)^2 + 12 \ln(R/a) - 5 \right)
  \right].
\end{equation}
\end{widetext}

\subsubsection{Numerical checks}
\label{sec:numerical-checks}

\begin{figure*}
  \centering
  \includegraphics[width=\linewidth]{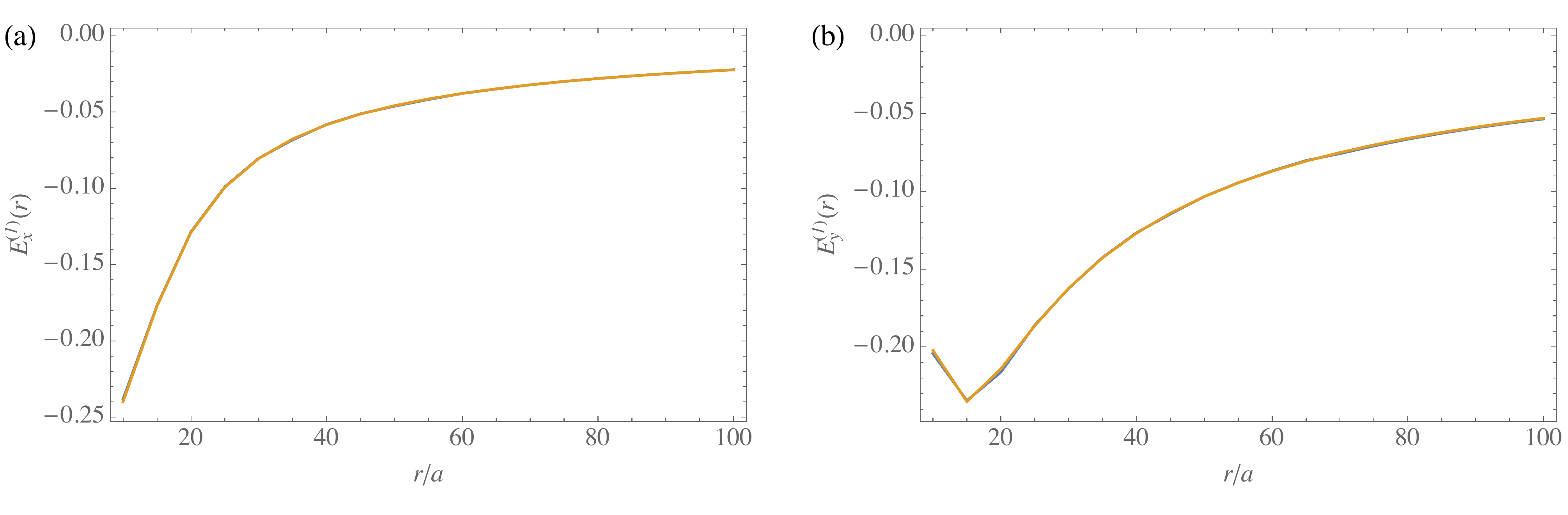}
  \includegraphics[width=\linewidth]{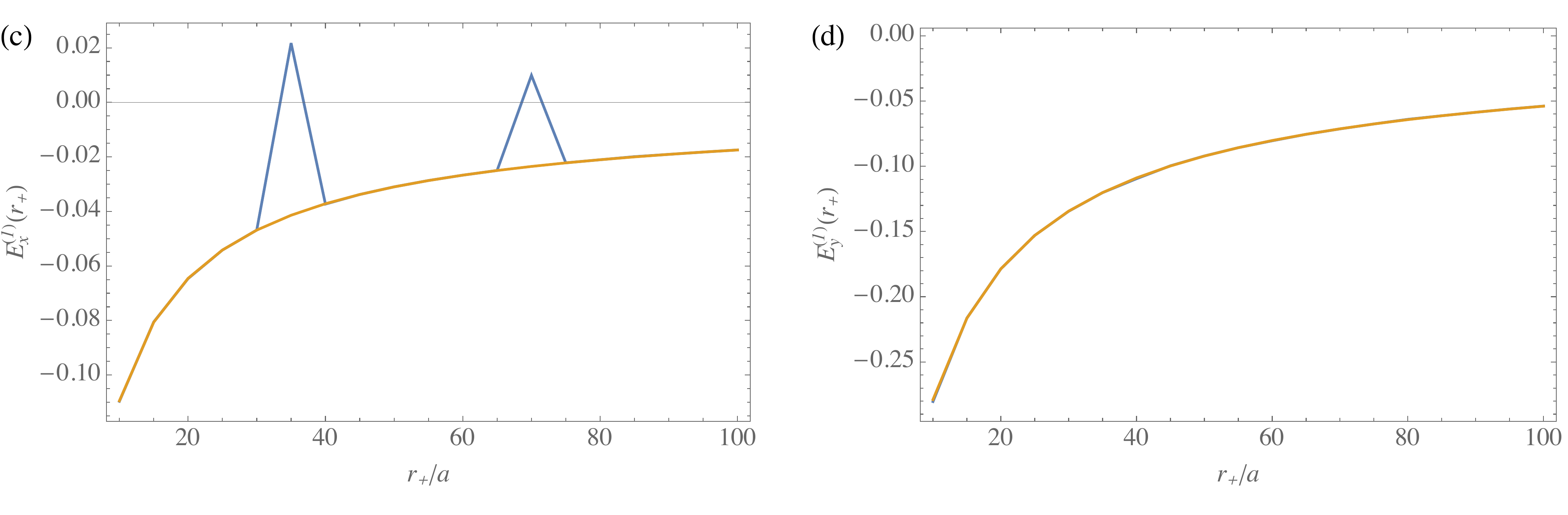}  
  \caption{Comparison between analytical (orange lines) and numerical (blue
    lines) results for the first-order correction to the electric field. (a) and
    (b) show the $x$ and $y$ components of $\mathbf{E}^{(1)}(\mathbf{r})$ for
    $\alpha_+ = 1, \alpha_- = 0.7, \mathbf{r} = r \left( \cos(\theta),
      \sin(\theta) \right)$
    with $\theta = 7 \pi/8$, and
    $\mathbf{r}_+ = r_+ \left( \cos(\theta_+), \sin(\theta_+) \right)$ with
    $r_+/a = 10, \theta_+ = \pi/5$. In (c) and (d) we plot the components of
    $\mathbf{E}^{(1)}(\mathbf{r}_+)$ for
    $\alpha_+ = 1, \alpha_- = -1.2, \theta_+ = 3 \pi/4$. For some apparently
    random values of $r_+$ in (c) the convergence of the numerical integration
    is relatively poor. In both plots, the negative charge is located at
    $\mathbf{r}_- = -\mathbf{r}_+$.}
  \label{fig:E1_numerics}
\end{figure*}

\begin{figure*}
  \centering  
  \includegraphics[width=\linewidth]{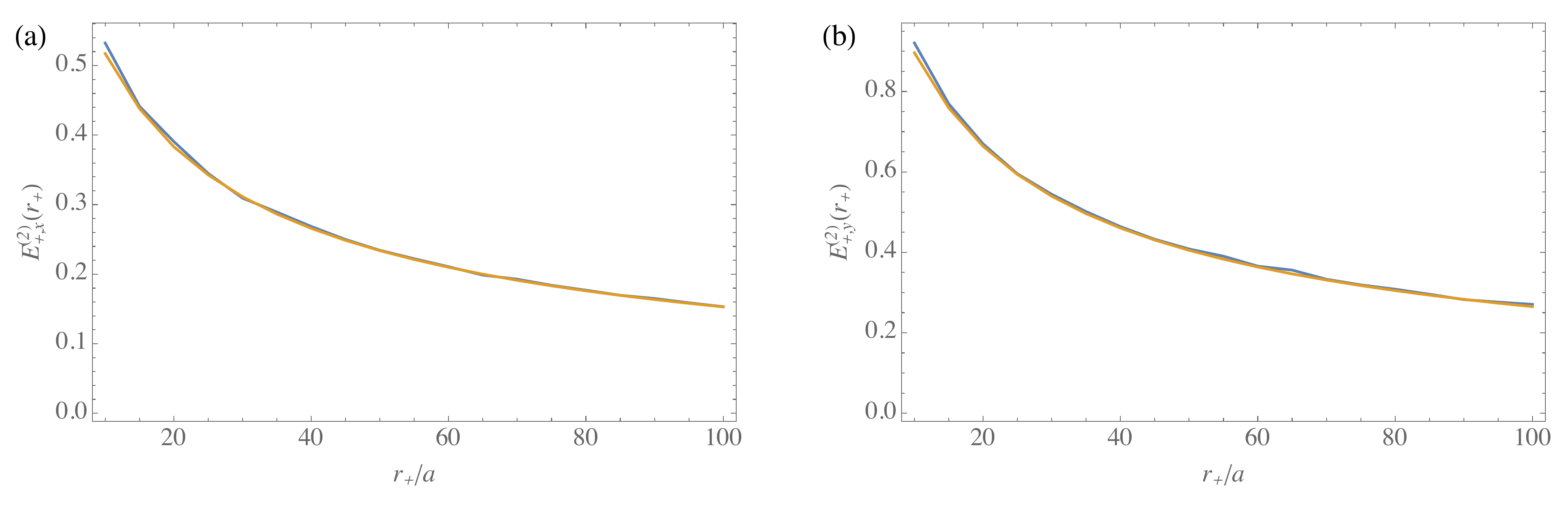}
  \includegraphics[width=\linewidth]{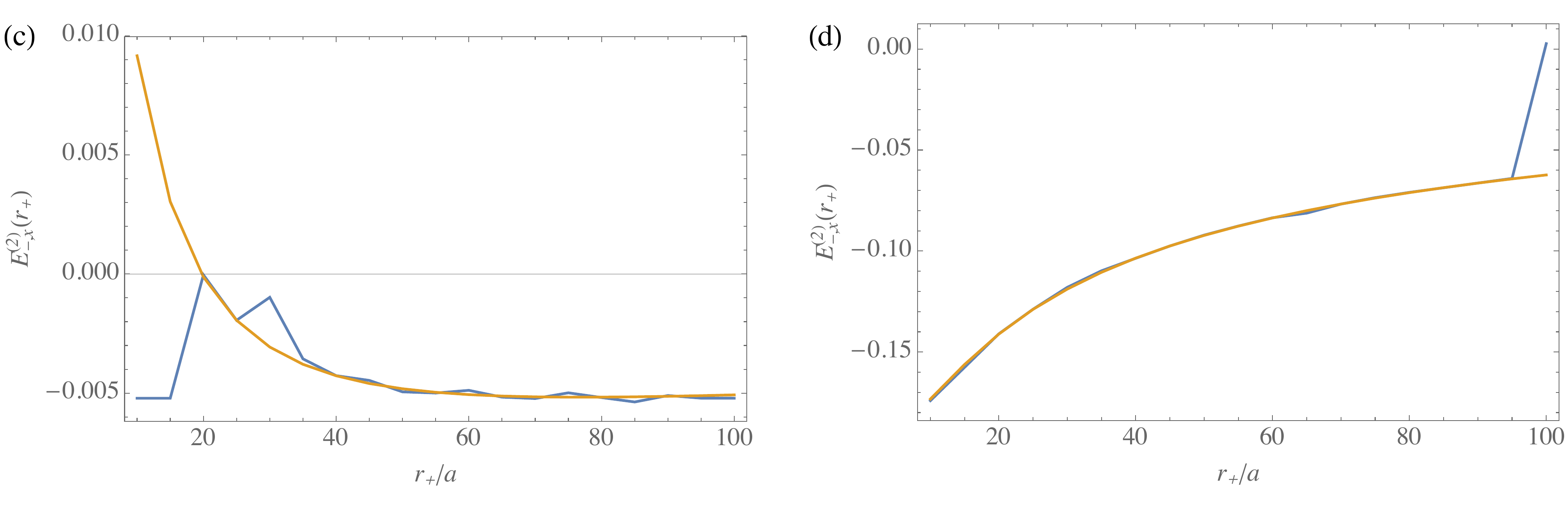}
  \includegraphics[width=\linewidth]{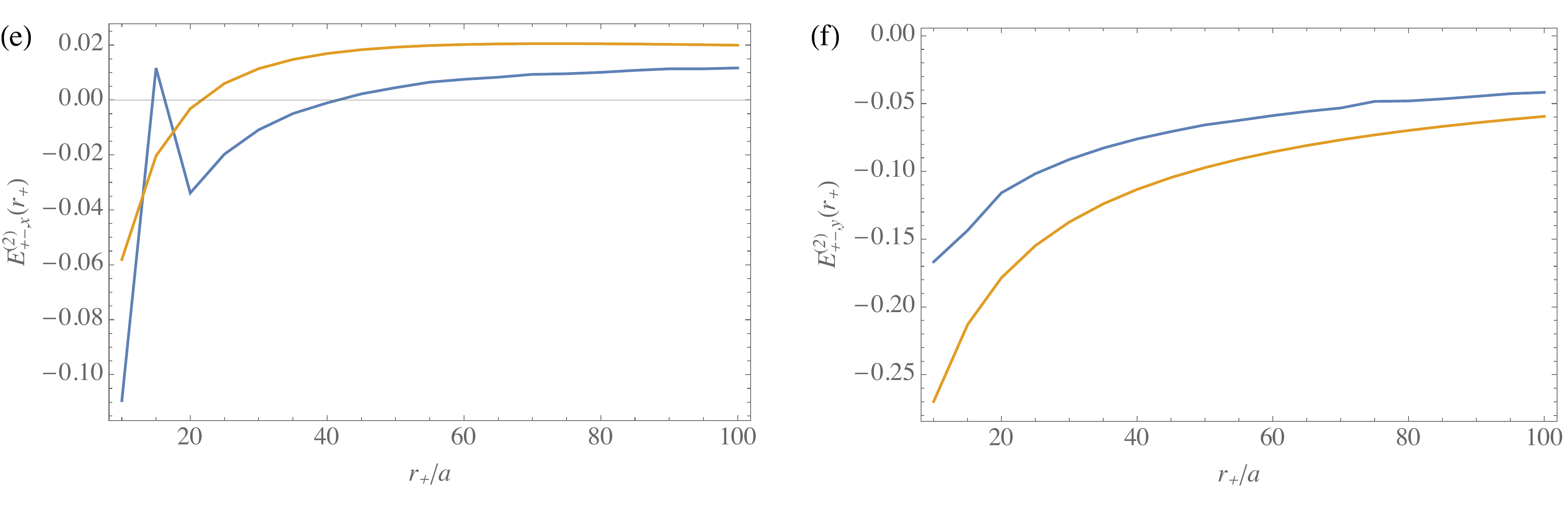}
  \caption{Comparison between analytical (orange lines) and numerical (blue
    lines) results for the second-order correction to the electric field. (a)
    and (b) show, respectively, the $x$ and $y$ components of
    $\mathbf{E}^{(2)}_+(\mathbf{r}_+)$ for
    $\mathbf{r}_+ = r_+ \left( \cos(\theta_+), \sin(\theta_+) \right)$ with
    $\theta_+ = \pi/3$. In (c) and (d) are the components of
    $\mathbf{E}^{(2)}_-(\mathbf{r}_+)$ for $\theta_+ = 2 \pi/5$. For some
    apparently random values of $r_+$ in (c) and (d) the convergence of the
    numerical integration is relatively poor. Finally, (e) and (f) are the
    components of $\mathbf{E}^{(2)}_{+-}(\mathbf{r}_+)$ for $\theta_+ = \pi/8$.
    In all plots, the negative charge is at $\mathbf{r}_- = -\mathbf{r}_+$.}
  \label{fig:E21_numerics}
\end{figure*}

\begin{figure*}
  \centering  
  \includegraphics[width=\linewidth]{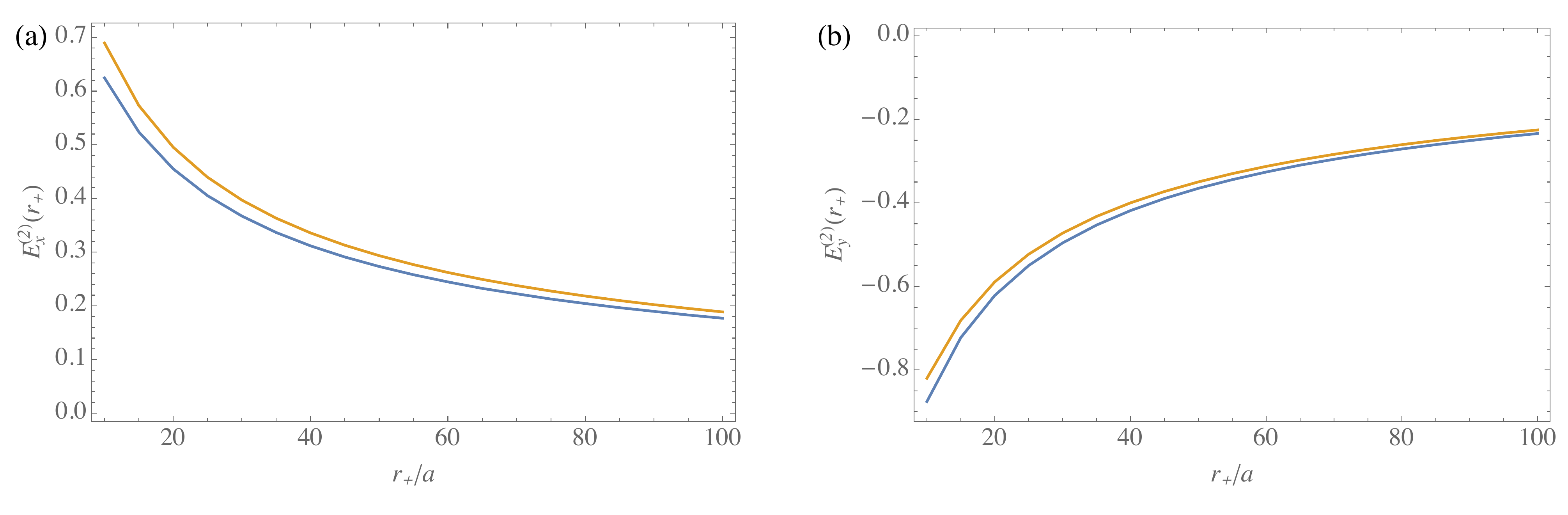}
  \caption{Comparison between analytical (orange lines) and numerical (blue
    lines) results for the second-order correction to the electric field. (a)
    and (b) show, respectively, the $x$ and $y$ components of
    $\mathbf{E}^{(2)}(\mathbf{r}_+)$ for
    $\alpha_+ = 1, \alpha_- = 0.6, \theta_+ = 5 \pi/3$. The negative charge is
    at $\mathbf{r}_- = - \mathbf{r}_+$.}
  \label{fig:E22_numerics}
\end{figure*}

To check the above calculations, we evaluated Eqs.~\eqref{eq:a^1_sigma},
\eqref{eq:a^2_sigma}, and~\eqref{eq:a^2_+-} numerically. The results for some
sample parameter values are shown in Figs.~\ref{fig:E1_numerics},
\ref{fig:E21_numerics}, and~\ref{fig:E22_numerics}. We find agreement between
analytics and numerics up to convergence problems of the numerical integration
for select values of $\mathbf{r}_+$ (the position of the positive charge). The
slight discrepancy between results for $\mathbf{E}^{(2)}_{+-}(\mathbf{r}_+)$
and, consequently, $\mathbf{E}^{(2)}(\mathbf{r}_+)$ shown in
Figs.~\ref{fig:E21_numerics} and~\ref{fig:E22_numerics}, respectively, can be
traced back to particularly slow convergence of the numerical integration for
integrals of the type of Eq.~\eqref{eq:m3}.

\section{Dynamics of vortices in the compact anisotropic KPZ equation}
\label{sec:dynam-topol-defects}

Having found the electric field acting upon the charges (i.e., vortices)
constituting a dipole, we proceed to study the motion of the charges under the
influence of Markovian noise. In particular, we are interested in the stationary
distribution, which forms the basis of the RG treatment below. The noise acting
on the charges originates from the one in the original caKPZ equation, i.e, Eq.~(1) in
the main text. Hence, the strengths of the noise sources in the caKPZ equation and the equation of motion of vortices
are related, but will be renormalized
differently~\cite{Sieberer2016b,Wachtel2016}. 

\subsection{Equations of motion}
\label{sec:equations-motion}

As above, we consider a dipole consisting of a positive charge (vortex) at
$\mathbf{r}_+$, and a negative charge (antivortex) at $\mathbf{r}_-$. The
equations of motion for the vortices read
\begin{equation}
  \label{eq:r_pm_eom}
  \frac{d \mathbf{r}_{\sigma}}{d t} = \sigma \mu \mathbf{E}(\mathbf{r}_{\sigma}) +
  \boldsymbol{\xi}_{\sigma},
\end{equation}
where $\sigma = \pm$, cf.~Eq.~(6) in the main text. As discussed there, the
noise correlations read
$\langle \xi_{\sigma, i}(t) \xi_{\sigma', j}(t') \rangle = 2 \mu T
\delta_{\sigma \sigma'} \delta_{i j} \delta(t - t'),$
where $\sigma, \sigma' = \pm$, and $\mu$ is the vortex mobility. Thus, the
relative and ``center-of-mass'' coordinates,
$\mathbf{r} = \mathbf{r}_+ - \mathbf{r}_-$ and
$\mathbf{R} = (\mathbf{r}_+ + \mathbf{r}_-)/2$, respectively, obey the following
equations of motion:
\begin{align}
  \label{eq:r_eom}
  \frac{d \mathbf{r}}{d t} & = \mu \left( \mathbf{E}(\mathbf{r}_+) +
                             \mathbf{E}(\mathbf{r}_-) \right) + \boldsymbol{\xi}, \\
  \label{eq:R_eom}
  \frac{d \mathbf{R}}{d t} & = \frac{\mu}{2} \left( \mathbf{E}(\mathbf{r}_+) -
                             \mathbf{E}(\mathbf{r}_-) \right) + \boldsymbol{\Xi},  
\end{align}
where $\boldsymbol{\xi} = \boldsymbol{\xi}_+ - \boldsymbol{\xi}_-$ and
$\boldsymbol{\Xi} = (\boldsymbol{\xi}_+ + \boldsymbol{\xi}_-)/2$. (To avoid
confusion, we note that in the above calculation of the electric field we
denoted the relative coordinate by $\mathbf{R}$.) The correlations of the noise
acting on the relative coordinate read
\begin{equation}  
  \langle \xi_i(t) \xi_j(t') \rangle = \sum_{\sigma, \sigma'} \sigma \sigma'
  \langle \xi_{\sigma, i}(t) \xi_{\sigma', j}(t') \rangle = 4 \mu T
  \delta_{ij} \delta(t - t').
\end{equation}
As we show in the following, the relative coordinate is only affected by the
zeroth and second order contributions to the electric field, whereas the first
order corrections induces motion of the center of mass.

To this end, for the bare Coulomb interaction, from Eqs.~\eqref{eq:E_012}
and~\eqref{eq:grad_phi} we find
$\mathbf{E}^{(0)}(\mathbf{r}_+) = \mathbf{f}(\mathbf{r}_+ - \mathbf{r}_-)$
(recall that $\mathbf{f}(0) = 0$ due to the short-distance cutoff) and hence
\begin{equation}  
  \mathbf{E}^{(0)}(\mathbf{r}_-) = -
  \mathbf{f}(\mathbf{r}_- - \mathbf{r}_+) =
  \mathbf{f}(\mathbf{r}_+ - \mathbf{r}_-) = \mathbf{E}^{(0)}(\mathbf{r}_+),
\end{equation}
where we used that $\mathbf{f}(-\mathbf{r}) = - \mathbf{f}(\mathbf{r})$ as can
be seen from Eq.~\eqref{eq:f}. Thus, the Coulomb interaction enters
Eq.~\eqref{eq:r_eom} with a factor of two but drops out of the difference in
Eq.~\eqref{eq:R_eom}.

\begin{widetext}
  The first order correction to the electric field is given by (cf.\
  Eqs.~\eqref{eq:E_012} and~\eqref{eq:a^1_sigma}; recall that the $\odot$
  product becomes the usual scalar product for $\sigma = +$)
\begin{equation}  
    \mathbf{E}_{\sigma}^{(1)}(\mathbf{r}_-)
     = \frac{1}{2 \pi} \int_{\mathbf{r}'} \unitvec{z} \times
    \mathbf{f}(\mathbf{r}_- - \mathbf{r}') \left( \mathbf{f}(\mathbf{r}' -
      \mathbf{r}_+) - \mathbf{f}(\mathbf{r}' - \mathbf{r}_-) \right)^{\odot 2}.
\end{equation}
With a change of the integration variable according to $\mathbf{r}' \to
-\mathbf{r}' + \mathbf{r}_+ + \mathbf{r}_-$, this can be written as
\begin{equation}  
  \mathbf{E}_{\sigma}^{(1)}(\mathbf{r}_-)
  = - \frac{1}{2 \pi} \int_{\mathbf{r}'} \unitvec{z} \times
  \mathbf{f}(\mathbf{r}_+ - \mathbf{r}') \left( \mathbf{f}(\mathbf{r}' -
    \mathbf{r}_+) - \mathbf{f}(\mathbf{r}' - \mathbf{r}_-) \right)^{\odot 2} = -
  \mathbf{E}_{\sigma}^{(1)}(\mathbf{r}_+).
\end{equation}
As stated above, we see the first order correction contributes to
Eq.~\eqref{eq:R_eom} but not to~\eqref{eq:r_eom}.

Finally, let us consider the second order contributions, and here first the
diagonal parts. According to Eqs.~\eqref{eq:E_012} and~\eqref{eq:a^2_sigma} it is
given by
\begin{equation}  
  \begin{split}
    \mathbf{E}^{(2)}_{\sigma}(\mathbf{r}_-) & = \frac{1}{\pi} \int_{\mathbf{r}'}
    \unitvec{z} \times \mathbf{f}(\mathbf{r}_- - \mathbf{r}') \left[ \left(
        \mathbf{f}(\mathbf{r}' - \mathbf{r}_+) - \mathbf{f}(\mathbf{r}' -
        \mathbf{r}_-) \right) \odot \left( \unitvec{z} \times
        \mathbf{a}_{\sigma}^{(1)}(\mathbf{r}') \right) \right] \\
    & = - \frac{1}{\pi} \int_{\mathbf{r}'} \unitvec{z} \times
    \mathbf{f}(\mathbf{r}_+ - \mathbf{r}') \left[ \left( \mathbf{f}(\mathbf{r}' -
        \mathbf{r}_+) - \mathbf{f}(\mathbf{r}' - \mathbf{r}_-) \right) \odot
      \left( \unitvec{z} \times \mathbf{a}_{\sigma}^{(1)}(-\mathbf{r}' +
        \mathbf{r}_+ + \mathbf{r}_-) \right)
    \right],
  \end{split}
\end{equation}
where in the second equality we performed the same change of the integration
variable $\mathbf{r}'$ as above. Then, from Eq.~\eqref{eq:a^1_sigma},
\begin{equation}  
  \mathbf{a}^{(1)}_{\sigma}(-\mathbf{r} + \mathbf{r}_+ + \mathbf{r}_-) =
  \frac{1}{2 \pi} \int_{\mathbf{r}'} \mathbf{f}(- \mathbf{r} + \mathbf{r}_+ +
  \mathbf{r}_- - \mathbf{r}') \left( \mathbf{f}(\mathbf{r}' - \mathbf{r}_+) -
    \mathbf{f}(\mathbf{r}' - \mathbf{r}_-) \right)^{\odot 2} = -
  \mathbf{a}^{(1)}_{\sigma}(\mathbf{r}),   
\end{equation}
and hence
$\mathbf{E}^{(2)}_{\sigma}(\mathbf{r}_-) =
\mathbf{E}^{(2)}_{\sigma}(\mathbf{r}_+)$.
Along the same lines, starting from Eqs.~\eqref{eq:E_012} and~\eqref{eq:a^2_+-}
it is straightforward to see that
$\mathbf{E}^{(2)}_{+-}(\mathbf{r}_-) =
\mathbf{E}^{(2)}_{+-}(\mathbf{r}_+)$. Thus, we find
\begin{align}
  \label{eq:r_eom_1}
  \frac{d \mathbf{r}}{d t} & = 2 \mu \left( \frac{1}{\varepsilon} \mathbf{E}^{(0)}(\mathbf{r}_+) +
  \frac{1}{\varepsilon^3} \mathbf{E}^{(2)}(\mathbf{r}_+) \right) +
  \boldsymbol{\xi}, \\ \label{eq:R_eom_1}
  \frac{d \mathbf{R}}{d t} & = \frac{\mu}{\varepsilon^2}
  \mathbf{E}^{(1)}(\mathbf{r}_+) + \boldsymbol{\Xi},
\end{align}
where
\begin{equation}
  \label{eq:E^1_E^2}
  \mathbf{E}^{(1)}(\mathbf{r}) = - \sum_{\sigma = \pm} \sigma \alpha_{\sigma}
  \mathbf{E}_{\sigma}^{(1)}, \qquad
  \mathbf{E}^{(2)} = - \sum_{\sigma = \pm} \alpha_{\sigma}^2 \mathbf{E}_{\sigma}^{(2)} + \alpha_+
  \alpha_- \mathbf{E}_{+-}^{(2)}.
\end{equation}
For convenience, we list again the various contributions to the electric field
obtained in the previous section:
\begin{equation}
  \begin{split}
  \mathbf{E}^{(0)}(\mathbf{r}_+) & = \mathbf{f}(\mathbf{r}) = -
  \frac{\mathbf{r}}{r^2}, \\
  \mathbf{E}_+^{(1)}(\mathbf{r}_+) & = \frac{1}{2} \unitvec{z} \times \mathbf{f}(\mathbf{r}) \left( 4 \ln(r/a) - 1
  \right), \\
  \mathbf{E}_-^{(1)}(\mathbf{r}_+) & = \frac{3}{2} \unitvec{z} \times \mathbf{f}(\mathbf{r}) \cos(2
  \theta) - \frac{1}{r^2}
  \begin{pmatrix}
    y \\ x
  \end{pmatrix}
  \left( \ln(r/a) - 1 \right), \\
  \mathbf{E}_+^{(2)}(\mathbf{r}_+) & = \frac{1}{4} \mathbf{f}(\mathbf{r}) \left( 8
    \ln(r/a)^2 + 4 \ln(r/a) - 1 \right), \\  
  \mathbf{E}_-^{(2)}(\mathbf{r}_+) & = - \frac{1}{16} \left[
      \mathbf{f}(\mathbf{r}) \left( 8 \ln(r/a)^2 - 20 \ln(r/a) + 15 - 8 \cos(4
        \theta) \right) - \frac{6}{r^2}
    \begin{pmatrix}
      x \\ - y
    \end{pmatrix}
  \cos(2 \theta) \left( 4 \ln(r/a) - 5 \right) \right], \\  
  \mathbf{E}^{(2)}_{+-}(\mathbf{r}_+) & = \frac{1}{16} \left[
     2 \mathbf{f}(\mathbf{r}) \cos(2 \theta) \left( 28 \ln(r/a) + 3 \right) + \frac{1}{r^2}
    \begin{pmatrix}
      x \\ - y
    \end{pmatrix}
  \left( 8 \ln(r/a)^2 + 12 \ln(r/a) - 5 \right)
  \right],
\end{split}
\end{equation}
\end{widetext}
where
$\mathbf{r} = \mathbf{r}_+ - \mathbf{r}_- = \left( x, y \right) = r \left(
  \cos(\theta), \sin(\theta) \right)$ is the dipole moment.

Let's consider first the ``isotropic'' corrections, $\mathbf{E}^{(1)}_+$ and
$\mathbf{E}^{(2)}_{++}$, which were previously obtained in
Ref.~\cite{Wachtel2016}. We note that the first-order correction is
perpendicular to the dipole moment and causes motion of the center of mass in
this direction, while the second-order term is a central force and adds to the
Coulomb force $\mathbf{E}^{(0)}$ affecting the relative motion. Both
$\mathbf{E}^{(0)}$ and $\mathbf{E}^{(2)}_{++}$ can be derived from a potential
by taking the derivative with respect to the relative coordinate
$\mathbf{r}$. The ``anisotropic'' and ``mixed'' second order corrections,
$\mathbf{E}^{(2)}_{--}$ and $\mathbf{E}^{(2)}_{+-}$, on the other hand, cannot
be derived from a potential. In addition to the central contributions
$\propto \mathbf{r}$, they include terms $\propto \left( x, - y \right)$ that
favor alignment of the dipole along the $x$ or $y$-axis (depending on the signs
of $\alpha_+$ and $\alpha_-$). As mentioned in the main text, this is in line
with numerical simulations of the anisotropic complex Ginzburg-Landau
equation~\cite{Faller1998,ROLANDFALLERandLORENZKRAMER1999}.

All types of contributions have the common structure of being power series in
logarithms. For this reason, perturbation theory is valid up to the scale
$L_v \sim a e^{1/\alpha_{\mathrm{max}}}$ where
$\alpha_{\mathrm{max}} = \max \{ \abs{\alpha_{\pm}} \}$. For distances $r$ which
are much larger than the microscopic cutoff but below $L_v$, $a \ll r \ll L_v$,
the second order corrections are dominated by the leading powers of logarithms,
$\ln(r/a)^2$. Remarkably, for both $\mathbf{E}^{(2)}_{++}$ and
$\mathbf{E}^{(2)}_{--}$ these contributions take the same form, i.e., they are
centrally symmetric and potential --- in spite of $\mathbf{E}^{(2)}_{--}$
originating from a fully anisotropic non-linearity in the caKPZ equation. The
crucial difference between the ``isotropic'' and ``anisotropic'' (according to
their origin) second order contributions is that the former gives a repulsive
correction to the Coulomb force, while the latter gives an \emph{attractive}
one.

\subsection{Stationary distribution of a dipole}
\label{sec:stat-distr}

We proceed to derive the stationary distribution of a dipole subject to the
Langevin equation~\eqref{eq:r_eom_1}. The associated Fokker-Planck equation
reads~\cite{Kamenev2011} 
\begin{equation}
  \label{eq:FP}
  \partial_t \mathcal{P} = - 2 \mu \nabla \cdot \left( \mathbf{F}
    \mathcal{P} - T \nabla \mathcal{P} \right),
\end{equation}
with the drift generated by the electric field:
\begin{equation}  
  \mathbf{F}(\mathbf{r}) = \mathbf{F}^{(0)}(\mathbf{r}) +
  \mathbf{F}^{(2)}(\mathbf{r}) = \frac{1}{\varepsilon}
  \mathbf{E}^{(0)}(\mathbf{r}_+) + \frac{1}{\varepsilon^3}
  \mathbf{E}^{(2)}(\mathbf{r}_+).
\end{equation}
All the physics below the microscopic cutoff scale $a$ is contained in a single
phenomenological parameter, the vortex fugacity $y$, which quantifies the
probability of finding a dipole at the separation $a$ and thus sets the boundary
condition for the stationary distribution of the dipole,
$\mathcal{P}(\mathbf{r}) = y^2$ for $r = a$. We seek a steady-state solution of
Eq.~\eqref{eq:FP} in the form
\begin{equation}
  \label{eq:P}
  \mathcal{P}(\mathbf{r}) \sim y^2 e^{- \Phi(\mathbf{r})/T}.
\end{equation}
This is the exact form of the solution in thermal equilibrium, when
$\nabla \Phi = \mathbf{F}^{(0)}$ and hence
\begin{equation}
  \Phi(\mathbf{r}) = \Phi^{(0)}(\mathbf{r}) = \left( 1/\varepsilon \right)
  \ln(r/a).
\end{equation}
Out of equilibrium, the ansatz~\eqref{eq:P} yields the leading behavior at low
noise strengths~\cite{Kamenev2011}. Inserting this ansatz in the Fokker-Planck
equation~\eqref{eq:FP} and expanding the potential as
$\Phi = \Phi^{(0)} + \Phi^{(2)}$, where $\Phi^{(0)}$ is the equilibrium
solution, results for $T \ll 1$ in
\begin{equation}
  \mathbf{F}^{(0)} \cdot \left( \mathbf{F}^{(2)} + \nabla \Phi^{(2)} \right) = 0.
\end{equation}
In order to solve this partial differential equation for $\Phi^{(2)}$ we apply
the method of characteristics, which yields the following system of ordinary
differential equations:
\begin{equation}  
  \begin{split}
    \frac{d \mathbf{r}}{d t} & = \mathbf{F}^{(0)}, \\
    \frac{d \Phi^{(2)}}{d t} & = - \mathbf{F}^{(0)} \cdot \mathbf{F}^{(2)}.
  \end{split}
\end{equation}
The integral curves of the first equation flow upstream against the equilibrium
part of the drift field, i.e., they are the activation trajectories of the
unperturbed equilibrium problem. Integrating the second equation and inserting
the first one for $\mathbf{F}^{(0)}$ we get
\begin{equation}  
  \Phi^{(2)}(\mathbf{r}) = - \int_{t_0}^t d t' \frac{d \mathbf{r}}{d t} \cdot \mathbf{F}^{(2)} = -
  \int_{a \unitvec{r}}^{\mathbf{r}} d \mathbf{s} \cdot \mathbf{F}^{(2)},
\end{equation}
where the line integral has to be taken along the activation trajectory of the
equilibrium problem that connects $\mathbf{r}(t_0) = a \unitvec{r}$ and
$\mathbf{r}$. The initial condition at $t_0$ is chosen to ensure
$\Phi^{(2)}(\mathbf{r}) = 0$ and thus $\mathcal{P}(\mathbf{r}) = y^2$ for
$r = a$. We thus find
\begin{widetext}
  \begin{multline}
  \label{eq:Phi^2}
  \Phi^{(2)}(\mathbf{r}) = - \frac{1}{\varepsilon^3} \left\{ \frac{1}{3} \left(
      2 \alpha_+^2 - \frac{\alpha_-^2}{2} + \frac{\alpha_+ \alpha_-}{2} \cos(2
      \theta) \right) \ln(r/a)^3 \right. \\ + \frac{1}{2} \left[ \alpha_+^2 +
    \frac{\alpha_-^2}{2} \left( 1 - \frac{3}{2} \cos(4 \theta) \right) -
    \frac{11 \alpha_+ \alpha_-}{4} \cos(2 \theta) \right] \ln(r/a)^2 \\ \left. -
    \frac{1}{4} \left( \alpha_+^2 - \frac{23 \alpha_-^2}{4} \cos(4 \theta) +
      \frac{11 \alpha_+ \alpha_-}{4} \cos(2 \theta) \right) \ln(r/a) \right\}.
\end{multline}
\end{widetext}

\section{RG flow}
\label{sec:rg-flow}

The ``macroscopic'' electrodynamics of the previous sections captures the
screening of the electric field due to bound vortex-antivortex pairs by
introducing the dielectric constant $\varepsilon$. The latter describes the
response of the dielectric medium of bound pairs to an electric field. This
definition as a response leads to an implicit equation for $\varepsilon$, since
the polarization of a single test dipole due to an external electric field is
determined by the balance between the external field and the screened Coulomb
interaction between the charges constituting the dipole --- with the screening
in turn determined by $\varepsilon$. The resulting implicit equation for
$\varepsilon$ can be solved by a renormalization group (RG) approach described
in the following. Section~\ref{sec:derivation-rg-flow} describes the derivation
of RG flow equations, which generalizes the one given in Ref.~\cite{Wachtel2016}
for the isotropic case. We study the phases and fixed points of the RG flow in
Sec.~\ref{sec:phases-fixed-point}. The rather peculiar divergence of the
correlation length at the critical point is discussed in
Sec.~\ref{sec:asympt-analys-RG}.

\subsection{Derivation of the RG flow equations}
\label{sec:derivation-rg-flow}

As outlined above, in the following we derive an implicit equation for the
dielectric constant $\varepsilon$ by calculating the polarization of a single
test dipole that is induced in linear order by an external electric field. This
implicit equation is the starting point from which we obtain a system of RG flow
equations.

Adding an external electric field $\mathbf{E}_{\mathrm{eff}}$ in the equations
of motion~\eqref{eq:r_pm_eom} modifies the potential as
$\Phi(\mathbf{r}) \to \Phi(\mathbf{r}) - \mathbf{E}_{\mathrm{ext}} \cdot
\mathbf{r}$.
To first order in the external field, the resulting average polarization of our
test dipole is given by
\begin{widetext}
\begin{equation}
  \label{eq:polarization}  
  \left\langle \mathbf{P} \right\rangle = \frac{1}{L^2} \int \frac{d^2
    \mathbf{R}}{a^2} \frac{d^2 \mathbf{r}}{a^2} \mathbf{r}
  \mathcal{P}(\mathbf{r}) \left( 1 + \frac{1}{T} \mathbf{E}_{\mathrm{ext}}
    \cdot \mathbf{r} \right) = \frac{1}{T} \int \frac{d^2
    \mathbf{r}}{a^2} \mathcal{P}(\mathbf{r}) \frac{\mathbf{r}
    \mathbf{r}^T}{a^2} \mathbf{E}_{\mathrm{ext}} = \chi
  \mathbf{E}_{\mathrm{ext}},
\end{equation}
which defines the scusceptibility tensor $\chi$. In an isotropic system, when
$\mathcal{P}(\mathbf{r})$ does not depend on the direction, the susceptibility
$\chi \propto \int d^2 \mathbf{r} \, \mathcal{P}(\mathbf{r}) \mathbf{r}
\mathbf{r}^T$
is proportional to the identity matrix, which can be seen by noting that
$\mathcal{P}(\mathbf{r})$ is symmetric under (each of the transformations)
$x \to -x, y \to -y,$ and $x \leftrightarrow y$. As can be seen in
Eq.~\eqref{eq:Phi^2}, the anisotropy we consider here leaves the reflection
symmetries under $x \to -x$ and $y \to -y$ intact, but breaks the symmetry under
the exchange $x \leftrightarrow y$ (note that
$\cos(2 \theta) = (x^2 - y^2)/r^2$). Hence, the susceptibility tensor $\chi$ is
still diagonal,
\begin{equation}
  \label{eq:chi_tensor}
  \chi =
  \begin{pmatrix}
    \chi_x & 0 \\
    0 & \chi_y
  \end{pmatrix},
\end{equation}
but in general its eigenvalues are distinct, $\chi_x \neq \chi_y$. More
specifically, we find
\begin{equation}
  \label{eq:chi_xy}
  \begin{pmatrix}
    \chi_x \\ \chi_y
  \end{pmatrix}
  = \frac{y^2}{T} \int_a^{\infty} \frac{d r}{a} \frac{r^3}{a^3} e^{-\Phi^{(0)}(r)/T} \int_0^{2 \pi} d \theta
  \begin{pmatrix}
    \cos(\theta)^2 \\
    \sin(\theta)^2
  \end{pmatrix}
  e^{-\Phi^{(2)}(\mathbf{r})/T}.
\end{equation}
We note that strictly speaking the integral over $r$ should be cut at the scale
$L_v$ at which the perturbative expansion of the vortex interaction and hence
the potential $\Phi(\mathbf{r})$ breaks down. This is assumed implicitly in the
following. To make progress with the expression for $\chi$, we expand the last
exponential in Eq.~\eqref{eq:chi_xy}. This is justified up to parametrically
large distances, for which
\begin{equation}
  \label{eq:L_T}
  \frac{\alpha_{\sigma} \alpha_{\sigma'}}{T} \ln(r/a)^3 \ll 1 \quad
  \Longleftrightarrow \quad r \ll L_T = a e^{\left[ T/(\alpha_{\sigma}
      \alpha_{\sigma'}) \right]^{1/3}}.
\end{equation}
where we use the bare value $\varepsilon = 1$ to estimate $L_T$. We note that
for distances below $L_T$, the temperature is always ``high'' with regard to the
terms $\propto \alpha_{\sigma} \alpha_{\sigma'}$, which means that by
noise-induced fluctuations the test dipole can explore all possible
orientations. Only at much larger distances $r \gg L_T$ the test dipole is
essentially restricted to the direction that minimizes $\Phi^{(2)}(\mathbf{r})$
with strongly suppressed fluctuations around this direction. Thus, we find
($i = x,y$)
\begin{equation}
  \label{eq:chi_i}
  \chi_i = \frac{\pi y^2}{T} \int_a^{\infty} \frac{d r}{a} \frac{r^3}{a^3}
  e^{-\Phi^{(0)}(r)/T} \left( 1 - \frac{1}{\varepsilon^3 T} \sum_{n = 1}^3
    \beta_{i, n} \ln(r/a)^n \right),
\end{equation}
with the coefficients
\begin{equation}
  \label{eq:beta}
  \begin{aligned}  
    \beta_{x, 1} & = \frac{1}{32} \alpha_+ \left( 8 \alpha_+ + 11 \alpha_-
    \right), & \beta_{x, 2} & = - \frac{1}{16} \left( 8 \alpha_+^2 - 11 \alpha_+
      \alpha_- + 4 \alpha_-^2 \right),
    & \beta_{x, 3} & = - \frac{1}{12} \left(  8 \alpha_+^2 + \alpha_+ \alpha_- - 2 \alpha_-^2 \right), \\
    \beta_{y, 1} & = \frac{1}{32} \alpha_+ \left(8 \alpha_+ -11 \alpha_-
    \right), & \beta_{y, 2} & = - \frac{1}{16} \left( 8 \alpha_+^2 + 11 \alpha_+
      \alpha_- + 4 \alpha_-^2 \right), & \beta_{y, 3} & = - \frac{1}{12} \left(
      8 \alpha_+^2 - \alpha_+ \alpha_- - 2 \alpha_-^2 \right).
  \end{aligned}
\end{equation}
Despite the noise-induced angular averaging, the coefficients
$\beta_{x,n} \neq \beta_{y,n}$ for $n = 1,2,3$ if both
$\alpha_+, \alpha_- \neq 0$. In consequence, the eigenvalues of the
susceptibility tensor are distinct, $\chi_x \neq \chi_y$, and a single
dielectric constant $\varepsilon$ --- as we have assumed in our derivation ---
is insufficient to describe the resulting anisotropic screening. Remarkably,
this complication does not arise in the fully anisotropic configuration in which
$\alpha_+ = 0$ (and, of course, also in an isotropic system with $\alpha_- = 0$
a single dielectric constant suffices). In the following, we focus on this
case. Then, we find $\chi_x = \chi_y = \chi$, where
\begin{equation}
  \label{eq:chi}
  \chi = \frac{\pi y^2}{T} \int_a^{\infty} \frac{dr}{a} \left( \frac{r}{a}
  \right)^{3 - \frac{1}{\varepsilon T}} \left[ 1 - \frac{\alpha_-^2}{6
      \varepsilon^3 T} \left( \ln(r/a)^3 - c \ln(r/a)^2 \right) \right].
\end{equation}
Anticipating renormalization of the coefficient of the last term in brackets, we
introduced a coupling $c$ with microscopic value $c = 3/2$. The crucial
difference between the above expression and the corresponding result in the
isotropic case~\cite{Wachtel2016} is the sign of the leading term
$\propto \ln(r/a)^3$. This difference brings about major qualitative changes in
the behavior of vortices. According to Eq.~\eqref{eq:displacement_field}, the
renormalized dielectric constant is then
\begin{equation}
  \label{eq:epsilon_R}
  \varepsilon_R = \varepsilon + 2 \pi \chi = \varepsilon + \frac{2
    \pi^2 y^2}{T} \int_a^\infty \frac{dr}{a} \left(
    \frac{r}{a}\right)^{3-\frac{1}{\varepsilon T}} \left[ 1 -
    \frac{\alpha_-^2}{6 \varepsilon^3 T} \left( \ln(r/a)^3 - c \ln(r/a)^2
    \right) \right].
\end{equation}
In the integral on the RHS of this relation, the dielectric constant should be
interpreted as the renormalized, scale-dependent value. Thus,
Eq.~\eqref{eq:epsilon_R} is an implicit integral equation for
$\varepsilon_R$. It can be solved by breaking the integral into small steps and
absorbing the contribution of each of them progressively in renormalized
coefficients. To derive RG differential equations that describe this procedure
in the limit of infinitesimal steps, we separate the integral into two parts,
\begin{equation}  
  \int_a^\infty=\int_a^{a \left( 1 + d \ell \right)} + \int_{a \left( 1 + d \ell \right)}^\infty.
\end{equation}
The first part is used to redefine $\varepsilon$ on a slightly larger cutoff
scale $a \left( 1 + d \ell \right)$,
\begin{equation}  
  \varepsilon' = \varepsilon + \frac{2 \pi^2 y^2}{T} d \ell.
\end{equation}
In the remaining integral, we rescale $r$ to restore the lower limit of
integration to $a$,
\begin{equation}  
  \begin{split}
    \varepsilon_R & = \varepsilon' + \frac{2 \pi^2 y^2}{T} \int_{a \left( 1 + d
        \ell \right)}^\infty \frac{dr}{a} \left(
      \frac{r}{a}\right)^{3-\frac{1}{\varepsilon T}} \left[ 1 -
      \frac{\alpha_-^2}{6 \varepsilon^3 T} \left( \ln(r/a)^3 - c \ln(r/a)^2
      \right) \right] \\ & = \varepsilon' + \frac{2 \pi^2 y^2}{T}
    \int_{a}^\infty \frac{dr}{a} \left(
      \frac{r}{a}\right)^{3-\frac{1}{\varepsilon T}} \left( 1 + d \ell
    \right)^{4 - \frac{1}{\varepsilon T}} \left[ 1 - \frac{\alpha_-^2}{6
        \varepsilon^3 T} \left( \ln \! \left( \frac{r \left( 1 + d \ell
            \right)}{a} \right)^3 - c \ln \! \left( \frac{r \left( 1 + d \ell
            \right)}{a} \right)^2 \right) \right].
  \end{split}
\end{equation}
Expanding in $d \ell$ we find
\begin{equation}  
  \begin{split}
    \left( 1 + d \ell \right)^{4 - \frac{1}{\varepsilon T}} & = 1 + \left( 4 -
      \frac{1}{\varepsilon T} \right) d \ell + O(d \ell^2), \\
    \left( \ln(r/a) + \ln(1 + d \ell) \right)^n & = \ln(r/a)^n + n \ln(r/a)^{n -
    1} d \ell + O(d \ell^2). \\    
  \end{split}
\end{equation}
A redefinition of the other coupling constants is required to bring the
expression for $\varepsilon_R$ to its original form, Eq.~\eqref{eq:epsilon_R},
\begin{equation}
  \label{eq:epsilon_R_rescaled}
  \begin{split}
    \varepsilon_R & = \varepsilon' + \frac{2 \pi^2 y^2}{T} \left[ 1 + \left( 4 -
        \frac{1}{\varepsilon T} \right) d \ell \right] \int_a^\infty
    \frac{dr}{a} \left(\frac{r}{a}\right)^{3 - \frac{1}{\varepsilon T} \left( 1
        - \frac{c \alpha_-^2}{3 \varepsilon^2} d \ell \right)} \left\{ 1 -
      \frac{\alpha_-^2}{6 \varepsilon^3 T} \left[ \ln(r/a)^3 - \left( c - 3 d
          \ell \right) \ln(r/a)^2 \right] \right\} \\ & = \varepsilon' + \frac{2
      \pi^2 y^{\prime 2}}{T'} \int_a^\infty \frac{dr}{a} \left( \frac{r}{a}
    \right)^{3 - \frac{1}{\varepsilon T'}} \left[ 1 - \frac{\alpha_-^2}{6
        \varepsilon^3 T} \left( \ln(r/a)^3 - c' \ln(r/a)^2 \right) \right].
  \end{split}
\end{equation}
In the last line, we identified the following renormalized coupling constants:
\begin{align}  
  \frac{y'^2}{T'} & = \frac{y^2}{T} \left[ 1 + \left( 4 - \frac{1}{\varepsilon
                    T} \right) d \ell \right]
  & \Rightarrow &
  & \frac{d}{d \ell} \left(
    \frac{y^2}{T} \right)
  & = \left( 4 - \frac{1}{\varepsilon T} \right) \frac{y^2}{T}, \\  
  \frac{1}{T'} & = \frac{1}{T} \left( 1 - \frac{c \alpha_-^2}{3 \varepsilon^2}
                 d \ell \right)
  & \Rightarrow & & \frac{d}{d \ell} \frac{1}{T}
  & = - \frac{c  \alpha_-^2}{3 \varepsilon^2 T}, \\
  c' & = c - 3 d \ell
  & \Rightarrow &
  & \frac{d c}{d \ell} & = - 3.
\end{align}
\end{widetext}
We note that at the given order of $y^2$ and $\alpha_-^2$, all couplings on the
RHS of Eq.~\eqref{eq:epsilon_R_rescaled} can be replaced by the renormalized
values. The last line can be integrated trivially and yields
\begin{equation}
  \label{eq:c}
  c = \frac{3}{2} \left( 1 - 2 \ell \right),
\end{equation}
where the bare value at $\ell = 0$ is $c = 3/2$ as indicated below
Eq.~\eqref{eq:chi}. The remaining flow equations read
\begin{equation}
  \label{eq:RG_flow}  
  \begin{split}
    \frac{d \varepsilon}{d \ell} & = \frac{2 \pi^2 y^2}{T}, \\
    \frac{d y}{d \ell} & = \frac{1}{2} \left( 4 - \frac{1}{\varepsilon T} +
      \frac{c \alpha_-^2}{3 \varepsilon^2} \right) y, \\
    \frac{d T}{d \ell} & = \frac{c \alpha_-^2 T}{3 \varepsilon^2}.  
  \end{split}
\end{equation}
This is the form reported in the main text. The appearance of the logarithmic
scale $\ell$ in the flow equations upon inserting Eq.~\eqref{eq:c} reflects that
our perturbative treatment of the non-linearity does not yield the true
large-distance behavior of the vortex interaction. For this reason, the RG flow
has to be cut when the perturbative corrections become large, i.e., at the scale
$L_v$ (or, if it is smaller, at $L_T$ given in Eq.~\eqref{eq:L_T} where the
angular averaging becomes invalid).

\subsection{Phases and fixed point of the RG flow}
\label{sec:phases-fixed-point}

\begin{figure}
  \centering
  \includegraphics[width=\linewidth]{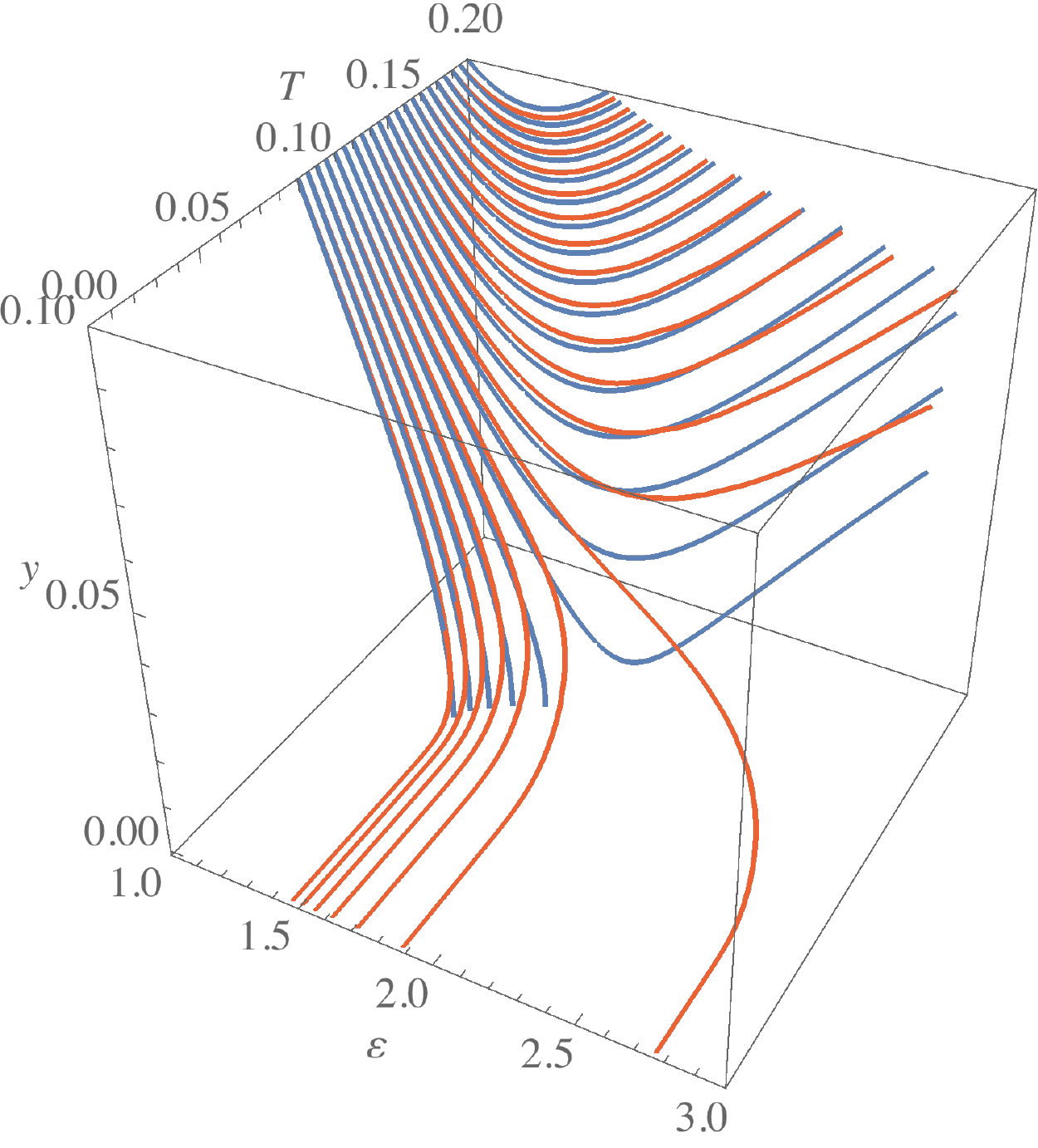}
  \caption{RG flow of $\varepsilon$, $y$, and $T$ with $\alpha_-^2 = 0.1$ as
    described by Eqs.~\eqref{eq:RG_flow} (red). Two phases are clearly
    distinguishable: At low temperatures $T < T_c \approx 0.13$, $y, T \to 0$
    and $\varepsilon \to \mathrm{const.}$, while at $T > T_c$,
    $y, \varepsilon \to \infty$ and $T \to \mathrm{const.}$ For comparison we
    show the equilibrium KT flow with $\alpha_-^2 = 0$ (blue). Here, the value
    of $T$ is conserved in the RG flow. In this figure, the microscopic value of
    the fugacity is $y = 0.1$, and the temperature is varied from $T = 0.1$ to
    $0.2$.}
  \label{fig:anisotropic_RG_flow_3D}
\end{figure}

To get a feeling for the RG flow described by Eqs.~\eqref{eq:RG_flow}, we
disregard for the moment that they are valid only up to parametrically large
distances, and integrate the flow for a sample of microscopic values. As can be
seen in Fig.~\ref{fig:anisotropic_RG_flow_3D} (and also Fig.~2 of the main
text), there is a critical temperature $T_c$ separating two phases with distinct
flow patterns.
In the low-temperature phase, the dielectric constant approaches a constant
value, while the temperature and the fugacity flow to zero. This is in stark
contrast to the corresponding phase in the equilibrium KT case, where the
temperature is conserved in the RG flow. In fully anisotropic non-equilibrium
systems, such a scale-invariant temperature is encountered only asymptotically
in the high-temperature phase, when
$\varepsilon \sim e^{4 \ell}, y \sim e^{2 \ell}$, and
$T \to T_{\infty} = \mathrm{const.}$ At large scales, the flow equations
simplify as
\begin{equation}
  \frac{d \varepsilon}{d \ell} \sim \frac{2 \pi^2 y^2}{T_{\infty}}, \qquad
  \frac{d y}{d \ell} \sim 2 y, \qquad
  \frac{d T}{d \ell} \to 0.
\end{equation}
This is just the usual KT flow with a renormalized temperature.

\begin{figure*}
  \centering
  \includegraphics[width=\linewidth]{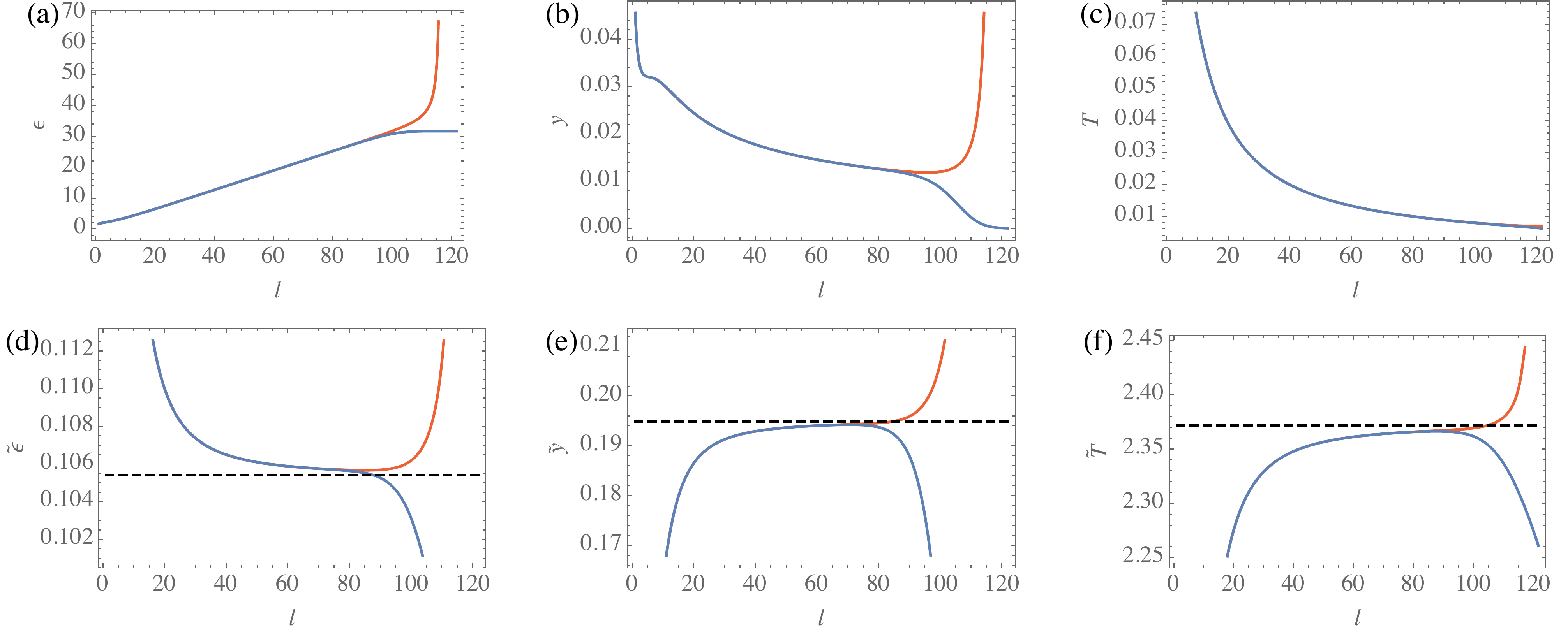}
  \caption{RG flow slightly above and below the critical temperature (red and
    blue solid lines, respectively). Panels (a-c) show the RG flow in terms of
    $\varepsilon$, $y$, and $T$, while in panels (d-f) the rescaled couplings
    defined in Eq.~\eqref{eq:rescaled_couplings} are plotted. Black dashed lines
    indicate the critical values given in Eq.~\eqref{eq:critical_point}.}
  \label{fig:close-to-crit_flow}
\end{figure*}

The existence of two distinct phases in the RG flow suggests there is a fixed
point separating these phases and controlling critical behavior at the
transition. In stark contrast to the usual case encountered in continuous phase
transitions, the flow equations~\eqref{eq:RG_flow} cannot have a true fixed
point since $c$ always grows logarithmically with the running cutoff (i.e.,
linearly in $\ell$). However, as we show in the following, there is nevertheless
a fixed point of the flow of a reduced set of logarithmically rescaled
variables. To this end, it is convenient to regard the couplings $\varepsilon$,
$y$, and $T$ as functions of $x = -c \in [-3/2, \infty)$ instead of $\ell$,
\begin{equation}  
  \begin{split}
    \frac{d \varepsilon}{d x} & = \frac{2 \pi^2 y^2}{3 T}, \\ \frac{d y}{d x} &
    = \frac{1}{6} \left( 4 - \frac{1}{\varepsilon T} -
      \frac{x \alpha_-^2}{3 \varepsilon^2} \right) y, \\
    \frac{d T}{d x} & = - \frac{x \alpha_-^2 T}{9 \varepsilon^2}.
  \end{split}
\end{equation}
To find the fixed point of these equations, we effect another change of
variables:
\begin{equation}
  \label{eq:rescaled_couplings}
  \tilde{\varepsilon} = \varepsilon/x, \qquad \tilde{y} = \sqrt{x} y, \qquad
  \tilde{T} = x T.
\end{equation}
Strictly speaking, the rescaled variables are ill-defined at the beginning of
the flow when $x < 0$. However, here we are concerned with the behavior of the
solutions to the flow equations for $x \to \infty$. The flow equations are then
recast as
\begin{equation}  
  \begin{split}
    \frac{d \tilde{\varepsilon}}{d x} & = \frac{1}{x} \left( \frac{2 \pi^2
      \tilde{y}^2}{3 \tilde{T}} - \tilde{\varepsilon} \right), \\
  \frac{d \tilde{y}}{d x} & = \frac{1}{6} \left[ 4 -
    \frac{1}{\tilde{\varepsilon} \tilde{T}} + \frac{1}{x} \left( 3 - \frac{\alpha_-^2}{3
        \tilde{\varepsilon}^2} \right) \right] \tilde{y}, \\
  \frac{d \tilde{T}}{d x} & = \frac{1}{3 x} \left( 3 - \frac{\alpha_-^2}{3
      \tilde{\varepsilon}^2} \right) \tilde{T}.
  \end{split}
\end{equation}
In this form, it is straightforward to see there is a fixed point at
$\tilde{\varepsilon}_{*}$, $\tilde{y}_{*}$, and $\tilde{T}_{*}$ determined by
\begin{equation}  
  \frac{2 \pi^2 \tilde{y}_{*}^2}{3 \tilde{T}_{*}} - \tilde{\varepsilon}_{*} = 0, \qquad 4 -
  \frac{1}{\tilde{\varepsilon}_{*} \tilde{T}_{*}} = 0, \qquad 3 - \frac{\alpha_-^2}{3
    \tilde{\varepsilon}_{*}^2} = 0.
\end{equation}
We find
\begin{equation}
  \label{eq:critical_point}
  \tilde{\varepsilon}_{*} = \frac{\abs{\alpha_-}}{3}, \qquad \tilde{y}_{*} =
  \sqrt{\frac{3}{8}} \frac{1}{\pi}, \qquad \tilde{T}_{*} = \frac{3}{4 \abs{\alpha_-}}.
\end{equation}
Flow trajectories close to criticality are shown in
Fig.~\ref{fig:close-to-crit_flow}, both for the original and rescaled
couplings~\eqref{eq:rescaled_couplings} (see panels (a-c) and (d-f),
respectively). The rescaled couplings are close to their fixed-point values in
the range $30 \lesssim \ell \lesssim 90$, during which the original ones evolve
according to Eq.~\eqref{eq:rescaled_couplings}. This logarithmic flow is the
origin of the peculiar singularity of the correlation length $\xi$ at the
critical point which is distinct from both the algebraic scaling at conventional
second order phase transitions, and the essential singularity at the equilibrium
KT transition. In the next section, we discuss how the singularity of the
correlation length can be inferred from the a linearization of the flow around
the fixed point~\eqref{eq:critical_point}.

\subsection{Asymptotic analysis of the linearized flow equations}
\label{sec:asympt-analys-RG}

As usual, we define the correlation length $\xi$ as the scale on which the
renormalized fugacity reaches the value $y_1 = 1$. We fix the microscopic value
$y_0$ and regard the temperature $T$ as the tuning parameter through the
transition. The origin of the singularity of $\xi$ as $T \to T_c$ is apparent
from panels (b) and (e) in Fig.~\ref{fig:close-to-crit_flow} which show the flow
of $y$ and $\tilde{y}$, respectively. This flow can be divided into three
stages: (i) for $0 \leq \ell < \ell_0$ the rescaled fugacity $\tilde{y}$
approaches its fixed-point value $\tilde{y}_{*}$ and (ii) stays close to this
value for $\ell_0 \leq \ell < \ell_1$; eventually, (iii) $\tilde{y}$ flows away
from the fixed point and $y$ grows strongly until it reaches $y_1 = 1$ for
$\ell_1 \leq \ell \leq \ell_2$. When $T \to T_c$, $\ell_1 \to \infty$, while
$\ell_0$ and $\ell_2 - \ell_1$ remain finite, and therefore
$\xi \sim a e^{\ell_2}$. Consequently, to determine the singularity of $\xi$, it
is sufficient to consider stage (ii) of the flow in the vicinity of the fixed
point where we can linearize the flow equations.

We collect the deviations from the fixed point in the variable
$\mathbf{a} = ( \delta \tilde{\varepsilon}, \delta \tilde{y}, \delta \tilde{T} )
= ( \tilde{\varepsilon} - \tilde{\varepsilon}_{*}, \tilde{y} - \tilde{y}_{*},
\tilde{T} - \tilde{T}_{*} )$. The linearized flow equations read
\begin{equation}
  \label{eq:linearized_flow}
  \frac{d \mathbf{a}}{d x} = A \mathbf{a}, \qquad A = A_0 + \frac{A_1}{x},
\end{equation}
where
\begin{equation}
  \begin{split}  
  A_0 & =
  \begin{pmatrix}
    0 & 0 & 0 \\
    - \sqrt{\frac{3}{2}} \frac{1}{\pi \alpha_-} & 0 & - \sqrt{\frac{2}{3}}
    \frac{2 \alpha_-}{3 \pi} \\
    0 & 0 & 0
  \end{pmatrix},
  \\ A_1 & =
  \begin{pmatrix}
    -1 & \sqrt{\frac{2}{3}} \frac{4 \pi \alpha_-}{3} & - \frac{4 \alpha_-^2}{9}
  \\
  \sqrt{\frac{3}{2}} \frac{3}{2 \pi \alpha_-} & 0 & 0 \\
  \frac{9}{2 \alpha_-^2} & 0 & 0
  \end{pmatrix}.
\end{split}  
\end{equation}
Note that these equations still depend on $x$ and thus cannot be solved
straightforwardly. However, since $\ell_2 \gg 1$ when $t = (T - T_c)/T_c \ll 1$,
we only need to know the asymptotic behavior of the solution for
$x \sim 3 \ell \to \infty$. To solve this problem, we closely follow the method
described in Ref.~\cite{Wasow2002}.

What makes finding an asymptotic expansion of the solution to
Eq.~\eqref{eq:linearized_flow} slightly complicated is first that $x = \infty$
is an irregular singular point of this equation and second that $A_0$, the
leading matrix for $x \to \infty$, has only one eigenvalue. In the following, we
apply a series of transformations to bring Eq.~\eqref{eq:linearized_flow} to a
form in which the leading matrix has three distinct eigenvalues. Then, the
leading matrix can be diagonalized, which results in three decoupled equations
that can be integrated straightforwardly.

\begin{figure*}
  \centering
  \includegraphics[width=\linewidth]{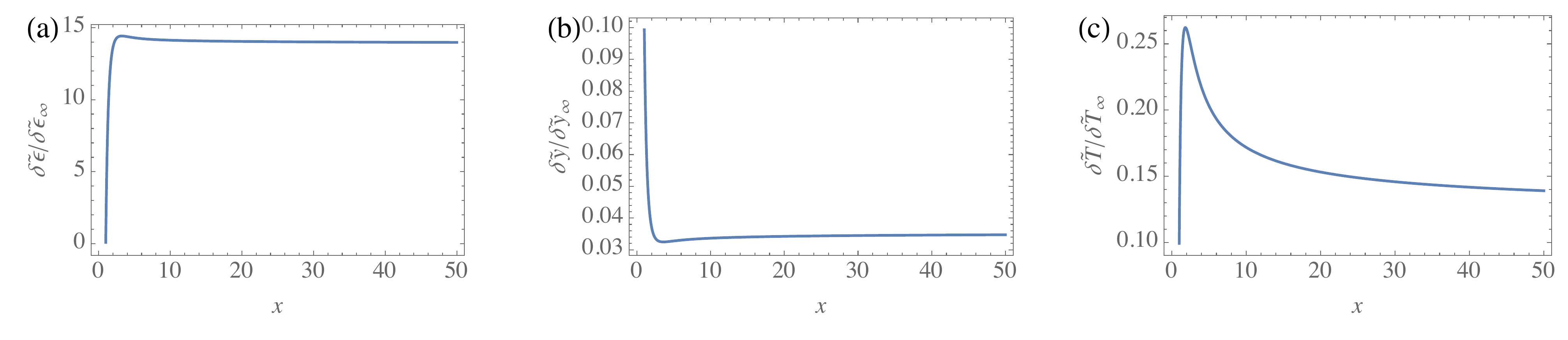}
  \caption{Comparison between the numerical solution of the linearized flow
    equations~\eqref{eq:linearized_flow} and the asymptotic
    expansion~\eqref{eq:linearized_flow_asymptotic_x}. For the numerical
    integration, we chose initial values
    $\delta \tilde{\varepsilon} = \delta \tilde{y} = \delta \tilde{T} = 1$.
    Division by the asymptotic expressions shows that the solution is well
    described by the asymptotic behavior already for small values of
    $x \lesssim 10$. Asymptotic analysis does not determine the prefactors in
    Eq.~\eqref{eq:linearized_flow_asymptotic_x}, hence we arbitrarily set them
    to one. (For the correct value, the plotted curves would approach $1$ for
    $x \to \infty$.)}
  \label{fig:asymptotic_expansion}
\end{figure*}

\begin{figure}
  \centering
  \includegraphics[width=\linewidth]{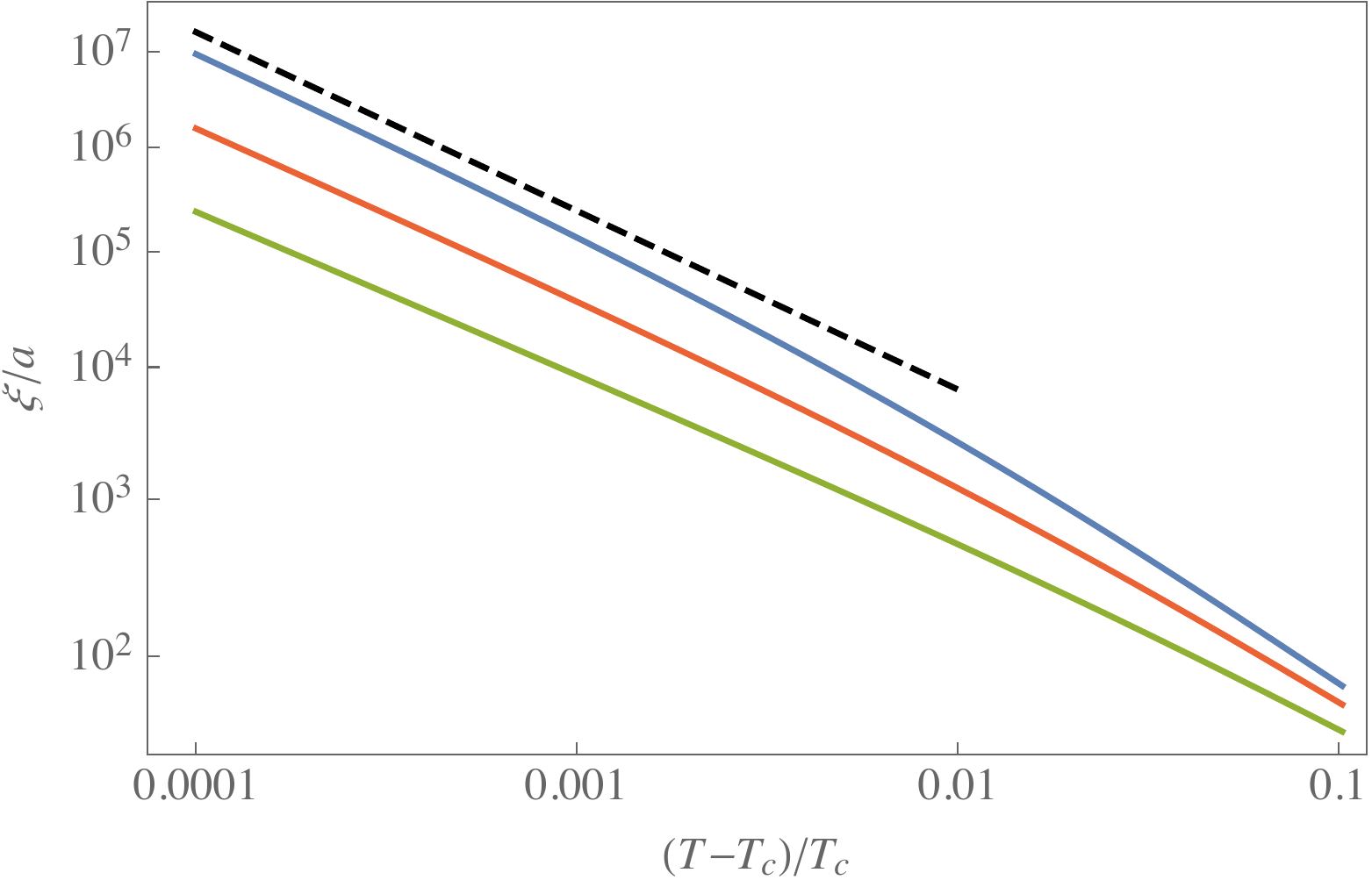}  
  \caption{Divergence of the correlation length at the critical temperature for
    $\alpha_-^2 = 0.1, 0.2, 0.5$ (top to bottom). The vertical axis is rescaled
    as $4 \sqrt{\ln(\xi/a)} + \ln(\ln(\xi/a))/4$ (cf.\ Eq.~\eqref{eq:xi_t}) so
    that the curves approach straight lines with slope $-1$ as $T \to T_c$ (for
    comparison shown as black dashed line).}
  \label{fig:corr_length}
\end{figure}

The first step is to bring $A_0$ to Jordan normal form by means of a
transformation $P_1$,
\begin{equation}  
  \mathbf{a}_1 = P_1^{-1} \mathbf{a}, \quad \frac{d \mathbf{a}_1}{d x} = A_1
  \mathbf{a}_1, \quad A_1 = P_1^{-1} A P_1 = A_{10} + \frac{A_{11}}{x},
\end{equation}
where
\begin{gather}
  P_1 = 
\begin{pmatrix}
  -\frac{4 \alpha_-^2}{9} & 0 & \sqrt{\frac{2}{3}} \pi  \alpha_- \\
  0 & 1 & 0 \\
  1 & 0 & 0 \\
\end{pmatrix},
\qquad A_{10} = 
\begin{pmatrix}
 0 & 0 & 0 \\
 0 & 0 & 1 \\
 0 & 0 & 0 \\
\end{pmatrix},
\\ A_{11} =
\begin{pmatrix}
  -2 & 0 & \sqrt{\frac{3}{2}} \frac{3 \pi}{\alpha_-} \\
  \sqrt{\frac{2}{3}} \frac{\alpha_-}{\pi } & 0 & \frac{3}{2} \\
  -\sqrt{\frac{2}{3}} \frac{4 \alpha_-}{3 \pi} & \frac{4}{3} & 1 \\
\end{pmatrix}.
\end{gather}
The matrix $A_{10} = H_1 \oplus H_2$ is the direct sum of two shifting matrices,
which are matrices with ones on the superdiagonal and zeroes elsewhere,
$H_1 = 0$ and $H_2 =
\begin{psmallmatrix}
  0 & 1 \\
  0 & 0
\end{psmallmatrix}
$.
We next apply a transformation $P_2$ to bring the sub-leading matrix $A_{11}$ to
a form in which the only non-zero entries occur in the rows corresponding to the
last rows of the blocks $H_{1,2}$, i.e.,
\begin{equation}
  \label{eq:a_2}
  \mathbf{a}_2 = P_2^{-1} \mathbf{a}_1, \quad \frac{d \mathbf{a}_2}{d x} = 
  A_2 \mathbf{a}_2, \quad A_2 = P_2^{-1} A_1 P_2 - P_2^{-1} \frac{d P_2}{d x},
\end{equation}
where the matrix $A_2$ has the structure
\begin{equation}
  \label{eq:A2}
  A_2 =
  \begin{pmatrix}
    A_{211} & A_{212} & A_{213} \\
    0 & 0 & 1 \\
    A_{231} & A_{232} & A_{233}
  \end{pmatrix}.
\end{equation}
Inserting in Eq.~\eqref{eq:a_2} the asymptotic ans\"atze
$P_2 \sim \sum_{r = 0}^{\infty} P_{2r}/x^r$ and
$A_2 \sim \sum_{r = 0}^{\infty} A_{2r}/x^r$ and identifying coefficients of the
same powers of $x$, we obtain $A_{10} P_{20} - P_{20} A_{20} = 0$ and
\begin{multline}
  \label{eq:A2_recursion}
  A_{10} P_{2r} - P_{2r} A_{20} \\ =
  \sum_{s = 0}^{r - 1} \left( P_{2s} A_{2, r - s} - A_{1, r - s} P_{2s} \right)
  - \left( r - 1 \right) P_{2, r - 1}.
\end{multline}
The first relation can be solved by setting $A_{20} = A_{10}$ and $P_{20} =
\id$; the second relation determines $A_{2r}$ and $P_{2r}$ for $r \geq 1$
recursively. We obtain the desired transformation by restricting the form of
$P_{2r}$ with $r \geq 1$ as
\begin{equation}
  \label{eq:P2r}
  P_{2r} =
  \begin{pmatrix}
    0 & 0 & 0 \\
    0 & P_{222r} & 0 \\
    P_{231r} & P_{232r} & 0
  \end{pmatrix}.
\end{equation}
Insertion of Eqs.~\eqref{eq:A2} and~\eqref{eq:P2r} in
Eq.~\eqref{eq:A2_recursion} yields a sequence of linear equations for the
elements of $A_2$ and $P_2$ that can be solved straightforwardly to any desired
order. We omit the explicit expressions. Next, we apply a first shearing
transformation,
\begin{equation}  
  \mathbf{a}_3 = P_3^{-1} \mathbf{a}_2, \quad \frac{d \mathbf{a}_3}{d x} = 
  A_3 \mathbf{a}_3,
\end{equation}
where
\begin{equation}  
  \begin{split}
    A_3 & = P_3^{-1} A_2 P_3 - P_3^{-1} \frac{d P_3}{d x}, \\ P_3 & =
  \diag(1, x^{-g_1}, x^{-2 g_1}).
  \end{split}
\end{equation}
We choose $g_1 = 1/3$ (this choice is, of course, not arbitrary, but is
determined by a well-defined procedure~\cite{Wasow2002}), and to bring the
equation back to a form that involves only integer powers of the variables, we
switch to $x = \alpha_1 y^{p_1}$, where $\alpha_1 = p_1^{1/(g_1 - 1)}$ and
$p_1 = 3$. This yields
\begin{equation}  
  \frac{1}{y} \frac{d \mathbf{a}_3}{d y} = \tilde{A}_3 \mathbf{a}_3, \qquad \tilde{A}_3
  = x^{g_1} A_3.
\end{equation}
The leading matrix in the last equation,
$\tilde{A}_{30} = \lim_{y \to \infty} \tilde{A}_3$, still has only one distinct
eigenvalue, and it seems as if we would not have gained anything. However, we
must not despair. Instead, we bring $\tilde{A}_{30}$ again to Jordan normal
form,
\begin{equation}  
  \mathbf{a}_4 = P_4^{-1} \mathbf{a}_3, \quad \frac{1}{y} \frac{d \mathbf{a}_4}{d y} = A_4
  \mathbf{a}_4, \quad A_4 = P_4^{-1} \tilde{A}_3 P_4,
\end{equation}
with
\begin{equation}
  P_4 = 
  \begin{pmatrix}
    0 & 0 & - \sqrt{\frac{3}{2}} \frac{3 \pi}{4 \alpha_-} \\
    1 & 0 & 0 \\
    0 & 1 & 0 \\
  \end{pmatrix},
\end{equation}
and perform a second shearing transformation,
\begin{equation}  
  \mathbf{a}_5 = P_5^{-1} \mathbf{a}_4, \quad \frac{1}{y} \frac{d \mathbf{a}_5}{d y} = 
  A_5 \mathbf{a}_5,
\end{equation}
where
\begin{equation}  
  \begin{split}
    A_5 & = P_5^{-1} A_4 P_5 - \frac{1}{y} P_5^{-1} \frac{d P_5}{d y}, \\ P_5
    & = \diag(1, y^{-g_2}, y^{-2 g_2}).
  \end{split}
\end{equation}
This time, we choose $g_2 = 1/2$, and another change of variables
$y = \alpha_2 z^{p_2}$ with $\alpha_2 = p_2^{1/(g_2 - 2)}$ and $p_2 = 2$ brings
us to
\begin{equation}  
  \frac{1}{z^2} \frac{d \mathbf{a}_5}{d z} = \tilde{A}_5 \mathbf{a}_5, \qquad
  \tilde{A}_5 = y^{g_2} A_5.
\end{equation}
Miraculously, $\tilde{A}_{50} = \lim_{z \to \infty} \tilde{A}_5$ has three
distinct eigenvalues, $0$ and $\pm 2/3^{1/4}$. Hence, we can now go ahead and
diagonalize $\tilde{A}_5$ order by order in $1/z$. The first step is to
diagonalize the leading matrix $\tilde{A}_{50}$,
\begin{equation}  
  \mathbf{a}_6 = P_6^{-1} \mathbf{a}_5, \quad \frac{1}{z^2} \frac{d \mathbf{a}_6}{d z}
  = A_6 \mathbf{a}_6, \quad A_6 = P_6^{-1} \tilde{A}_5 P_6,
\end{equation}
where
\begin{equation}
  \label{eq:P6}
  P_6 = 
\begin{pmatrix}
  -\frac{3^{1/4}}{2} & \frac{3^{1/4}}{2} & -\frac{\sqrt{3}}{4} \\
  1 & 1 & 0 \\
  0 & 0 & 1 \\
\end{pmatrix}.
\end{equation}
We move on to diagonalize the sub-leading parts of $A_6$ in two steps,
\begin{gather}  
  \mathbf{a}_7 = P_7^{-1} \mathbf{a}_6, \quad \frac{1}{z^2} \frac{d \mathbf{a}_7}{d z} = 
  A_7 \mathbf{a}_7, \\ A_7 = P_7^{-1} A_6 P_7 - \frac{1}{z^2} P_7^{-1} \frac{d P_7}{d z},
\end{gather}
and
\begin{gather}
  \label{eq:linearized_flow_8}
  \mathbf{a}_8 = P_8^{-1} \mathbf{a}_7, \quad \frac{1}{z^2} \frac{d \mathbf{a}_8}{d z} = 
  A_8 \mathbf{a}_8, \\ A_8 = P_8^{-1} A_7 P_8 - \frac{1}{z^2} P_8^{-1} \frac{d P_8}{d z}.
\end{gather}
The matrices $A_{7,8}$ and $P_{7,8}$ can be found order by order in $1/z$ by
recursion relations similar to Eq.~\eqref{eq:A2_recursion}. They take the forms
\begin{gather}
  A_7 =
  \begin{pmatrix}
    A_{711} & 0 & 0 \\
    0 & A_{722} & A_{723} \\
    0 & A_{732} & A_{733} \\
  \end{pmatrix}, \quad
  P_7 =
  \begin{pmatrix}
    0 & P_{712} & P_{713} \\
    P_{721} & 0 & 0 \\
    P_{731} & 0 &
  \end{pmatrix}, \\
  P_8 =
  \begin{pmatrix}
    1 & 0 & 0 \\
    0 & 0 & P_{823} \\
    0 & P_{832} & 0
  \end{pmatrix},
\end{gather}
and $A_8$ is diagonal. The general solution to Eq.~\eqref{eq:linearized_flow_8}
is thus
\begin{equation}
  \label{eq:asymptotic_solution_8}
  \mathbf{a}_8(z) = e^{\int_{z_0}^z d z' \, z^{\prime 2} A_8(z')} \mathbf{a}_{80}
  \sim e^{\int_{z_0}^z d z' \, z^{\prime 2} A_8^{\infty}(z')} D_8^{\infty} \mathbf{a}_{80}.
\end{equation}
In the last equation, in $A_8^{\infty}$ we keep terms in the asymptotic
expansion of $A_8$ up to order $O(1/z^3)$ so that the lowest order term in the
exponent is $O(\ln(z))$ --- this is the order to which we have to perform all
the above transformations to find the leading asymptotic
behavior. $D_8^{\infty}$ is a constant diagonal matrix which could only be found
be carrying out the above analysis \emph{exactly} (not just asymptotically)
since every order of $1/z$ in $A_8$ contributes to $D_8^{\infty}$. However, the
precise value of $D_8^{\infty}$ is not important for our purposes. From
Eq.~\eqref{eq:asymptotic_solution_8}, we can reconstruct $\mathbf{a}(x)$ by
undoing all transformations:
\begin{equation}  
  \begin{split}
    \mathbf{a}(x) & = P_1 P_2 P_3 P_4 P_5 P_6 P_7 P_8 \mathbf{a}_8(z) \\ & \sim
    P_1 P_2 P_3 P_4 P_5 P_6 P_7 P_8 e^{\int_{z_0}^z d z' \, z^{\prime 2}
      A_8^{\infty}(z')} D_8^{\infty} \mathbf{a}_{80},
  \end{split}
\end{equation}
where
\begin{equation}
  z = \left[ \frac{1}{\alpha_2} \left( \frac{x}{\alpha_1} \right)^{1/p_1} \right]^{1/p_2}.
\end{equation}
We find, for $x \to \infty$,
\begin{equation}
  \label{eq:linearized_flow_asymptotic_x}
  \begin{split}
    \delta \tilde{\varepsilon} & \sim x^{1/4} e^{4 \sqrt{x/3}}, \\ \delta
  \tilde{y} & \sim x^{3/4} e^{4 \sqrt{x/3}}, \\ \delta \tilde{T} & \sim x^{-1/4}
  e^{4 \sqrt{x/3}}.
  \end{split}
\end{equation}
As we demonstrate in Fig.~\ref{fig:asymptotic_expansion}, these asymptotic
expressions give an excellent approximation to the exact solution of the
linearized flow equation~\eqref{eq:linearized_flow} even at relatively small
values of $x$.
The corresponding asymptotic behavior of the original couplings follows from
Eq.~\eqref{eq:rescaled_couplings} and $x = 3/2 \left( 2 \ell - 1 \right) \sim 3 \ell$,
\begin{equation}
  \label{eq:asymtotic_expansion}
  \begin{split}
    \varepsilon & \sim \delta \varepsilon_{\infty} = \delta \varepsilon_{\infty,0} \ell^{5/4} e^{4
    \sqrt{\ell}}, \\  y & \sim \delta y_{\infty} = \delta y_{\infty,0} \ell^{1/4} e^{4
    \sqrt{\ell}}, \\ T & \sim \delta T_{\infty} = \delta T_{\infty,0} \ell^{-5/4} e^{4
    \sqrt{\ell}}.
  \end{split}
\end{equation}
At $T = T_c$, the couplings flow to the fixed point, and
$\delta \varepsilon_{\infty, 0} = \delta y_{\infty, 0} = \delta T_{\infty, 0} =
0$.
Close to criticality, when $t = (T - T_c)/T_c \ll 1$, we expect that the
amplitudes $\varepsilon_{\infty, 0}$, $\delta y_{\infty, 0}$, and
$\delta T_{\infty, 0}$ in Eq.~\eqref{eq:asymtotic_expansion} are proportional to
$t$. In particular, setting $\delta y_{\infty} \propto t$, the correlation
length is $\xi = a e^{\ell}$ where
\begin{multline}
  \label{eq:xi_t}
  4 \sqrt{\ell} + \frac{1}{4} \ln(\ell) \sim - \ln(t) \\ \Rightarrow \quad
  \ln(\xi/a) \sim \frac{1}{16} \ln(t) \left( \ln(t) + \ln( \abs{\ln(t)}) \right).
\end{multline}
This type of singularity is between true scaling behavior encountered at a
second order phase transition and the essential singularity of $\xi$ at the
equilibrium KT transition:
\begin{equation}
  \begin{aligned}
    & \text{true scaling:} & \frac{1}{\nu} \ell & \sim - \ln(t) & \Rightarrow & &
    \xi/a & \sim
    t^{-\nu}, \\
    & \text{equilibrium KT:} & 2 \ln(\ell) & \sim - \ln(t) & \Rightarrow & &
    \xi/a & \sim e^{C/\sqrt{t}}.
  \end{aligned}
\end{equation}
A comparison showing the good agreement between our analytical
prediction~\eqref{eq:xi_t} and the correlation length obtained from a numerical
integration of the flow equations is shown in Fig.~\ref{fig:corr_length}.

\bibliography{anisotropic_KPZ}